\RequirePackage{fix-cm}
\documentclass[natbib,smallextended]{svjour3}       
\smartqed  
\usepackage{graphicx}
\usepackage{mathptmx}      
\usepackage{amssymb}                    
\usepackage{aas_macros}
\usepackage[colorlinks=true,citecolor=blue,urlcolor=blue,breaklinks]{hyperref}
\journalname{Living Reviews in Solar Physics}
\begin{document}

\title{Flare-productive active regions}


\author{Shin Toriumi         \and
        Haimin Wang 
}


\institute{S. Toriumi \at
              Institute of Space and Astronautical Science (ISAS)/Japan Aerospace Exploration Agency (JAXA), 3-1-1 Yoshinodai, Chuo-ku, Sagamihara, Kanagawa 252-5210, Japan\\
              \email{toriumi.shin@jaxa.jp}\\
              National Astronomical Observatory of Japan, 2-21-1 Osawa, Mitaka, Tokyo 181-8588, Japan \\
           \and
           H. Wang \at
              Space Weather Research Laboratory, New Jersey Institute of Technology, University Heights, Newark, New Jersey 07102-1982, USA\\
              Big Bear Solar Observatory, New Jersey Institute of Technology, 40386 North Shore Lane, Big Bear City, California 92314-9672, USA\\
              \email{haimin.wang@njit.edu}
}

\date{Received: date / Accepted: date}

\maketitle

\begin{abstract}
Strong solar flares and coronal mass ejections,
here defined not only as the bursts of electromagnetic radiation
but as the entire process in which magnetic energy is released
through magnetic reconnection and plasma instability,
emanate from active regions (ARs)
in which high magnetic non-potentiality resides in a wide variety of forms.
This review focuses on the formation and evolution of flare-productive ARs
from both observational and theoretical points of view.
Starting from a general introduction of the genesis of ARs and solar flares,
we give an overview of the key observational features
during the long-term evolution in the pre-flare state,
the rapid changes in the magnetic field associated with the flare occurrence,
and the physical mechanisms behind these phenomena.
Our picture of flare-productive ARs is summarized as follows:
subject to the turbulent convection,
the rising magnetic flux in the interior deforms into a complex structure
and gains high non-potentiality;
as the flux appears on the surface,
an AR with large free magnetic energy and helicity is built,
which is represented by $\delta$-sunspots,
sheared polarity inversion lines, magnetic flux ropes, etc;
the flare
occurs when sufficient magnetic energy has accumulated,
and the drastic coronal evolution affects magnetic fields even in the photosphere.
We show that the improvement of observational instruments and modeling capabilities
has significantly advanced our understanding in the last decades.
Finally, we
discuss the outstanding issues and future perspective
and further broaden our scope to the possible applications of our knowledge
to space-weather forecasting,
extreme events in history,
and corresponding stellar activities.
\end{abstract}

\newpage
\setcounter{tocdepth}{3}
\tableofcontents

\section{Introduction}
\label{sec:intro}

Ever since sunspot observations with telescopes started
in the beginning of 17th century,
vast amounts of observational data have been collected.
Triggered by the momentous discovery of solar flares
by \citet{1859MNRAS..20...13C} and \citet{1859MNRAS..20...15H}
and by the report of the existence of magnetic fields in sunspots
by \citet{1908ApJ....28..315H},
the close relationship
between the production of solar flares
and the magnetism of active regions (ARs)
has been extensively argued.

Advances in ground-based and space-borne telescopes
have accelerated this trend.
In recent decades,
new instruments such as Hinode \citep{2007SoPh..243....3K},
Solar Dynamics Observatory \citep[SDO;][]{2012SoPh..275....3P},
and the Goode Solar Telescope \citep[GST;][]{2010AN....331..636C}\footnote{The GST was formerly called the New Solar Telescope (NST).}
have delivered rich observational information
and enabled us to study flares and ARs
in unprecedented detail.
Moreover,
the ever-increasing capability of numerical simulations
performed on supercomputers
has improved the advanced modeling of these phenomena
and deepened our understanding
of their physical background.

From experience
we know that there are flare-productive and flare-quiet ARs.
Then, some of the key questions are:
\begin{itemize}
\item What are the important morphological and magnetic properties
  of the flare-productive ARs that differentiate these from flare-quiet ARs?
\item What are the key observational features that are created
  during the course of large-scale, long-term AR evolution?
\item What subsurface dynamics and physical mechanisms produce
  such observed properties and features?
\item What rapid changes occur in magnetic fields during the flare eruptions?
\end{itemize}

The understanding of the flaring of ARs
is not only motivated by academic curiosity
but also desired by the practical demand of space weather forecasts
that is growing more rapidly than ever before.
Needless to say,
the flaring activity of our host star
directly affects the condition of the near-Earth environment
through emitting coronal mass ejections (CMEs),
electromagnetic radiation, and high energy particles.\footnote{This is why
a study report on the future of solar physics,
published by the Next Generation Solar Physics Mission (NGSPM)'s Science Objectives Team (SOT),
chartered by NASA, JAXA, and ESA,
cites the formation mechanism of flare-productive ARs
as one of the most important science targets. At the time of this writing, the report is available at
 {\url{https://hinode.nao.ac.jp/SOLAR-C/SOLAR-C/Documents/NGSPM_report_170731.pdf}}. Also, observation and modeling of such ARs is recognized as an important target in the International Space-weather Roadmap \citep{2015AdSpR..55.2745S}.}
As the successful detection of stellar flares and starspots
of solar-like stars
is now increasing more and more,
it is a key remaining issue
for solar physicists
to reveal the conditions of strong flare eruptions
based on the rich information
of solar ARs and flares.

Therefore,
we set as primary aim of this review article
the summary of the current understanding
of the formation and evolution of flare-productive ARs
that has been brought about through decades of effort of
observational and theoretical investigations.
For this aim,
we first highlight key observational properties of flaring ARs
during the course of long-term and large-scale evolution.
We then proceed to the theoretical studies
that try to understand the physical origins of these observed properties.
We switch our focus to the drastic evolution
during the main stage of the flare
and discuss the possibility that the changes in coronal fields
affect the photospheric conditions.
After we summarize what we have learned so far,
especially in the age with Hinode, SDO, and GST,
our discussion extends further to the possibilities
of space weather forecasting and historical data analysis
and even to the connection with stellar flares and CMEs.
Although we carefully avoid stepping into the details too much,
we provide references to excellent reviews
since the main topic of this article,
i.e., the development of flaring ARs,
is closely related to a wide spectrum of phenomena from solar dynamo,
flux emergence and AR formation to sunspots, flares and CMEs.

The rest of this article is structured as follows.
Sect.~\ref{sec:ar_sf} provides the general introduction
to the AR formation, solar flares and CMEs, and their relationships.
Sect.~\ref{sec:longterm} reviews
the key morphological and magnetic properties of flare-productive ARs
that are observed during the long-term and large-scale evolution.
Then, in Sect.~\ref{sec:num},
we show the theoretical and numerical attempts
to model and understand how these properties are created.
Sect.~\ref{sec:change} is dedicated to the discussion
on rapid changes associated with flare eruptions.
Finally, the summary and discussion are given
in Sects.~\ref{sec:summary} and \ref{sec:discussion}, respectively.

\section{Active regions and solar flares}
\label{sec:ar_sf}

\begin{figure*}
  \centering
  \includegraphics[width=1.\textwidth]{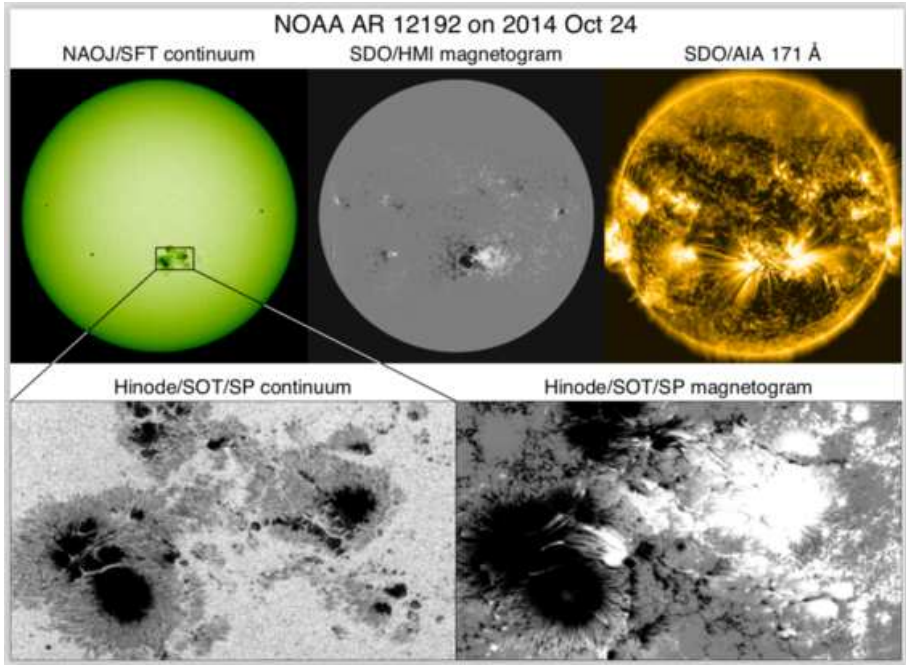}
  \caption{Huge flare-productive AR NOAA 12192.
    Images are obtained by the SDO and Hinode satellites
    as well as the Solar Flare Telescope
    in NAOJ.}
\label{fig:20141024}
\end{figure*}

Figure~\ref{fig:20141024} shows example images of the Sun.
In the southern hemisphere,
one may find a large sunspot group (top left: surrounded by a box),
in which the magnetic field is strongly concentrated
\citep[top middle: magnetogram by SDO's Helioseismic and Magnetic Imager (HMI);][]{2012SoPh..275..207S,2012SoPh..275..229S}
and the bright loop structures are clearly seen in the EUV image
\citep[top right: 171 {\AA} channel of SDO's Atmospheric Imaging Assembly (AIA);][]{2012SoPh..275...17L}.
This region, numbered 12192 by National Oceanic and Atmospheric Administration (NOAA),
appeared in October 2014 as one of the largest spot groups ever observed
with a maximum spot area of 2750 MSH\footnote{Millionths of the solar hemisphere. $1\ {\rm MSH}\sim 3\times 10^{6}\ {\rm km}^{2}$.}
and produced numerous solar flares
including six X-class events
on the Geostationary Operational Environmental Satellite (GOES) scale.
These centers of activity are called ARs
\citep[see][for the history of the definition of ARs]{2015LRSP...12....1V}.
In the simplest cases,
ARs take a form of a simple bipole structure.
However,
as the detailed observation by Hinode's Solar Optical Telescope
\citep[SOT;][]{2008SoPh..249..167T} shows,
ARs are sometimes composed of a number of magnetic elements
of various size scales (bottom panels),
and the flare productivity is known to increase
with the ``complexity'' of the ARs.

In this section,
we introduce the present knowledge
of how the ARs and sunspots are generated,
how they become unstable and produce flares and CMEs,
and how these features, i.e., the spots and flares, are related.

\subsection{Flux emergence and AR formation}
\label{subsec:fe}

It is generally thought that ARs
are created as a result of the emergence
of toroidal magnetic flux
from the deeper convection zone
\citep[flux emergence:][]{1955ApJ...121..491P,1961ApJ...133..572B}.
In most dynamo models
\citep{2010LRSP....7....3C,2017LRSP...14....4B},
the toroidal flux is generated and amplified
by turbulence and shear in the tachocline,
the thin shear layer
at the base of the solar convection zone.
There are alternative possibilities
such as the dynamo working
in the near surface shear layer
\citep{2005ApJ...625..539B}
and the amplification of advected horizontal fields
by convection \citep{2012ApJ...753L..13S}.
Magnetic flux systems created through these processes
emerge to the solar surface and eventually generate ARs.

Below we introduce the emergence processes in the interior and to the atmosphere
from both theoretical and observational viewpoints.
For more comprehensive discussion,
interested readers may also consult the review papers
by \citet{2000SoPh..192..119F}, \citet{2010LRSP....7....3C}, and \citet{2017LRSP...14....4B}
that are specialized in magnetism in the solar interior,
\citet{1985SoPh..100..397Z} and \citet{2015LRSP...12....1V} for observational properties,
and \citet{2008JGRA..113.3S04A},
\citet{2009LRSP....6....4F}, \citet{2014LRSP...11....3C}, and \citet{2014SSRv..186..227S}
that elaborate on theories and models of flux emergence.

\subsubsection{Emergence in the interior: theory}
\label{subsubsec:fe_interior_theory}

\begin{figure*}
  \centering
  \includegraphics[width=1.\textwidth]{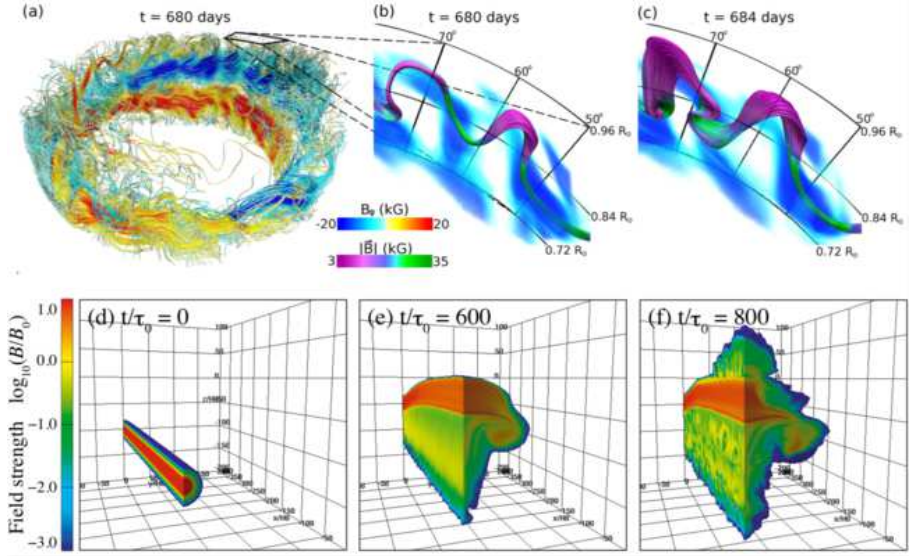}
  \caption{(a--c) Emergence of buoyant $\Omega$-loops
    from a magnetic wreath self-consistently generated in an anelastic dynamo model.
    Panels (b) and (c) demonstrate the local evolution within a domain
    extending from $0.72R_{\odot}$ ($-195$ Mm from the solar surface)
    to $0.96R_{\odot}$ ($-28$ Mm),
    with volume rendering indicating the toroidal field strength.
    {Image reproduced by permission from \citet{2013ApJ...762...73N},
    copyright by AAS.}
    (d--f) Flux emergence simulation in a single computational domain
    that seamlessly covers from the convection zone to the corona
    with a vertical extent from $-40$ to $+50$ Mm
    (here shown up to $+20$ Mm).
    The rising flux tube, initially placed at $-20\ {\rm Mm}$,
    decelerates and expands horizontally
    before it appears on the photosphere and erupts into the corona.
    Normalizing units are $H_{0}=200\ {\rm km}$ for length,
    $\tau_{0}=25\ {\rm s}$ for time,
    and $B_{\rm 0}=300\ {\rm G}$ for magnetic field strength. 
    {Image reproduced by permission from \citet{2012A&A...539A..22T},
    copyright by ESO.}
    }
\label{fig:toriumi2012}
\end{figure*}

\citet{1955ApJ...121..491P} demonstrated that a horizontal flux tube,
a horizontal bundle of magnetic field lines,
will rise due to magnetic buoyancy.
Let us assume pressure balance
between inside and outside the
thin flux tube,
\begin{eqnarray}
  p_{\rm e}=p_{\rm i}+\frac{B^{2}}{8\pi},
\end{eqnarray}
where $p_{\rm i}$ and $p_{\rm e}$ are
the pressure inside and outside the flux tube,
whose average field strength is $B$.
When the plasma is
in local thermodynamic equilibrium,
i.e., $T_{\rm e}=T_{\rm i}=T$,
the above equation can be rewritten as
\begin{eqnarray}
  \rho_{\rm e}=\rho_{\rm i}+\frac{B^{2}}{8\pi}\frac{m}{k_{\rm B}T},
\end{eqnarray}
where $\rho$ is the density, $m$ mean molecular mass,
and $k_{\rm B}$ the Boltzmann constant.
It is obvious from this equation that
the flux tube is buoyant ($\rho_{\rm i}<\rho_{\rm e}$),
and the buoyancy per unit volume is
\begin{eqnarray}
  f_{\rm B}=(\rho_{\rm e}-\rho_{\rm i})g
  =\frac{B^{2}}{8\pi}\frac{mg}{k_{\rm B}T}
  =\frac{B^{2}}{8\pi H_{\rm p}},
\end{eqnarray}
where $H_{\rm p}=k_{\rm B}T/(mg)$ is the local pressure scale height.

In most parts of the interior,
the plasma-$\beta$ ($\equiv 8\pi p/B^{2}$) is (much) greater than unity.
For a magnetic flux at the base of the convection zone
with a field strength of $10^{5}$ G,
which is 10 times stronger than the field strength
that is in equipartition with the local kinetic energy density,
the plasma-$\beta$ is of the order of $10^{5}$
\citep[e.g.,][]{2009LRSP....6....4F}.
In such a situation,
the rising flux can still be affected
by external flow fields of thermal convection.

A large number of numerical models
have been developed and revealed
various physical mechanisms of
flux emergence and
observed AR characteristics.
For example,
magnetohydrodynamic (MHD) simulations show that a horizontal magnetic layer
at the base of the convection zone in mechanical equilibrium
can break up and develop into buoyant magnetic flux tubes
through the magnetic buoyancy instability
\citep{1988JFM...196..323C,1995ApJ...448..938M,2001ApJ...546..509F}.
In order to keep the flux tube coherent,
it was suggested that the flux tube needs twist,
i.e., the azimuthal component of the magnetic field
should wrap around the tube's axis
\citep{1979cmft.book.....P,1996ApJ...464..999L,1996ApJ...472L..53M}.
\citet{2000ApJ...540..548A} found that, in 3D simulations,
the amount of twist necessary for the tube to retain its coherency
is reduced substantially comparing to the 2D limit.

The effect of the Coriolis force on the rising flux tube,
including the asymmetry between the leading and following spots
of bipolar ARs,
has been studied by simulations
with the assumption that the flux tube is thin enough
that the cross sectional evolution can be neglected
\citep[thin flux tube approximation: e.g.,][]{1981A&A....98..155S,1987ApJ...316..788C,1993ApJ...405..390F,1993A&A...272..621D,1995ApJ...441..886C}.
The emergence in the convective interior
and its interaction with the flow fields
have been considered in simulations
that apply the anelastic MHD approximation
\citep[e.g.,][]{1969JAtS...26..448G,2003ApJ...582.1206F,2008ApJ...676..680F,2009ApJ...701.1300J,2011ApJ...739L..38N,2011ApJ...741...11W,2013ApJ...762....4J}.
The top panels of Fig.~\ref{fig:toriumi2012} illustrate
the anelastic simulation by \citet{2013ApJ...762...73N},
who modeled the buoyant rise of $\Omega$-shaped loops
generated self-consistently from a bundle of toroidal flux (magnetic wreath).

However,
these assumptions become inappropriate
in the uppermost convection zone
above a depth of about $20\ {\rm Mm}$ \citep{2009LRSP....6....4F}.
This difficulty motivated \citet{2010ApJ...714..505T,2011ApJ...735..126T}
to conduct fully-compressible MHD simulations
that seamlessly connect the different atmospheric layers
from a depth of $40\ {\rm Mm}$ in the interior to the solar corona.
They found that,
as illustrated in 3D models in Fig.~\ref{fig:toriumi2012}(d--f),
the rising flux tube, starting at $-20\ {\rm Mm}$,
temporarily slows down and undergoes horizontal expansion (pancaking)
while generating escaping plasma flows
before it resumes emergence into the photosphere and beyond.
This process, termed ``two-step emergence,''
is widely observed in the larger-scale models
from the interior to the atmosphere
\citep[see Sect. 3.3.5 of][]{2014LRSP...11....3C}.
As an alternative approach,
\citet{2003ApJ...582..475A} and \citet{2017ApJ...846..149C}
joined global-scale anelastic models and local MHD simulations
from the near-surface layer upwards
and investigated fuller history of emergence.

\subsubsection{Emergence in the interior: observation}

Several attempts have been made to detect
the subsurface emerging magnetic flux
using local helioseismology \citep[see review by][]{2005LRSP....2....6G}.
One of the earliest works, \citet{1995ASPC...76..250B},
reported on the p-mode scattering
starting about two days before the spot formation
in the emerging AR NOAA 5247.
The following case studies mainly focused on the wave-speed perturbation
and subsurface flow fields before the flux appearance:
\citet{1999ApJ...526L..53C}, \citet{2001ApJ...553L.193J},
\citet{2008ApJ...672.1254K}, \citet{2008ASPC..383...59K}, \citet{2008SoPh..251..369Z},
and \citet{2009SSRv..144..175K}.
However, in most cases,
it was difficult to detect significant seismic signatures
associated with the emerging flux,
probably because of
the fast rising motion and accordingly short observation time,
which leads to low signal-to-noise ratio.

A recent observation by \citet{2011Sci...333..993I}, however,
detected strong seismic perturbations in NOAA 10488 at depths between 42 and 75 Mm,
up to two days before the photospheric flux reaches its maximum flux growth rate.
The estimated rising speed from 65 Mm to the surface is about $0.6\ {\rm km\ s}^{-1}$
\citep[see also][]{2012Sci...336..296B,2013ApJ...777..138I,2013SoPh..287..229K,2018vsss.book...15K}.
Statistical studies by \citet{2009SoPh..258...13K,2011SoPh..268..407K,2012SoPh..277..205K}
showed indications of upflows, rotations, and increased vorticity
in the subsurface layer.
\citet{2013ApJ...762..130L}, \citet{2013ApJ...762..131B}, and \citet{2014ApJ...786...19B}
analyzed more than 100 emerging regions and found that
there are statistically significant seismic signatures
in average subsurface flows and the apparent wave speed,
at least one day prior to the emergence,
although their individual samples did not show discernible signal
greater than the noise level.

Other possible precursors
of flux emergence on the surface
are the reduction in acoustic oscillation power
\citep{2011SoPh..268..321H,2013ApJ...770L..11T},
f-mode amplification \citep{2016ApJ...832..120S},
and horizontal divergent flows
\citep{2012ApJ...751..154T,2014ApJ...794...19T}.

\subsubsection{Birth of ARs: observation}

\begin{figure*}
  \centering
  \includegraphics[width=1.\textwidth]{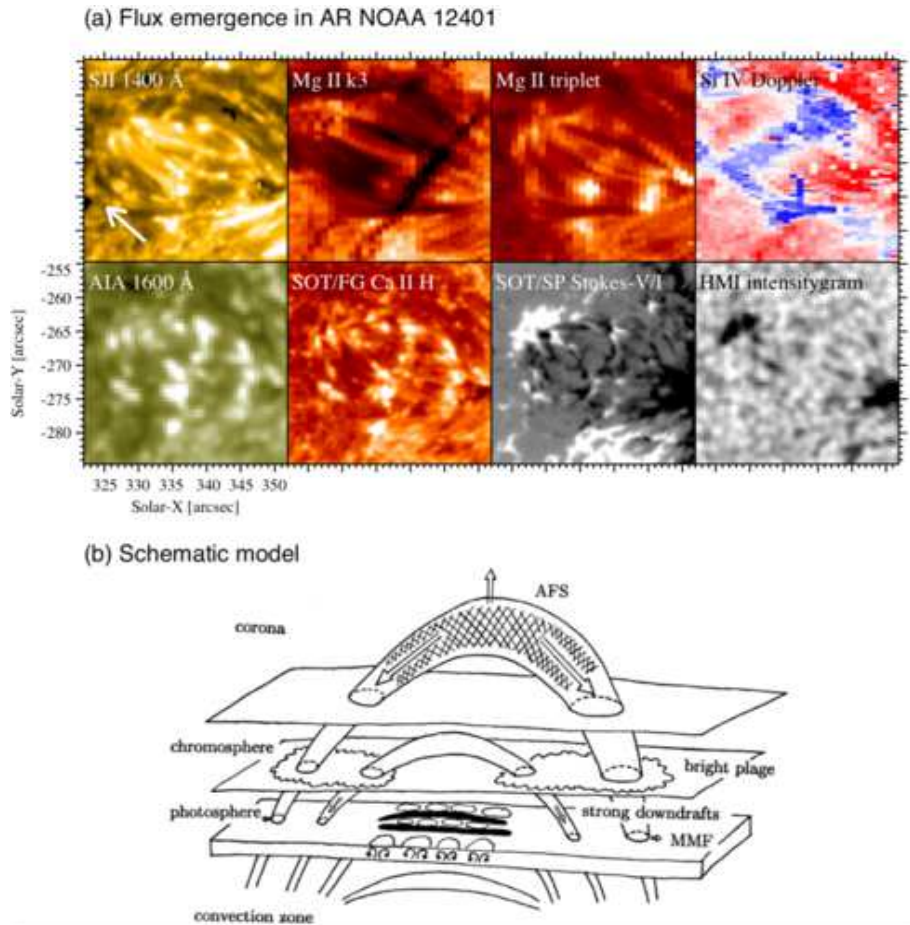}
  \caption{(a) ``Textbook'' flux emergence
    in AR NOAA 12401
    observed simultaneously by Hinode,
    the Interface Region Imaging Spectrograph \citep[IRIS;][]{2014SoPh..289.2733D},
    and SDO (2015 August 19).
    From top left to bottom right are the IRIS slit-jaw image of 1400 {\AA},
    raster-scan intensitygram at the Mg\,{\sc ii} k line core
    (k3: 2796 {\AA}),
    intensitygram at the Mg\,{\sc ii} triplet line (2798 {\AA}),
    Dopplergram produced from the Si\,{\sc iv} 1403 {\AA} spectrum
    (blue, white, and red correspond to
    $-10$, 0, and $+40\ {\rm km\ s}^{-1}$, respectively),
    SDO/AIA 1600 {\AA}, Hinode/SOT/FG Ca\,{\sc ii} H, SOT/SP Stokes-V/I,
    and SDO/HMI intensitygram.
    The white arrow in the top left panel indicates
    the direction of the disk center.
    In the accompanying movie,
    the Ca\,{\sc ii} H and Stokes-V/I maps
    are replaced by the AIA 1700 {\AA} image and HMI magnetogram,
    respectively. 
    (For movie see Electronic Supplementary Material.) 
    {Image and movie reproduced by permission from \citet{2017ApJ...836...63T},
    copyright by AAS.}
    (b) Schematic model of flux emergence.
    {Image reproduced by permission from \citet{1989ApJ...345..584S},
    copyright by AAS.}
    The original version of this illustration appeared in Shibata's review note in 1979.
    }
\label{fig:shibata1989}
\end{figure*}

As the rising magnetic flux reaches the photosphere,
it starts to build up an AR
if the flux is sufficiently large.
Figure~\ref{fig:shibata1989}(a) and its accompanying movie
show various aspects of a newly emerging flux region.
In a magnetogram (Stokes-V/I map),
the emerging flux is scattered throughout the region
as a number of small-scale magnetic elements
of positive and negative polarities.
These elements merge with and cancel each other
in the middle of the region
and gradually form pores and,
if the emerged flux is sufficient,
they eventually create sunspots
\citep{1978SoPh...60..213Z}.
\citet{1985SoPh..100..397Z} introduced the hierarchy
of magnetic elements.
Sunspots with a flux of
$5\times 10^{20}\ {\rm Mx}$ or more have a penumbra
and the umbral field is $2900$--$3300\ {\rm G}$,
sometimes exceeding $4000\ {\rm G}$,
while the flux of pores is
$2.5\times 10^{19}$--$5\times 10^{20}\ {\rm Mx}$
and the field strength is $\sim 2000\ {\rm G}$.
If the flux is less than $10^{20}\ {\rm Mx}$,
the emerging regions do not develop
beyond ephemeral regions
\citep{1973SoPh...32..389H}.

From the observation of repeated emergence and cancellation
of photospheric magnetic elements,
\citet{1996A&A...306..947S} and \citet{1999ApJ...527..435S} suggested that
this behavior is due to the rising
of undulatory (sea-serpent) field lines.
\citet{2002ApJ...575..506G}, \citet{2002SoPh..209..119B}, and \citet{2004ApJ...614.1099P}
suggested that Ellerman bombs,
the bursty intensity enhancements
in H$\alpha$ line wings
\citep{1917ApJ....46..298E},
are located at the dipped parts,
at which magnetic reconnection takes place
to disconnect emerged flux from un-emerged, mass-laden parts of the flux tube
(resistive emergence model).
UV bursts in the transition region lines
are similarly found at the cancellation sites
\citep{2014Sci...346C.315P,2018SSRv..214..120Y}.
Brightenings seen in 1400 {\AA}, 1600 {\AA}, and Ca\,{\sc ii} H
of Fig.~\ref{fig:shibata1989}(a)
correspond to Ellerman bombs and UV bursts.

Soon after the magnetic flux shows up,
an arch filament system (AFS) appears
as parallel dark fibrils,
probably the manifestation
of rising magnetic fields
\citep[][see Mg\,{\sc ii} k3 image of Fig.~\ref{fig:shibata1989}(a)]{1967SoPh....2..451B,1969SoPh....8...29B}.
Bipolar plages are observed
in the chromospheric Ca\,{\sc ii} H and K lines
at the footpoints of the AFS
\citep[][brightenings above the pores in Fig.~\ref{fig:shibata1989}(a)]{1976SoPh...46..125K}.
The Hinode analysis of AFS by \citet{2007PASJ...59S.649O,2010PASJ...62..893O}
shows the horizontal expansion and upward acceleration of emerging flux,
which strongly supports the ``two-step emergence'' scenario
(Sect.~\ref{subsubsec:fe_interior_theory}).
The observational characteristics of emerging flux regions
are schematically summarized by \citet{1989ApJ...345..584S}
as an illustration in Fig.~\ref{fig:shibata1989}(b).

\subsubsection{Birth of ARs: theory}
\label{subsubsec:fe_photosphere_theory}

The MHD modeling of flux emergence
from the photospheric layer to the corona
was pioneered by \citet{1989ApJ...345..584S},
who simulated the 2D emergence
due to the Parker instability,
the undular mode of the magnetic buoyancy instability
\citep{1979cmft.book.....P}.
They successfully reproduced
the observed dynamical features
such as rising motion of the AFS
and the strong downflow along the field lines.
Since then,
the flux emergence process has been widely studied
both in 2D and 3D
\citep[e.g.,][]{1990ApJ...351L..25S,1990ApJ...358..698K,1992ApJS...78..267N,2001ApJ...549..608M,1992PASJ...44..167M,1993ApJ...414..357M,2001ApJ...554L.111F,2001ApJ...559L..55M,2004A&A...426.1047A,2005Natur.434..478I,2006A&A...460..909M}.

\begin{figure*}
  \centering
  \includegraphics[width=1.\textwidth]{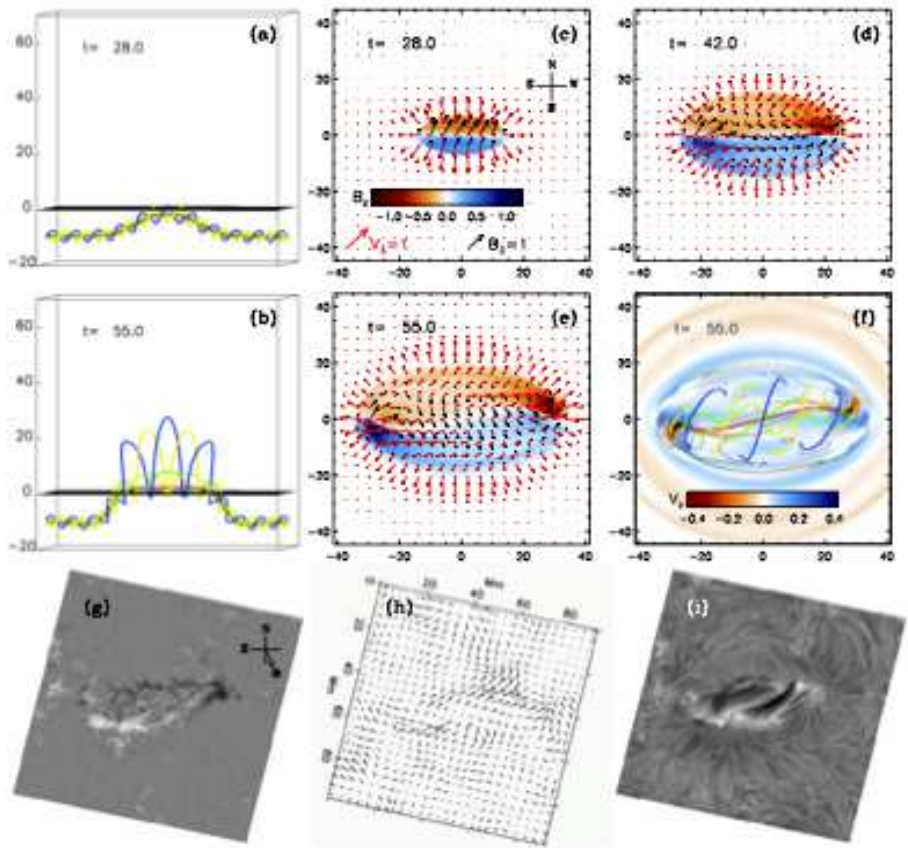}
  \caption{3D flux emergence simulation from around the photospheric height.
    (a and b) Selected field lines of the emerging flux tube.
    (c--e) Vertical magnetic field $B_{z}$, the horizontal magnetic field (black arrows),
    and the horizontal velocity field (red arrows).
    (f) Top-down view of panel (b) with vertical velocity $v_{z}$.
    (g--i) Line-of-sight (LOS) magnetic field, horizontal velocity,
    and H$\alpha$ image of NOAA AR 5617, respectively.
    {Image reproduced by permission from \citet{2001ApJ...554L.111F},
    copyright by AAS.}
    }
\label{fig:fan2001}
\end{figure*}

Figure~\ref{fig:fan2001} shows a typical example
of flux emergence simulations by \citet{2001ApJ...554L.111F},
which models the buoyant rise of a twisted flux tube
from just beneath the photosphere ($-1.5\ {\rm Mm}$) and upwards.
The initial flux tube, which is horizontal and endowed with a density deficit
at the middle with respect to the surroundings,
starts rising due to the magnetic buoyancy and deforms into an $\Omega$-loop (panel a).
As the flux tube penetrates into the upper atmosphere,
a ying-yang pattern of positive and negative polarities (vertical field $B_{z}$)
is produced in the photosphere (panels c--e),
which resembles the polarity layout in the actual AR (panel g).
Due to the initial twist,
magnetic field lines in the atmosphere show a twisted structure,
which also mimics the observed helical nature of the AFS (panel i).

\citet{1984SoPh...94..315F} and \citet{1995Natur.375...42Y,1996PASJ...48..353Y}
investigated the interaction 
between emerging flux and the preexisting coronal loop
\citep[the model proposed by][]{1977ApJ...216..123H}
and successfully reproduced jet ejections
\citep[see also][]{2003ApJ...593L.133M,2008ApJ...673L.211M,2008ApJ...683L..83N,2009A&A...494..329M,2010A&A...512L...2A,2013PASJ...65...62T,2013ApJ...771...20M}.
Magnetic flux cancellation at the emerging undular fields
and the resultant production of Ellerman bombs
were modeled by \citet{2007ApJ...657L..53I} in 2D
and \citet{2009A&A...508.1469A} in 3D.

With the growing ability of computation resources,
simulations have become more realistic
and now take into account
the effect of thermal convection
on flux emergence.
For instance,
\citet{2008ApJ...687.1373C} performed 3D radiative MHD simulations
of the emergence of an initially horizontal flux tube
in the granular convection.
They found that,
due to vigorous convective flows
at the top of the convection zone,
the rising tube is highly structured
by the surface granulation pattern,
which is well in agreement with the Hinode/SOT observations.
The series of numerical simulations
of similar setups
consistently showed
that the granular cells are expanded and elongated
as the horizontal flux approaches
and that the surface convection makes
undular field lines
(dipped field at the downflow lanes),
which reconnect with each other
and drain down the plasma from the surface layer
\citep{2007ApJ...665.1469A,2007A&A...467..703C,2008ApJ...679L..57I,2008ApJ...679..871M,2009ApJ...702..129M,2009A&A...507..949T,2010ApJ...714.1649F}.
The realistic modeling
by \citet{2014ApJ...788L...2A} and \citet{2017ApJ...839...22H}
successfully reproduced the small-scale reconnection events
at the dipped fields
and showed that they can be observed
as Ellerman bombs or UV bursts
depending on the reconnection heights.
Throughout these processes,
the magnetic elements grow larger
and, eventually, the sunspots are formed
\citep{2010ApJ...720..233C,2014ApJ...785...90R}.

\subsection{Solar flares and CMEs}
\label{subsec:flares}

\begin{figure*}
  \centering
  \includegraphics[width=0.75\textwidth]{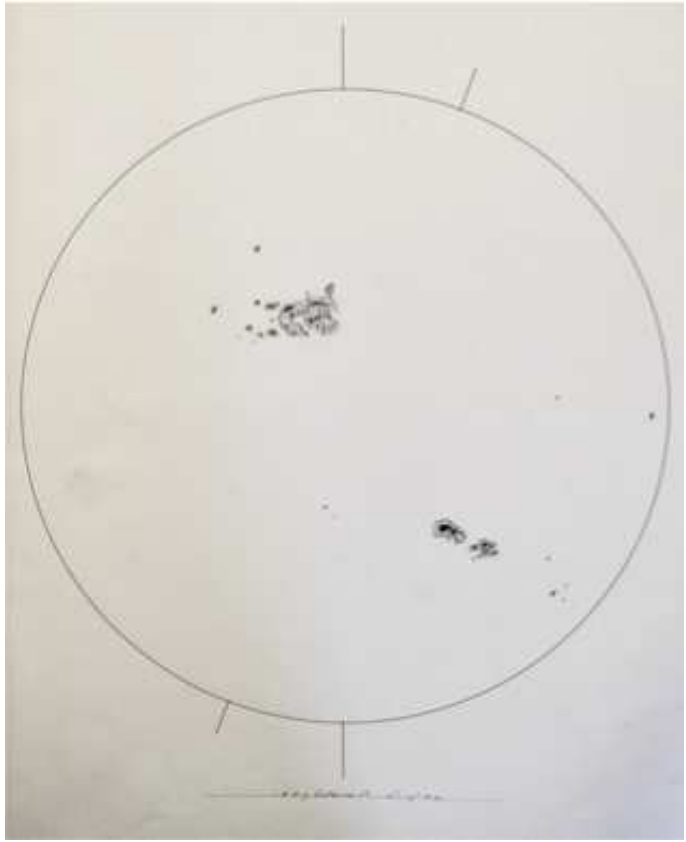}
  \caption{Carrington's original whole-disk drawing on 1859 September 1.
    \citet{1859MNRAS..20...13C} and \citet{1859MNRAS..20...15H} observed the white light flare
    in the large sunspot region in the northern hemisphere.
    This manuscript is currently preserved in the archive
    of the Royal Astronomical Society (RAS)
    as RAS MSS Carrington 3.2:
    Drawings of sunspots, showing the whole of the Sun's disk,
    v.2, f.313a.
    For a better visualization, the thickness of the limb and axes is enhanced.
    {Image reproduced by permission from
    \citet{2018ApJ...869...57H},
    copyright by AAS and RAS.}
    }
\label{fig:hayakawa2018_a}
\end{figure*}

In most astronomical contexts,
the term ``flare'' refers to the abrupt increase in intensity of electromagnetic waves,
and the flares on the Sun are detected over a wide range of spectrum
such as X-rays, (E)UV, radio, and even white light.
In fact, the discovery of flares was made
as a remarkable intensity enhancement in white light
\citep[Carrington event on 1859 September 1;][]{1859MNRAS..20...13C,1859MNRAS..20...15H}.
Figure~\ref{fig:hayakawa2018_a} is the original whole-disk drawing by Carrington,
which shows a large spot group that produced the strong white light flare.
Nowadays, flare strengths are grouped by peak soft X-ray flux
over 1--8 {\AA}, measured by GOES,
into logarithmic classes A, B, C, M, X,
corresponding to $10^{-8}$, $10^{-7}$, $10^{-6}$, $10^{-5}$, $10^{-4}\ {\rm W\ m}^{-2}$ at Earth,
respectively,
so X1.2 and M3.4 represent $1.2\times 10^{-4}\ {\rm W\ m}^{-2}$
and $3.4\times 10^{-5}\ {\rm W\ m}^{-2}$, respectively.
The Carrington flare is arguably considered as the most powerful event ever
with the estimated magnitude of X45 ($\pm 5$)
and bolometric energy of $5\times 10^{32}\ {\rm erg}$
\citep{2003JGRA..108.1268T,2004SoPh..224..407C,2006AdSpR..38..159B,2013JSWSC...3A..31C}.

\begin{figure*}
  \centering
  \includegraphics[width=1.\textwidth]{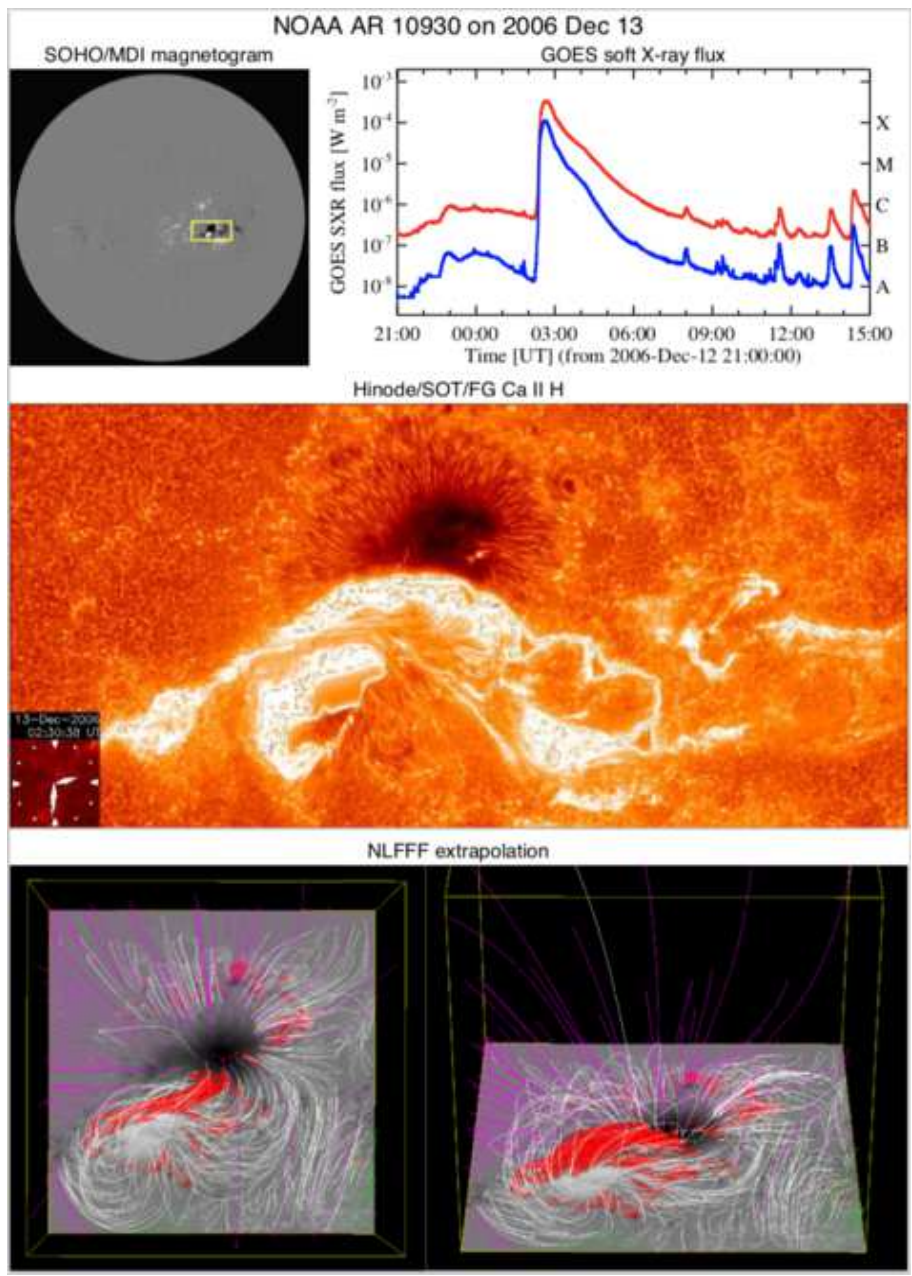}
  \caption{X3.4-class flare in AR NOAA 10930.
    The panels show full-disk magnetogram from Michelson Doppler Imager (MDI)
    aboard the Solar and Heliospheric Observatory (SOHO),
    GOES soft X-ray light curves for 1--8 {\AA} (red) and 0.5--4.0 {\AA} (blue),
    and Hinode/SOT/FG Ca\,{\sc ii} H image (see also the accompanying movie),
    whose FOV is indicated by a yellow box in the magnetogram.
    Hinode image courtesy of Joten Okamoto (ISAS/JAXA and NAOJ).
    The bottom panel displays the computationally extrapolated magnetic field lines
    before the X3.4 flare using the NLFFF method.
    The red isosurface shows where the electric current is highest.
    {Image reproduced by permission from \citet{2008ApJ...675.1637S},
    copyright by AAS.}
    }
\label{fig:20061213}
\end{figure*}

Solar flares are now considered as the conversion process
of (free) magnetic energy to kinetic and thermal energy as well as particle acceleration,
most probably through magnetic reconnection.
Figure~\ref{fig:20061213} shows the GOES X3.4-class flare in AR NOAA 10930.
From this figure and the corresponding movie,
one may find that
the flare occurs between the two major sunspots,
particularly at the polarity inversion line (PIL: also called the neutral line),
where the vertical field $B_{z}$
or the line-of-sight (LOS) field $B_{\rm LOS}$ remains zero
and the sign flips across it.
The most pronounced feature is the pair of flare ribbons
that spreads along and away from the PIL
\citep{1964ApJ...140..746B,2004ApJ...611..557A}.
The magnetic field in the corona,
which is computationally extrapolated from the photospheric magnetogram
using the non-linear force-free field (NLFFF) method
(Sect.~\ref{subsubsec:num_data_extrapolation}),
shows a helical topology above the PIL.
Such a highly non-potential, twisted magnetic structure called a magnetic flux rope
is often observed in soft X-rays prior to the flare occurrence
(see Sect.~\ref{subsubsec:sigmoids}).

\begin{figure*}
  \centering
  \includegraphics[width=0.6\textwidth]{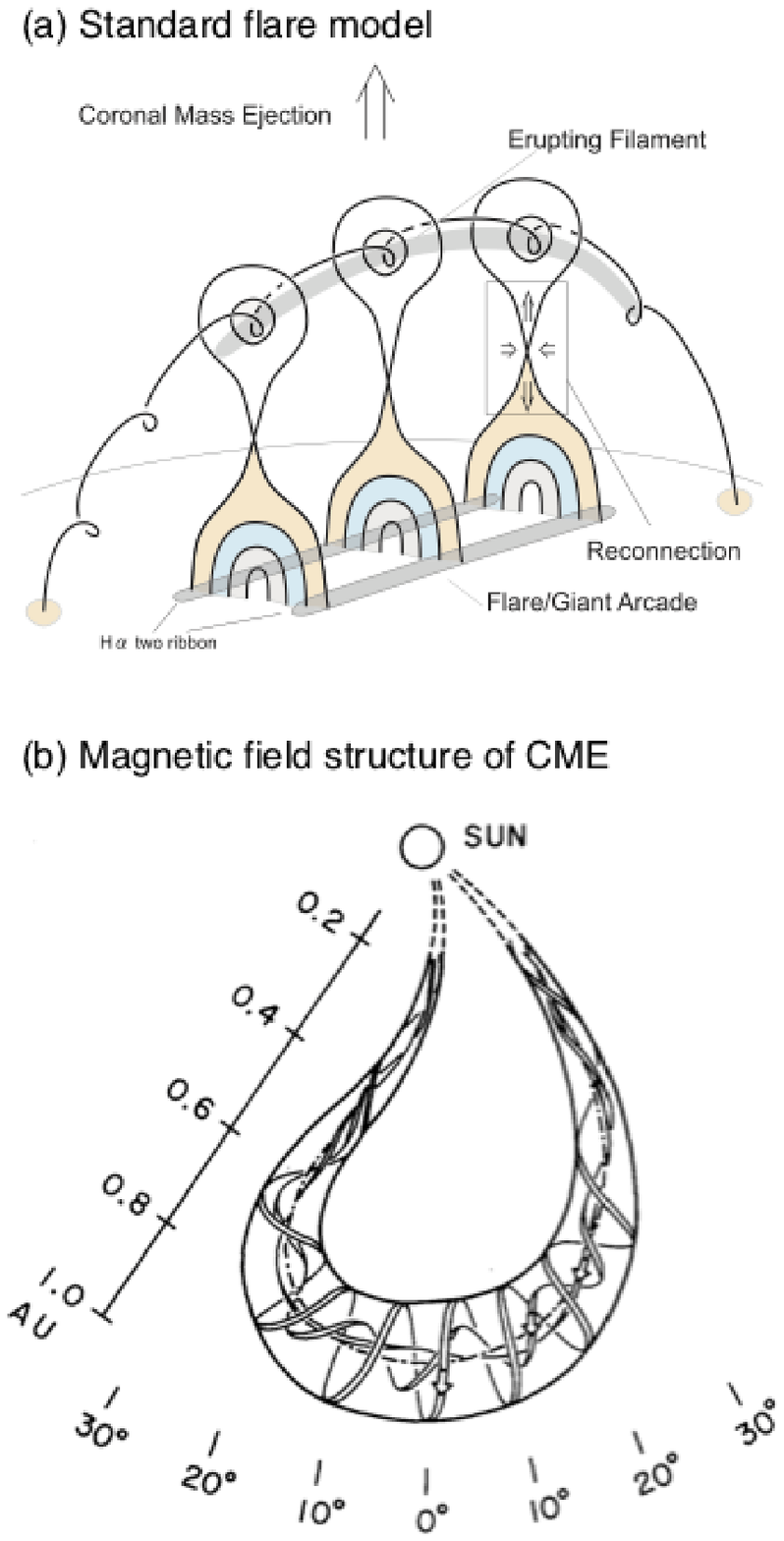}
  \caption{(a) Schematic illustration of the standard flare model.
    {Image reproduced by permission from \citet{2005ApJ...634..663S},
    copyright by AAS.}
    The thick solid lines represent magnetic field lines.
    Shaded, hatched, and dotted regions display
    the features observed in soft X-rays, EUV, and H$\alpha$, respectively.
    (b) Observationally inferred magnetic field structure of CMEs
    in the interplanetary space.
    {Image reproduced by permission from \citet{1989SSRv...51..197M},
    copyright by ***.}
    }
\label{fig:cshkp}
\end{figure*}

Various observational characteristics of the flares,
not only the ribbons and the flux rope
but also the cusp-shaped loops seen in soft X-rays \citep{1992PASJ...44L..63T},
hard X-ray loop-top source \citep{1994Natur.371..495M},
inflows toward a current sheet \citep{2001ApJ...546L..69Y},
etc.,
altogether lend support to the well-established flare model
based on the magnetic reconnection scenario,
referred to as the standard model,
or the CSHKP model after its major contributors
\citep[][see Fig.~\ref{fig:cshkp}(a)]{1964NASSP..50..451C,1966Natur.211..695S,1974SoPh...34..323H,1976SoPh...50...85K}.
In this paradigm and its updated versions
\citep[e.g.,][]{1986ApJ...302L..67F,1995ApJ...451L..83S,2012A&A...543A.110A,2013A&A...555A..77J},
the key features are explained as follows.
The magnetic flux rope becomes unstable and erupts into the higher atmosphere,
entraining the overlying coronal field.
The legs of the coronal field are drawn
into a current sheet underneath the flux rope as inflows
and reconnect with each other.
The outflows from the reconnection region further boost the flux rope eruption.
The post-reconnection field lines form a cusp structure,
while the accelerated electrons from the reconnection site
precipitate along the field lines and heat the chromosphere
to produce flare ribbons.

The flux rope, if ejected successfully,
expands and develops into the magnetic skeleton of a CME
that travels through interplanetary space.
This is well demonstrated by {\it in-situ} observations of magnetic fields
at vantage points, e.g., in front of the Earth \citep{1981JGR....86.6673B,1982JGR....87..613K,1986AdSpR...6..335M}.
Fig.~\ref{fig:cshkp}(b) shows a schematic illustration of the inferred topology.
The helical nature of the magnetic field of the CMEs
is strongly suggestive of their solar origins.

Regarding the onset of flux rope eruption
and subsequent ejection of CMEs,
various theories have extensively been proposed and investigated,
such as flux emergence \citep{1977ApJ...216..123H},
breakout \citep{1999ApJ...510..485A,2008ApJ...680..740D},
tether-cutting \citep{2001ApJ...552..833M},
emerging-flux trigger \citep{2000ApJ...545..524C},
kink instability \citep{2005ApJ...630L..97T,2007ApJ...668.1232F},
and torus instability \citep{2006PhRvL..96y5002K},
along with a more recent concept of the double-arc instability
\citep{2017ApJ...843..101I}.
In any case, there appears to be a consensus, at least, that
the flare/CME occurrence is caused through the dynamical coupling
between the unstable eruption of a flux rope (ideal MHD process)
and magnetic reconnection of surrounding arcades (resistive MHD process).

It should be noted, however, that not all the stronger flares
are accompanied by CMEs \citep[e.g.,][]{2006ApJ...650L.143Y}.
The best example is the giant AR NOAA 12192 (Fig.~\ref{fig:20141024}).
Throughout the disk passage,
this AR produced numerous energetic flares including the six X-class ones,
but surprisingly none of them were CME-eruptive.
\citet{2015ApJ...804L..28S} showed that in this AR,
the decay index $n=-\partial \ln{B_{\rm h}}/\partial \ln{z}$,
which measures the decreasing rate
of the horizontal magnetic field $B_{\rm h}$ with height $z$,
remains below the critical value $n_{\rm c}\approx 1.5$
for the torus instability
until a large altitude
and thus only failed eruptions took place
\citep{2016ApJ...818..168I,2016NatCo...711522J,2018Natur.554..211A}.
The confinement of flux rope eruption by strong overlying field
is also shown by the statistical studies on a number of ARs
\citep{2017ApJ...843L...9W,2018ApJ...860...58V,2018ApJ...864..138J}.
The same mechanism explains the observed result by \citet{2017ApJ...834...56T}
that the ratio of reconnected flux (in the flare ribbons) to the total AR flux
is, on average, smaller for failed events than eruptive cases.
\citet{2018ApJ...861..131D} showed that X-class flares
located near coronal fields that are open to the heliosphere
are eruptive at a higher rate than those lacking access to open fields.

The topics we have discussed above are only
the most representative aspects of the flares and CMEs.
In order to keep our primary focus on
the formation and evolution of flare-productive ARs, however,
we stop the discussion at this point
and yield the rest to reviews by, e.g.,
\citet{2009AdSpR..43..739S}, \citet{2011SSRv..159...19F}, and \citet{2017LRSP...14....2B}
for observational overviews
and \citet{2002A&ARv..10..313P}, \citet{2006SSRv..123..251F}, \citet{2011LRSP....8....1C}, \citet{2011LRSP....8....6S}, and \citet{2015SoPh..290.3425J}
for theoretical and modeling aspects.

\subsection{Categorizations of sunspots and flare productivity}
\label{subsec:spots}

\begin{figure*}
  \centering
  \includegraphics[width=1.\textwidth]{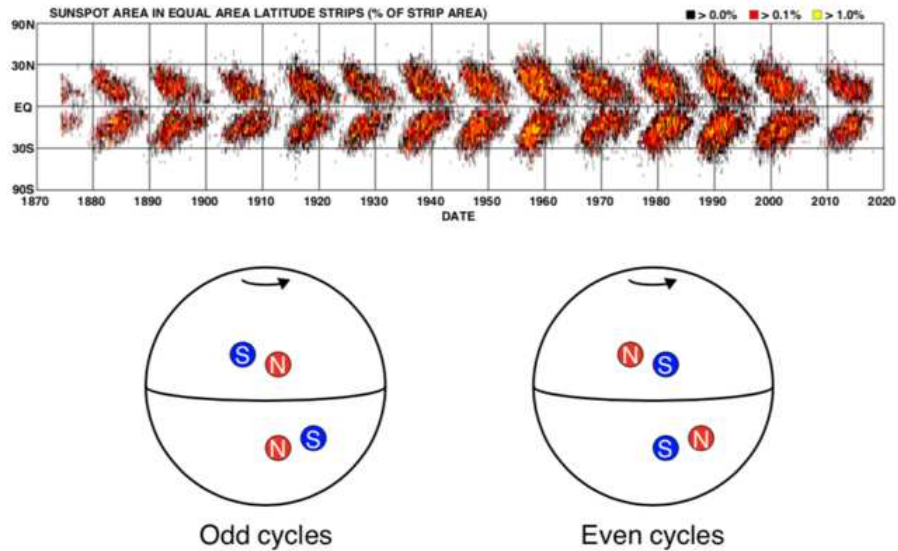}
  \caption{(Top) Sunspot butterfly diagram
    showing the total spot area as a function of time and latitude.
    {Image courtesy of Hathaway.}
    In each cycle, the latitudes of ARs shifts to the equator
    (Sp\"{o}rer's law).
    (Bottom) Schematic diagram showing the polarity alignments.
    The preceding spots appear closer to the equator
    than the following spots (Joy's rule).
    In each cycle,
    the preceding polarities on one hemisphere are the same
    and are opposite to those on the other hemisphere,
    and the order of the polarities
    reverses in the successive cycle
    (Hale-Nicholson rule).
    These are merely the overall trends
    and there exist many exceptional ARs.
  }
\label{fig:butterfly}
\end{figure*}

The number of sunspots varies with the 11 year solar activity cycle
\citep{1843AN.....20..283S,2015LRSP...12....4H}.
Early in a cycle,
the spots appear in higher latitudes
up to $40^{\circ}$
and, throughout the cycle,
the latitude gradually drifts
lower to the equator
\citep[Sp\"{o}rer's law:][]{1858MNRAS..19....1C}.
This behavior is illustrated
by the Maunder butterfly diagram
(Fig.~\ref{fig:butterfly} top).
In each bipolar AR,
the preceding spot tends to appear
closer to the equator
than the following spot
\citep[Joy's rule:][]{1919ApJ....49..153H}.
As the magnetic observation started
in the beginning of 20th century
\citep{1908ApJ....28..315H},
Hale's polarity rule was discovered:
for each cycle,
the bipolar ARs are aligned
in the east-west orientation
with opposite preceding magnetic polarities
on the opposite hemispheres.
Soon, they also noticed that
the polarities of the preceding spots
alternate between successive cycles
and these features are now altogether
called Hale-Nicholson rule
\citep[Fig.~\ref{fig:butterfly} bottom:][]{1925ApJ....62..270H}.

\begin{figure*}
  \centering
  \includegraphics[width=1.\textwidth]{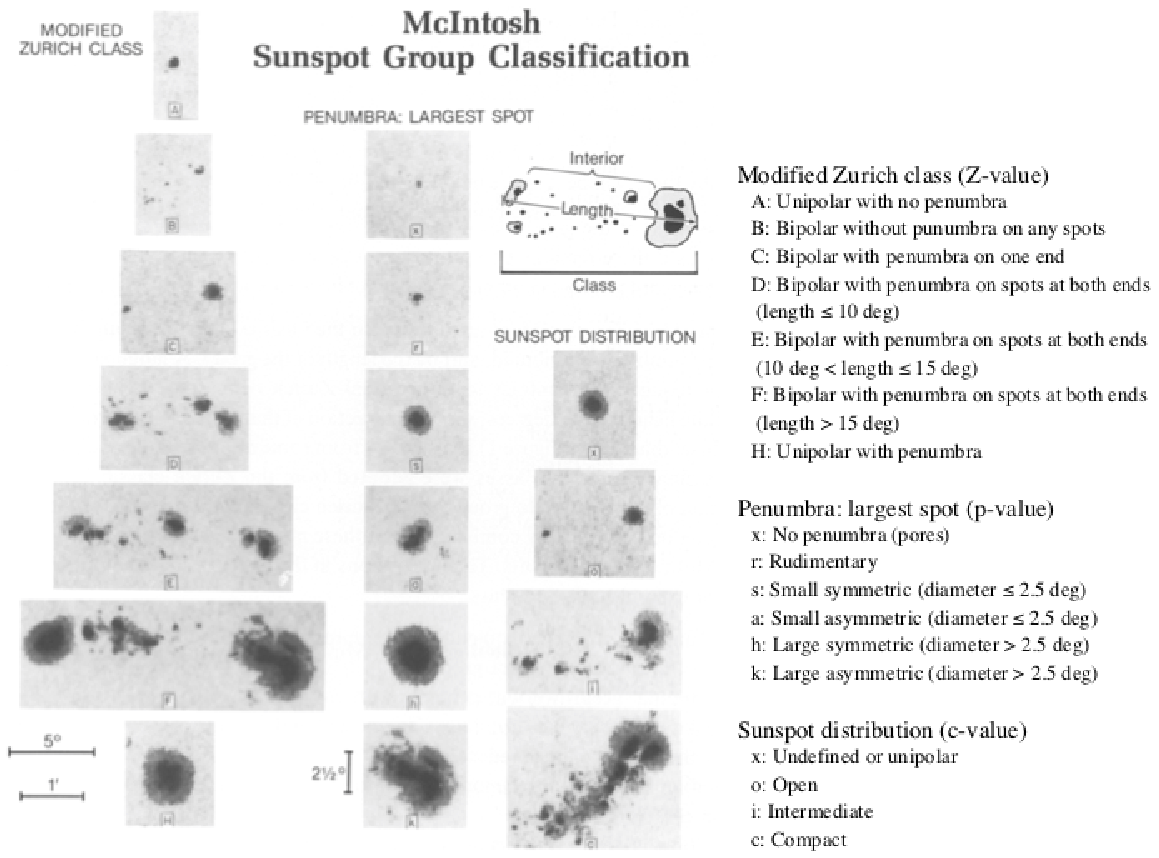}
  \caption{Example spot images for the three indices
    of McIntosh classification.
    {Image reproduced by permission from \citet{1990SoPh..125..251M},
    copyright by ***.}}
\label{fig:mcintosh1990}
\end{figure*}

Along with such long-term characteristics,
which impose strong constraints
on dynamo models,
the structure of each sunspot group
is also recognized as an important factor
\citep[see reviews by][]{2003A&ARv..11..153S,2011LRSP....8....4B}.
One method of categorizing the sunspots
is the Zurich classification
\citep{1901ApJ....13..260C,1938ZA.....16..276W},
which was further developed
as the McIntosh classification
\citep{1990SoPh..125..251M}.
The McIntosh classification uses
three letters to describe
the white-light properties of the spots,
which are the size,
penumbral type,
and distribution (see Fig.~\ref{fig:mcintosh1990}).
The combination of the three letters shows
the {\it morphological complexity} of ARs and,
according to \citet{1994SoPh..150..127B},
the flare production rate increases
along the diagonal line
in the 3D parameter space
from the simplest corner ``A/B/Hxx''
to the most complex end ``Fkc''.
Other studies show essentially a consistent result:
morphologically complex ARs produce more flares
\citep[e.g.,][]{1987Ap&SS.129..203A,2002SoPh..209..171G,2006AN....327...36T,2011SoPh..269..111N,2012SoPh..281..639L,2016SoPh..291.1711M}.
The primary advantage of this method is that
the spots are categorized simply from the white light observation
and thus it requires no magnetic measurement.\footnote{\citet{1990SoPh..125..251M} mentioned that {\it ``[r]arely will the measured magnetic class conflict with''} his definitions of unipolar and bipolar groups.}

\begin{figure*}
  \centering
  \includegraphics[width=1.\textwidth]{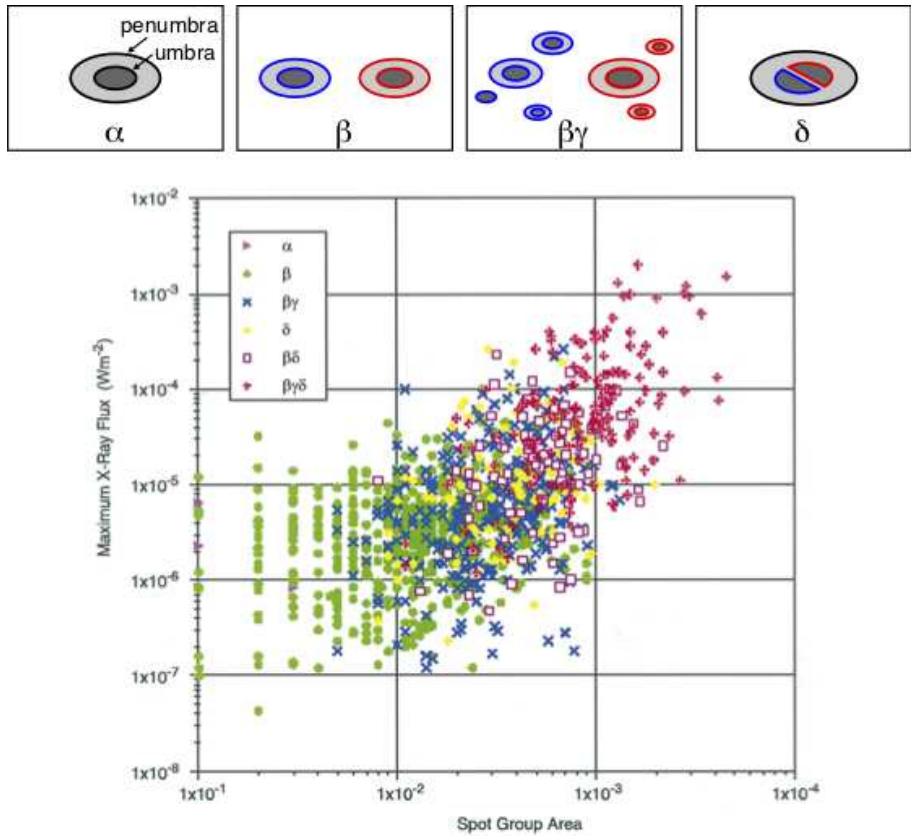}
  \caption{(Top) Sample diagrams of the Mount Wilson classification.
    (Bottom) Peak flare magnitudes as a function of
    maximum sunspot area.
    {Image reproduced by permission from \citet{2000ApJ...540..583S},
    copyright by AAS.}
    Note that the tick marks of the horizontal axis
    should be corrected as, from left to right,
    $1\times 10^{-5}$, $1\times 10^{-4}$, $1\times 10^{-3}$, and $1\times 10^{-2}$
    in the unit of the hemisphere,
    or equivalently, 10, 100, 1000, and 10,000 MSH.}
\label{fig:sammis2000}
\end{figure*}

Another categorization method is
the Mount Wilson classification,
which refers to the {\it magnetic} structures of ARs.
The original scheme of this method
has the following three identifiers
\citep[Fig.~\ref{fig:sammis2000} top:][]{1919ApJ....49..153H,1938mosn.book.....H}:
\begin{itemize}
\item $\alpha$, a unipolar spot group;
\item $\beta$, a simple bipolar spot group
  of both positive and negative polarities; and
\item $\gamma$, a complex spot group in which spots of both polarities
  are distributed so irregularly as to prevent classification as a $\beta$ group.
\end{itemize}
Often more than one identifier is appended to each AR
to indicate even more complex structures,
such as $\beta\gamma$, a bipolar spot group
which is so complex that preceding or following spots are accompanied by minor polarities.
It was shown that the flare productivity is related to this categorization.
\citet{1939ApJ....89..555G} found that the probability of the flare eruption
is proportional to the spot area
and it increases with the spot complexity
(in the order of $\alpha$, $\beta$, $\beta\gamma$, and $\gamma$).
Consistent results were reported by
\citet{1953BAICz...4....9K}, \citet{1959SCoA....3...25B},
and \citet{1963MNRAS.126..123G}.

Later, the $\delta$ group,
a spot group in which umbrae of opposite polarities
are separated by less than 2$^{\circ}$
and situated within the common penumbra,
was added to the Mount Wilson classification
by \citet{1960AN....285..271K,1965AN....288..177K}.
In this scheme,
the most complex ARs are
the spots appended with $\beta\gamma\delta$.
Ever since \citet{1960AN....285..271K} showed that
the $\delta$-spots are highly flare-productive,
a number of statistical investigations have been carried out
and showed consistent results
\citep[e.g.][]{1985SoPh...96..293M,2000ApJ...540..583S,2002SoPh..209..361T,2006AN....327...36T,2014MNRAS.441.2208G,2017ApJ...834...56T,2017ApJ...834..150Y}.
The bottom panel of Fig.~\ref{fig:sammis2000} is a diagram
of the peak GOES soft X-ray flux
vs. the maximum sunspot area
for various ARs
by \citet{2000ApJ...540..583S}.
Here, one may easily find
the clear positive correlation that
the flare magnitude increases with the spot area.
However, this diagram also shows that
more complex regions produce stronger flares.
For example,
all $\geq$X4-class flares occur in ARs
of area greater than 1000 MSH
and classified as the most complex $\beta\gamma\delta$.
Other studies show the correlations and associations
between the $\delta$-spots and the production of proton flares
\citep[here meaning that flares that emit energetic protons:][]{1966ApJ...145..215W,1970P&SS...18...33S},
white-light flares
\citep{1983SoPh...88..275N},
$\gamma$-ray flares
\citep{1991AcASn..32...36X},
and fast CMEs
\citep{2008ApJ...680.1516W}.

Yet another important finding is that
the {\it inverted} or {\it anti-Hale} spot groups,
i.e.,  the ARs violating Hale's polarity rule,
are flare productive
\citep{1968IAUS...35...33S,1970SoPh...14..328Z,1982SoPh...75..179T}.
In most cases,
polarities of ARs follow the Hale-Nicholson rule
described earlier in this subsection
and the spot groups violating this rule
are very small in number
\citep[appearance rate being 3--9\%;][]{1948ApJ...107...78R,1989SoPh..124...81W,2009ARep...53..281K,2012ApJ...745..129S,2014ApJ...797..130M}.
However,
it is known that
once this structure is created,
an AR tends to produce strong flares.
For example,
\citet{2002SoPh..209..361T} selected the 25 most violent ARs
in Cycles 22 and 23
based on five criteria: the largest spot area $>1000\ {\rm MSH}$;
X-ray flare index (related to the sum of peak flare intensities) $>5.0$;
10.7 cm radio flux $>1000\ {\rm s.f.u.}$;
proton flux ($>10\ {\rm MeV}$) $>400\ {\rm p.f.u.}$;
and geomagnetic $A_{p}$ index $>50$.
They found that most of them (68\%) violate the Hale-Nicholson rule.
Surveying 104 $\delta$-spots,
\citet{2005SoPh..229...63T} showed that about 34\% violate the Hale's rule
but follow
the hemispheric current helicity rule,
which describes the dominance
of negative (positive) current helicity in the northern (southern) hemisphere
\citep[e.g.,][see also Sect. \ref{subsubsec:seismology}]{1995ApJ...440L.109P}.
\citet{2005SoPh..229...63T}
found that such ARs have a much stronger tendency to produce X-class flares.

\begin{figure*}
  \centering
  \includegraphics[width=0.9\textwidth]{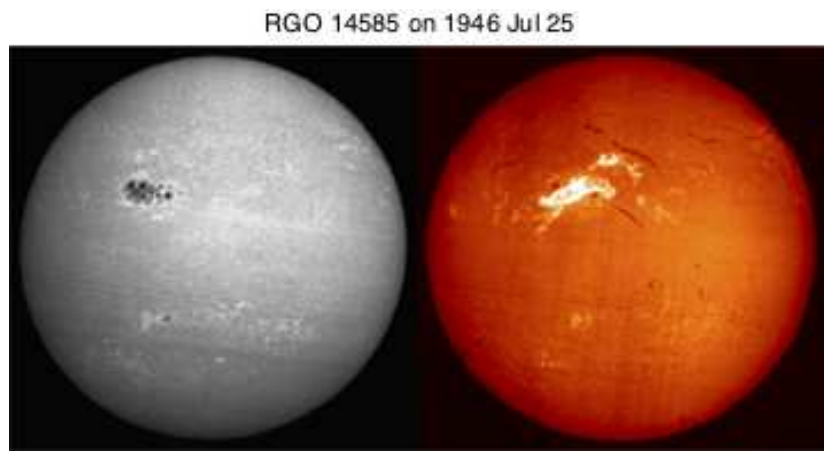}
  \caption{Great flare event in 1946 July 25 in RGO 14585,
    the fourth largest sunspot group since the late 19th century.
    A gorgeous two-ribbon flare breaks out in the huge, compact sunspot region.
    (Left) Sunspots observed in Ca\,{\sc ii} K1v.
    (Right) Very large flare ribbons observed in H$\alpha$.
    {Image reproduced by permission from \citet{2017ApJ...834...56T},
    copyright by AAS and Paris Observatory.}
    }
\label{fig:toriumi2017a_2}
\end{figure*}

In this subsection,
we reviewed several schemes of sunspot categorization
and showed that
ARs producing larger flares tend to have: a larger spot area;
morphological and magnetic complexity,
which is qualitatively indicated by McIntosh and Mount Wilson schemes;
and anti-Hale alignment.
However, for producing strong flares,
probably it is not enough to satisfy just one of these conditions.
For example, the largest-ever sunspot since the late 19th century,
RGO (Royal Greenwich Observatory) 14886 on April 1947
(maximum spot area of 6132 MSH),
is reported as flare quiet.
The spot image shown in Fig. 3 of \citet{2013A&A...549A..66A} indicates that
this region has a simple bipolar structure ($\beta$-spot).
On the other hand,
the fourth largest in history,
RGO 14585 on July 1946 (4279 MSH) as in Fig.~\ref{fig:toriumi2017a_2},
produced great flares and geomagnetic storms
with a ground-level enhancement
\citep[][]{1946MNRAS.106..500E,1946PhRv...70..771F,1949ApJ...110..242D}.
The spot image reveals that this region is strongly packed
as if it is a $\delta$-spot and,
judging from the Mount Wilson drawing,
it is very likely true.
Therefore, it is important to find
if there exist critical conditions for the strong flares and,
if so, what they are,
by conducting observational and theoretical studies of any kinds
to investigate the magnetic structure of flaring ARs and their evolution.

\section{Long-term and large-scale evolution: observational aspects}
\label{sec:longterm}

Observationally,
the changes of magnetic fields that are associated with flares
are often divided into two regimes:
the long-term, gradual evolution of large-scale fields
and the rapid changes associated with
(i.e., in the time scales comparable to) the flare occurrence.
In what follows (Sects.~\ref{sec:longterm} and \ref{sec:num}),
we review the first topic, the long-term evolution,
which is essentially related to the energy build-up process
in the pre-flare state.

\subsection{Formation and development of $\delta$-spots}
\label{subsec:deltaspots}

The role of long-term magnetic development in flare production
was first recognized by \citet{1968SoPh....5..187M},
who pointed out that the flares are often associated with evolving magnetic structures
(Structure magn\'{e}tique \'{e}volutive)
of opposite polarities,
in which
one is growing and the other decreasing.
Through accumulating a vast amount of observational data,
observers gradually found certain regularities
of flare-productive ARs.
After 18 years of observations
at Big Bear Solar Observatory (BBSO),
\citet{1987SoPh..113..267Z} summarized and classified
the formation of $\delta$-spots
that produce great flares
in three ways:
\begin{itemize}
\item Type 1: A complex of spots
  emerging all at once with different dipoles intertwined.
  This type is tightly packed with a large umbra
  and called ``island $\delta$ sunspot'';
\item Type 2: A single $\delta$-spot produced by emergence
  of satellite spots near large older spots; and
\item Type 3: A $\delta$-configuration formed by collision
  between two separate but growing bipoles.
  The overall polarity layout is quadrupolar
  and the preceding spot of one bipole collides
  with the following spot of the other.
\end{itemize}

\begin{figure*}
  \centering
  \includegraphics[width=1.\textwidth]{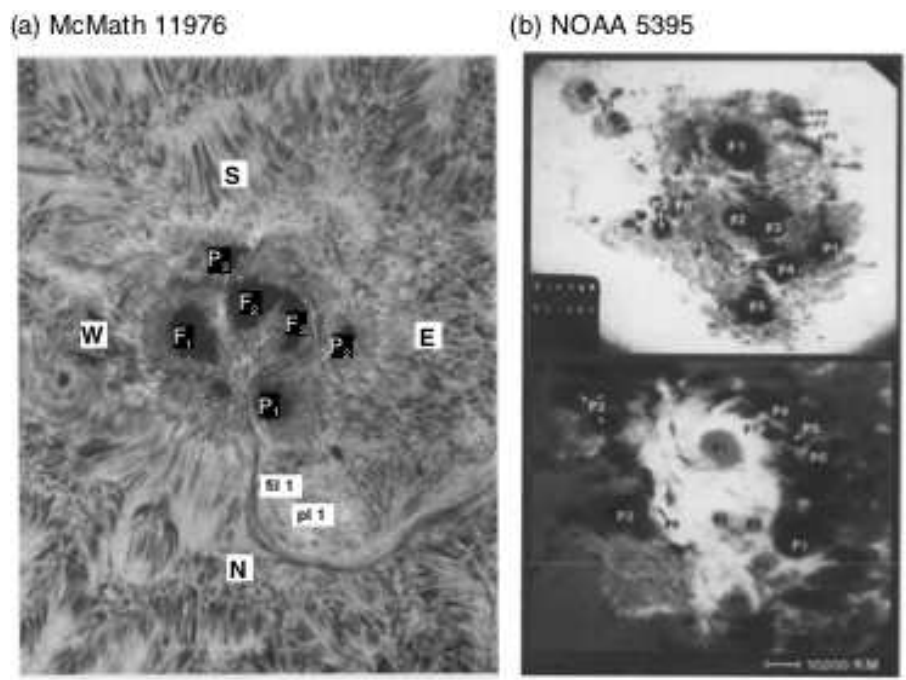}
  \caption{Examples of Type 1 $\delta$-spots.
    (a) AR McMath 11976 in August 1972.
    H$\alpha -0.5$ {\AA} image on August 3.
    Umbrae numbered F1, F2, F3, P1, P2, and P3
    all share a common penumbra.
    {Image reproduced by permission from \citet{1973SoPh...32..173Z},
    copyright by ***.}
    (b) NOAA 5395 in March 1989.
    He D3 image and magnetogram on March 10.
    {Image reproduced by permission from \citet{1991ApJ...380..282W},
    copyright by AAS.}
    }
\label{fig:zirintanaka1973}
\end{figure*}

Figure~\ref{fig:zirintanaka1973} shows
two typical examples of Type 1.
The AR in Fig.~\ref{fig:zirintanaka1973}(a), McMath 11976,
appeared in August 1972 and produced great flares \citep{1973SoPh...32..173Z}.
This region emerged as a tight complex of sunspots
with inverted magnetic polarity (i.e., anti-Hale region).
The negative spot P1 pushed into the positive spots (F1, F2, and F3)
and caused steep magnetic gradient on the central PIL.
The filament on the north (fil 1),
which may be the extension of the central PIL,
repeatedly erupted due to the continuous spot motion.
Another example is NOAA 5395 in March 1989
\citep[Fig.~\ref{fig:zirintanaka1973}(b):][]{1991ApJ...380..282W}.
This region also had a closely packed structure of multiple spots
and produced great flares including X4.5 (March 10) and X10 (March 12).
This region is known to produce the geomagnetic storm
that triggered the severe power outage in Quebec, Canada, on March 13 to 14
\citep[e.g.,][]{1989EOSTr..70.1479A,2013JSWSC...3A..31C}.
The analysis shows that,
at one edge of the large positive spot F1,
negative polarities successively emerged
and moved around the main spots,
creating a clockwise spiraling penumbral fields around it
\citep{1991ApJ...380..282W,1993SoPh..143..107T,1998ApJ...499..898I}.
The series of strong flares occurred
along the PIL surrounding the main positive spots.
Similar island-$\delta$ sunspots are observed
to show significant flaring activity,
such as flares in McMath 13043 (July 1974),
X20 event in NOAA 5629 (August 1989),
X13 in NOAA 5747 (October 1989),
and X12 in NOAA 6659 (June 1991)
\citep{1991SoPh..136..133T,1993SoPh..143..107T,1994SoPh..150..199S}.

\begin{figure*}
  \centering
  \includegraphics[width=1.\textwidth]{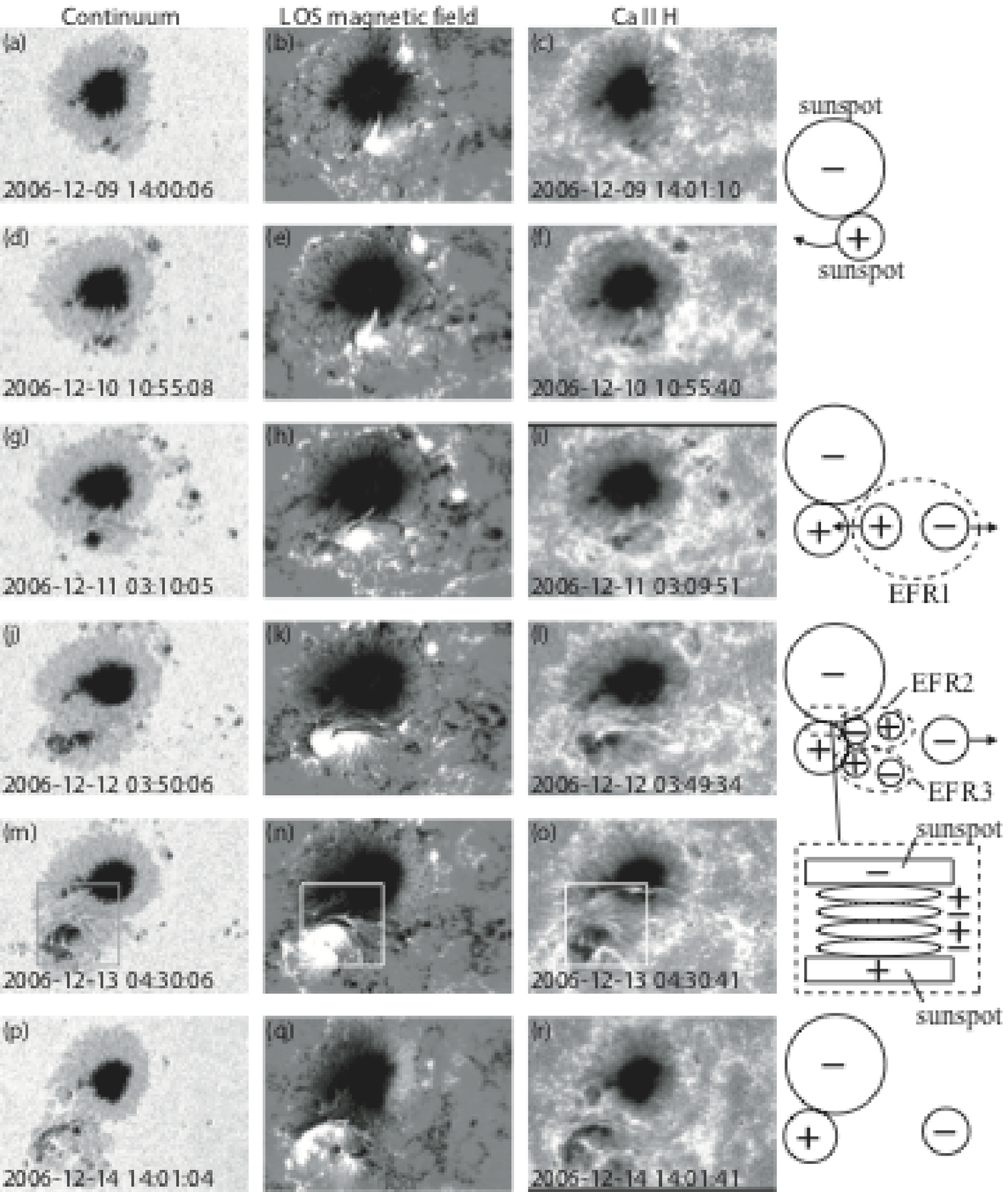}
  \caption{AR NOAA 10930 in December 2006
    as the example of Type 2 $\delta$-spot
    obtained by Hinode/SOT.
    Daily evolution of continuum, magnetic fields, and Ca\,{\sc ii} H is shown
    over the field of view of $128''\times 96''$.
    {Images reproduced by permission from \citet{2007PASJ...59S.779K},
    copyright by ***.}
    }
\label{fig:kubo2007}
\end{figure*}

Type 2 events are the flare eruptions
caused by the newly emerging satellite spots
in the penumbra of an existing spot
\citep{1968IAUS...35...77R},
and \citet{1987SoPh..113..267Z} classified
spot groups Mount Wilson 19469 and 20130 into this category
\citep{1981ApJ...243L..99P,1983SoPh...89...43T}.
Figure~\ref{fig:kubo2007} shows a clear example
of this type,
NOAA 10930 in December 2006
\citep{2007PASJ...59S.779K}.
Within the southern penumbra of the main negative spot,
a positive spot appears
and drifts around to the east
with showing a counter-clockwise rotation.
As a result,
an X3.4-class flare occurred
on December 13
at the PIL between the main and the satellite spots
(also refer to Fig.~\ref{fig:20061213}
and its corresponding movie).

\begin{figure*}
  \centering
  \includegraphics[width=1.\textwidth]{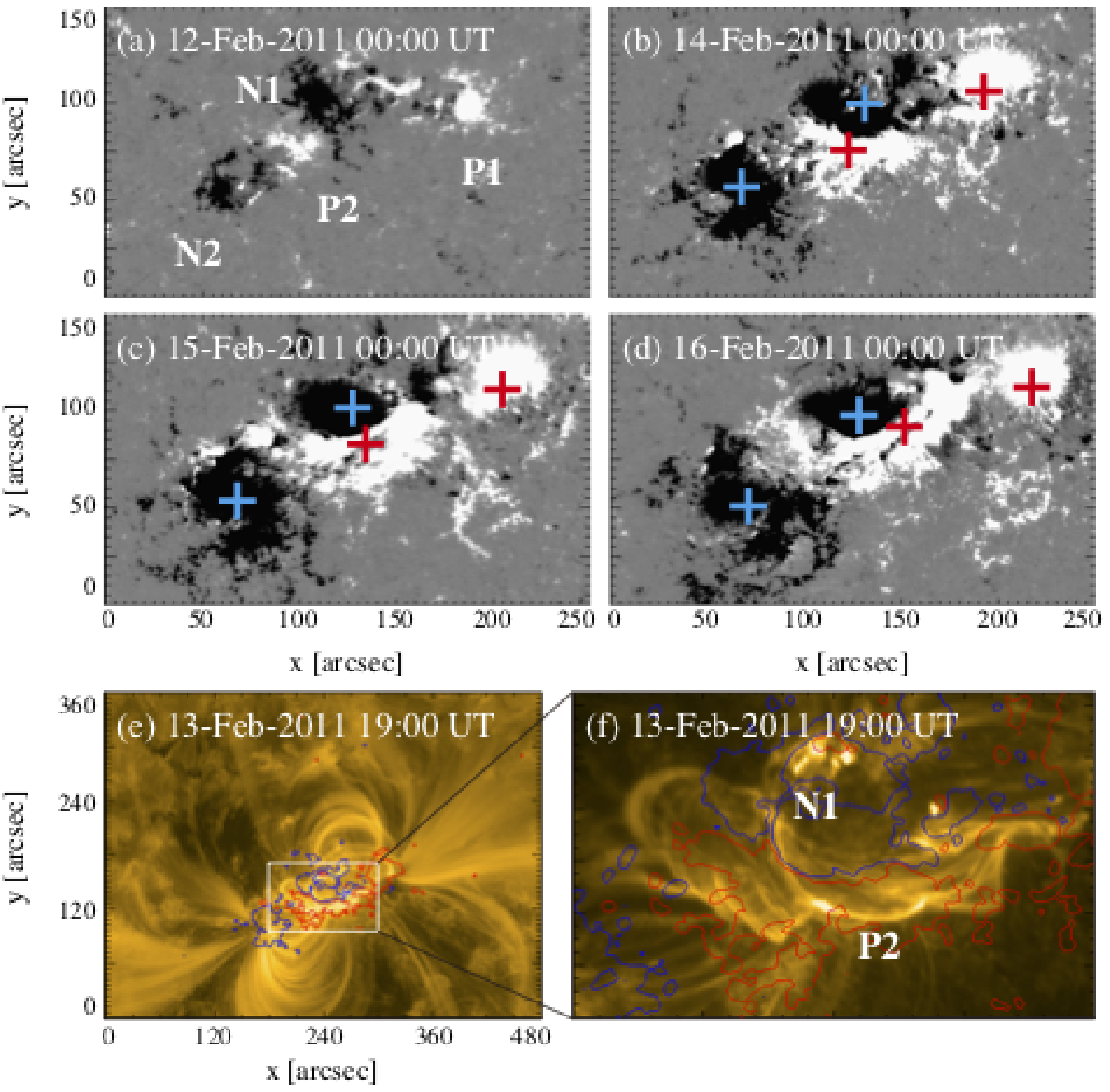}
  \caption{AR NOAA 11158 in February 2011
    as the example of Type 3 $\delta$-spot.
    {Image reproduced by permission from \citet{2014SoPh..289.3351T},
    copyright by ***.}
    Two emerging bipoles P1--N1 and P2--N2 collide against each other
    and produced a sheared PIL
    within a $\delta$-spot at the region center.
    The series of flares occur
    at the extended PIL between N1 and P2.
    Plus signs indicate the magnetic flux-weighted centroids
    of the four polarities.
    EUV images (panels e and f) show the field connectivity between N1 and P2.
    }
\label{fig:toriumi2014}
\end{figure*}

Figure~\ref{fig:toriumi2014} shows NOAA 11158 in February 2011,
the typical case of Type 3 $\delta$-spot
\citep{2014SoPh..289.3351T}.
Because of the collision of two emerging bipoles P1--N1 and P2--N2,
a highly sheared PIL with steep magnetic gradient
is produced in the central $\delta$-spot (N1 and P2)
and a series of flares including the X2.2-class event (February 15) occur.
Similar structures are seen in a variety of ARs,
such as NOAA 8562/8567, 6850, 7220/7222, 10314, and 10488
\citep{2000A&A...364..845V,2001A&A...371..731K,2005ASPC..346..317M,2013AdSpR..51.1834P,2006SoPh..234...21L}.

\begin{figure*}
  \centering
  \includegraphics[width=0.8\textwidth]{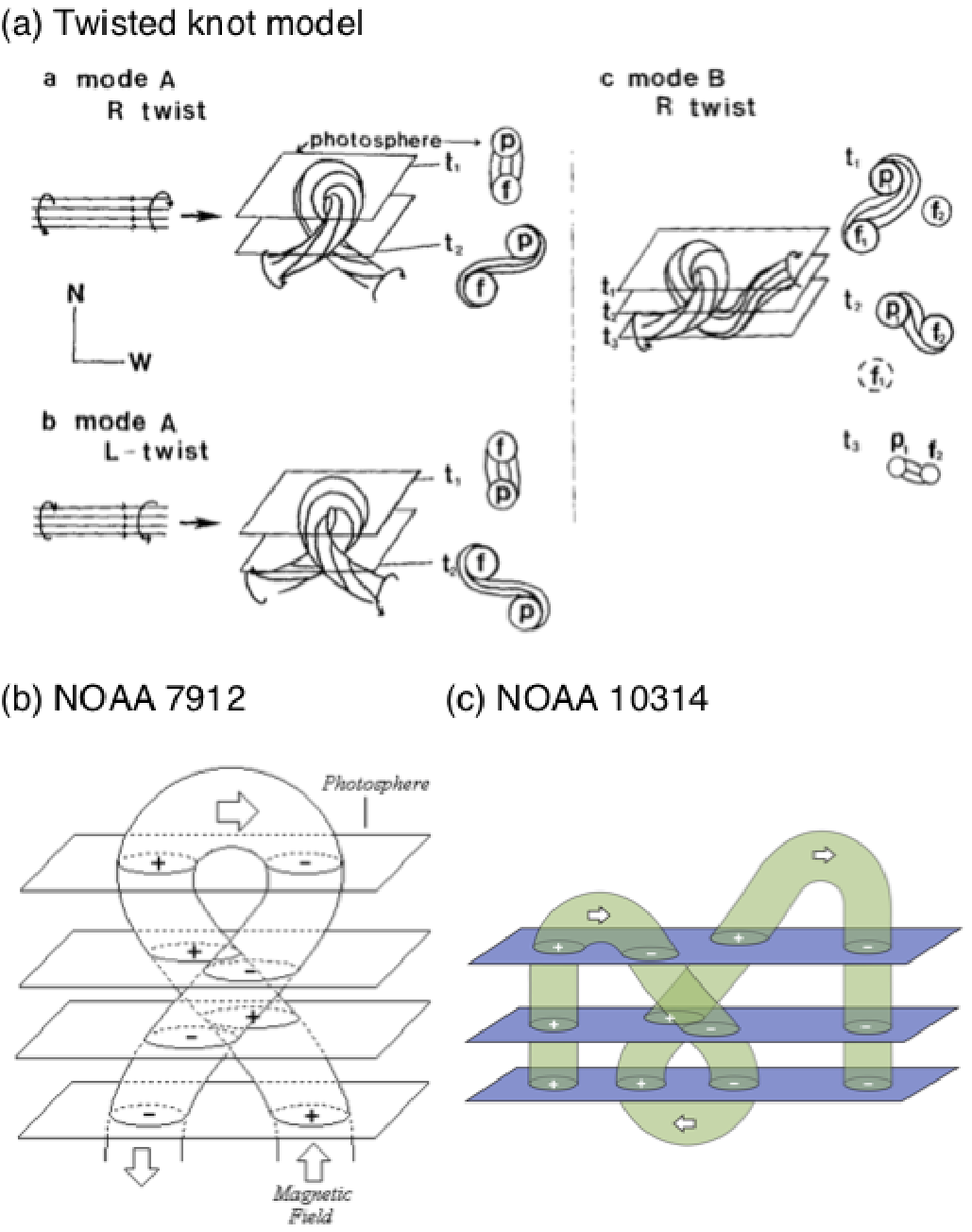}
  \caption{(a) Evolution patterns responsible for great flare occurrence
    and their explanations by an emerging twisted knot model.
    Mode A is a shearing process with spot growth
    and Mode B is an unshearing process with spot disappearance.
    Intersections represent the photosphere at times $t_{1}$, $t_{2}$ and $t_{3}$.
    {Image reproduced by permission from \citet{1991SoPh..136..133T},
    copyright by ***.}
    (b and c) Inferred 3D topologies for NOAA 7912 and 10314.
    {Images reproduced by permission from
    \citet{2000ApJ...544..540L} and \citet{2013AdSpR..51.1834P},
    copyrights by *** and ***, respectively.}
    }
\label{fig:tanaka1991}
\end{figure*}

\begin{figure*}
  \centering
  \includegraphics[width=0.9\textwidth]{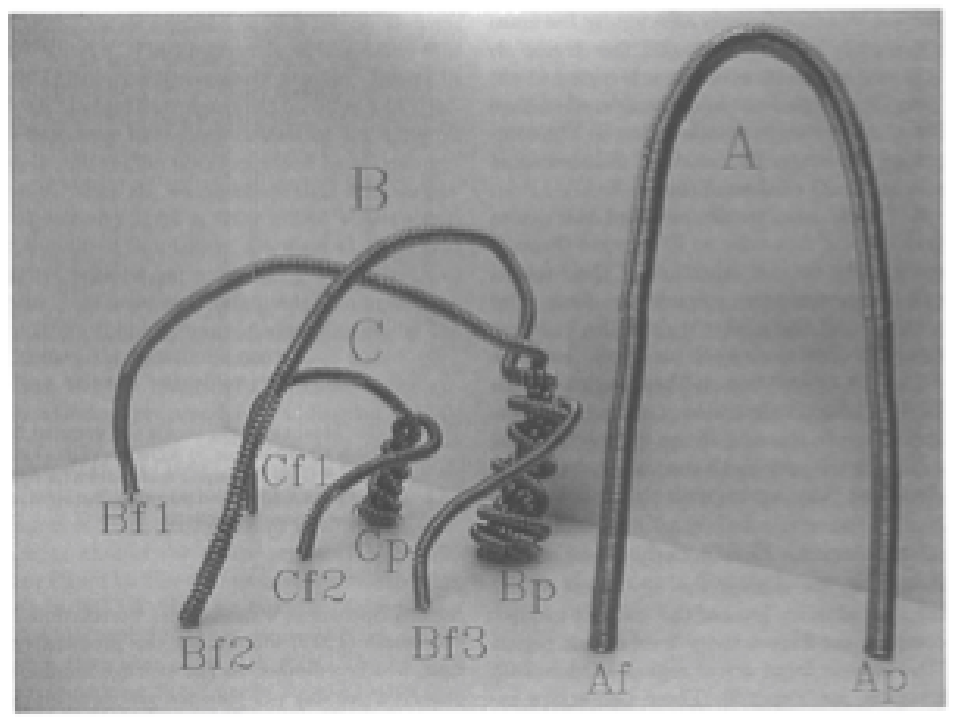}
  \caption{3D model made of flexible wires
    for explaining the evolution of NOAA 4021.
    {Image reproduced by permission from \citet{2000PASJ...52..337I},
    copyright by ***.}
    }
\label{fig:ishii2000}
\end{figure*}

How are these complex structures formed?
\citet{1987SoPh..113..267Z} mentioned that
{\it ``because Types 1 and 2 erupt in the same place,
and Type 3 requires large dipoles that are not close by mere accident,
the $\delta$ configuration must be the product of
a subsurface phenomenon.''}
However, we cannot directly observe below the surface.

One way to reconstruct the 3D topology
of emerging magnetic fields
is to study it using sequential images
(e.g., white light and magnetograms).
For example,
\citet{1991SoPh..136..133T} studied the evolution
of flare-active Type 1 $\delta$-spots McMath 13043 and 11976
and explained the observed proper motions,
the non-Hale spots turning to obey it,
by the emergence of knotted twisted flux tubes
(twisted knot model: Fig.~\ref{fig:tanaka1991}(a)).
This scenario was supported by many successive researchers
(e.g., Fig.~\ref{fig:tanaka1991}(b))
and it was suggested that
the deformation of emerging $\Omega$-loops
is due to the helical kink instability
\citep[e.g.,][]{1995ApJ...446..877L,1996ApJ...462..547L,2000ApJ...544..540L,2003A&A...397..305L,2004ApJ...611.1149H,2005SoPh..229...63T,2005SoPh..229..237T,2006JGRA..11112S01N,2015SoPh..290.2093T}
(see Sect.~\ref{subsubsec:kink} for theoretical investigations
on the kink instability
and Appendix~\ref{app:kink} for the story of the original advocates
of this instability as the formation mechanism of the $\delta$-spots).
\citet{2013AdSpR..51.1834P} explained
the formation of Type 3 $\delta$-spot NOAA 10314
as the ascent of a single large $\Omega$-loop
whose top is curled downward
and has a U-loop below the photosphere
\citep[Fig.~\ref{fig:tanaka1991}(c); see also][]{1998ApJ...508..908P,2000A&A...364..845V,2015SoPh..290.2093T}.
\citet{2000PASJ...52..337I} and \citet{2002ApJ...572..598K}
even used flexible wires to manually model
the inferred 3D configurations (Fig.~\ref{fig:ishii2000}).
From vertically stacked sequential magnetograms,
\citet{2013ApJ...764L...3C} inferred the subsurface topology of NOAA 11158
(Fig.~\ref{fig:toriumi2014}).
These observations consistently show that
the emerging flux tubes of $\delta$-spots
do not have a simple $\Omega$-shape
but are deformed
within the convection zone, prior to emergence.

\begin{figure*}
  \centering
  \includegraphics[width=1.\textwidth]{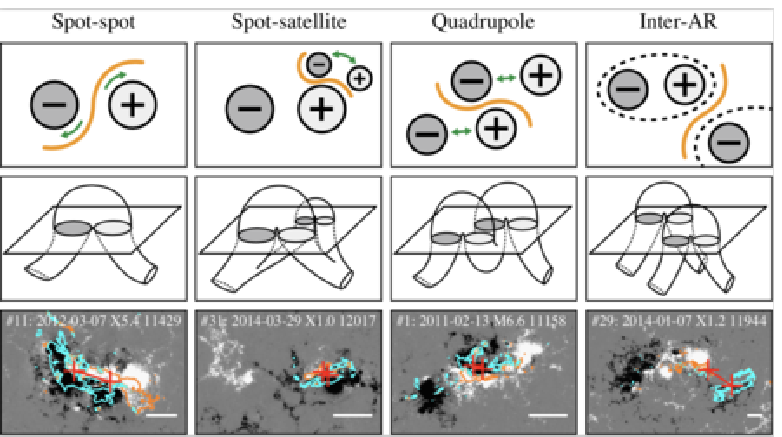}
  \caption{Classification of flaring ARs.
    {Image reproduced by permission from \citet{2017ApJ...834...56T},
    copyright by AAS.}
    (Top) Polarity distributions.
    Magnetic elements (spots) are indicated by circles with plus and minus signs.
    The PIL or filament involved in the flare is shown with an orange line,
    while proper motions of the polarities are indicated with green arrows.
    (Middle) Possible 3D structures of magnetic fields.
    Solar surface is indicated with a horizontal slice.
    (Bottom) Sample events.
    Gray scale shows magnetogram,
    overlayed by temporally stacked flare ribbons (orange and turquoise).
    Red plus signs show the area-weighted centroids of the ribbons.
    The white lines at the bottom right indicate the length of $50''$.
    }
\label{fig:toriumi2017a}
\end{figure*}

\citet{2017ApJ...834...56T} surveyed all $\geq$M5-class flares
within 45$^{\circ}$ from disk center
for six years from May 2010
and classified the host ARs into four groups
depending on their developments
(Fig.~\ref{fig:toriumi2017a}):
(1) Spot-spot, a complex, compact $\delta$-spot,
in which a large long, sheared PIL extends across the whole AR
(equivalent to Type 1 $\delta$-spot);
(2) Spot-satellite, in which a newly emerging bipole
appears in the vicinity of a preexisting main spot
(i.e., Type 2);
and (3) Quadrupole, a $\delta$-spot is created
by the collision of two bipoles (i.e., Type 3).
However, they also noticed that
even X-class events do not require $\delta$-spots or strong-gradient PILs.
Instead, some events occur between two independent ARs,
situations called (4) Inter-AR events \citep{1970SoPh...13..401D}.
For example, the X1.2 event on 2014 January 7
occurred between NOAA 11944 and 11943
\citep{2015NatCo...6E7135M,2015ApJ...814...80W}.
Figure~\ref{fig:toriumi2017a} also provides possible 3D topologies,
which were later modeled by numerical simulations
(see Sect.~\ref{subsubsec:unified}).

Through the analysis of Mount Wilson classifications from 1992 to 2015,
\citet{2016ApJ...820L..11J} discussed
the possible production mechanism of complex ARs.
They found that while the fractions of $\alpha$- and $\beta$-spots
remain constant over cycles
(about 20\% and 80\%, respectively),
that of complex ARs appended with $\gamma$ and/or $\delta$
increases drastically from 10\% at solar minimum
to more than 30\% at maximum.
According to the authors,
this may indicate that complex ARs are produced
by the collision of simpler ARs around the surface layer
through the higher rate of flux emergence during solar maximum.
This idea may be related to
the successive emergence model
\citep{1987SoPh..113..259K}
and perhaps to the concepts of
``complexes of activities'' and ``sunspot nests''
\citep{1965ApJ...141.1502B,1983ApJ...265.1056G,1986SoPh..105..237C,1994ApJ...422..883G}.

\subsection{Photospheric features}
\label{subsec:photo}

\subsubsection{Strong-field, strong-gradient, highly-sheared PILs and magnetic channels}
\label{subsubsec:photo_pil}

\begin{figure*}
  \centering
  \includegraphics[width=0.8\textwidth]{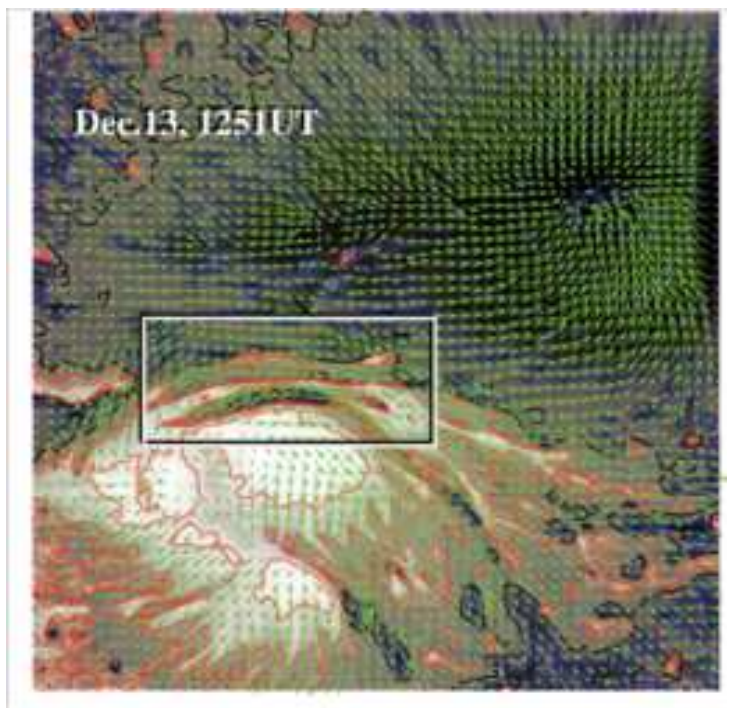}
  \caption{Hinode/SOT/SP vector magnetogram of AR NOAA 10930,
    which produced the X3.4-class flare
    (see Fig.~\ref{fig:20061213} and~\ref{fig:kubo2007}).
    The image shows the LOS magnetic fields (gray scale),
    transverse fields (green arrows),
    positive and negative polarities (red and blue contours),
    and the PILs (black contours).
    The FOV is $66''\times 66''$.
    The area around the sheared PIL is marked with a rectangular box.
    {Image reproduced by permission from \citet{2008ApJ...687..658W},
    copyright by AAS.}
    }
\label{fig:wang2008}
\end{figure*}

Because flares are the release of magnetic energy via magnetic reconnection,
it is natural that these events are observed around the PILs,
where the electric currents are strongly enhanced
(see, e.g., Fig.~\ref{fig:20061213}).
Since this fact was first pointed out by \citet{severny1958},
the importance of the PILs in the flare occurrence
has been repeatedly emphasized
\citep[e.g.,][]{1973SoPh...32..173Z,1984SoPh...91..115H,1996ApJ...456..861W,2007ApJ...655L.117S}.
The photospheric characteristics of the flaring PILs
are summarized as follows.
\begin{description}
\item[Strong field:] Both the vertical fields
  surrounding the PIL and the transverse fields
  along the PIL are very strong.
  \citet{1991SoPh..136..133T} and \citet{1993SoPh..144...37Z}
  reported on the detection
  of strong transverse fields of up to 4300 G
  \citep[see also][]{2016ApJ...818...81J,2018RNAAS...2a...8W}.
  \citet{2006SoPh..239...41L} also pointed out that part of the exceptionally strong fields
  they found are likely related to the transverse fields
  in light bridges of $\delta$-spots (i.e., PILs).
  \citet{2018ApJ...852L..16O} noticed the fields as high as 6250 G in a PIL,
  which is probably the highest value ever measured on the Sun
  including the sunspot umbrae.
\item[Strong gradient:] The horizontal gradient of the vertical field
  across the PIL is steep,
  indicating that positive and negative polarities
  are tightly pressed against each other
  \citep{1968SoPh....3..282M,1991ApJ...380..282W,1994SoPh..155...99W}.
  The gradient is sometimes up to several $100\ {\rm G\ Mm}^{-1}$
  \citep{1998ASPC..150...98W,2006ApJ...644.1273J,2009SoPh..254..101S}.
\item[Strong shear:] The transverse field is directed almost parallel to the PIL.
  The shear angle is often measured in the frame
  where $0^{\circ}$ is the azimuth of a potential field
  \citep{1984SoPh...91..115H,1993SoPh..148..119L},
  and large shears of $80^{\circ}$ to $90^{\circ}$
  are observed at flaring PILs
  \citep{1990ApJS...73..159H,1990MmSAI..61..337H}.
  Figure~\ref{fig:wang2008} clearly shows that the transverse fields
  at the PIL of NOAA 10930
  are along the direction of the PIL (marked by the box).
\end{description}

The strong-field, strong-gradient, highly-sheared PILs
may be the direct manifestation of non-potentiality of magnetic fields
and, therefore, these features are often used for the prediction of flares and CMEs.
\citet{2002ApJ...569.1016F,2006ApJ...644.1258F} measured
the lengths of PILs of, e.g., strong transverse field ($>150\ {\rm G}$),
large shear angle ($>45^{\circ}$),
and steep gradient ($> 50\ {\rm G\ Mm}^{-1}$)
and demonstrated that these parameters predict the occurrence of CMEs.
\citet{2007ApJ...655L.117S} evaluated the total unsigned flux
near the strong-gradient PILs
and showed that it gives the upper limit of possible GOES flare class.

\begin{figure*}
  \centering
  \includegraphics[width=1.\textwidth]{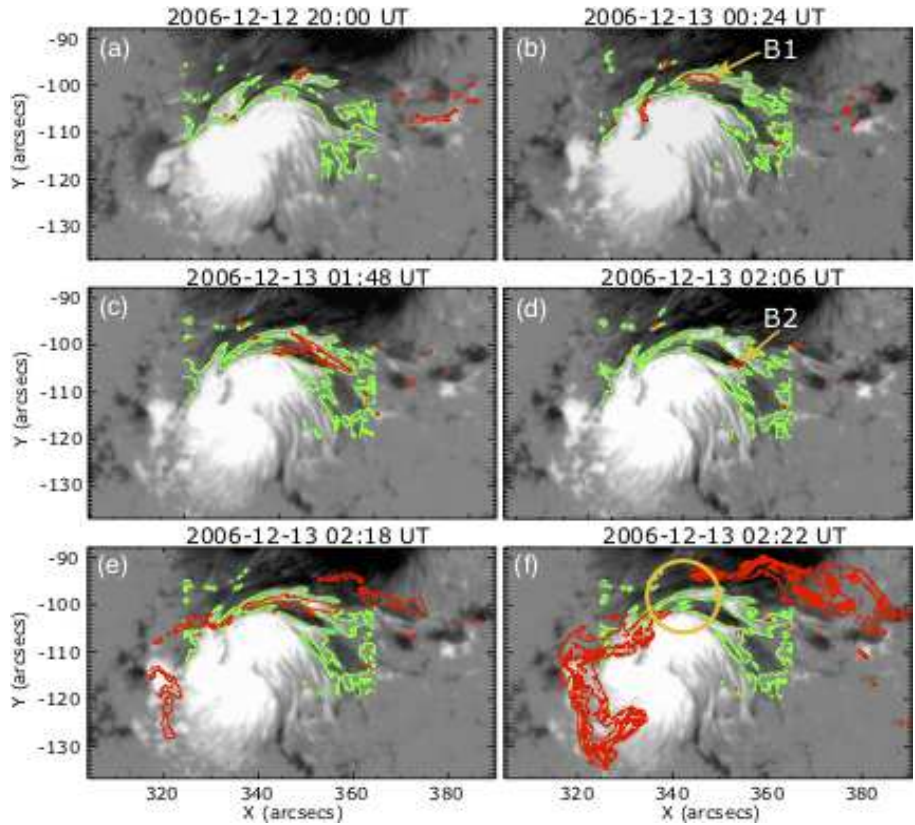}
  \caption{Temporal evolution of the X3.4-class flare in AR NOAA 10930.
    Background shows the LOS magnetogram,
    over which the PILs are plotted with green lines.
    The red contours show the Ca\,{\sc ii} H line enhancement.
    The pre-flare brightening (such as B1) continuously occurs
    around the central PIL (yellow circle).
    The flare ribbons originate and expand from this region
    (see, e.g., progenitor brightening of B2).
    {Image reproduced by permission from \citet{2013ApJ...778...48B},
    copyright by AAS.}
    }
\label{fig:bamba2013}
\end{figure*}

Another important feature of the flaring PILs
is the ``magnetic channel'',
which is an alternating pattern of
elongated positive and negative polarities
\citep{1993Natur.363..426Z,2002ApJ...569.1026W}.
Figure~\ref{fig:wang2008} displays the magnetic channel
in NOAA 10930 (see PIL marked by the box).
\citet{2008ApJ...687..658W} and \citet{2010ApJ...719..403L} showed that
high resolution with high polarimetric accuracy
is needed to adequately resolve such small-scale structures
(width $\lesssim 1''$).
Figure~\ref{fig:bamba2013} clearly shows that
the pre-flare brightening continues around this structure
and the flare ribbons originate from here
(see also the movie of Fig.~\ref{fig:20061213}).
From these observations, \citet{2013ApJ...778...48B} suggested
that such fine-scale magnetic structures galvanize
the whole system into producing flare eruptions
\citep{2013ApJ...773..128T,2017ApJ...838..134B,2018ApJ...856...43B}.

\begin{figure*}
  \centering
  \includegraphics[width=0.95\textwidth]{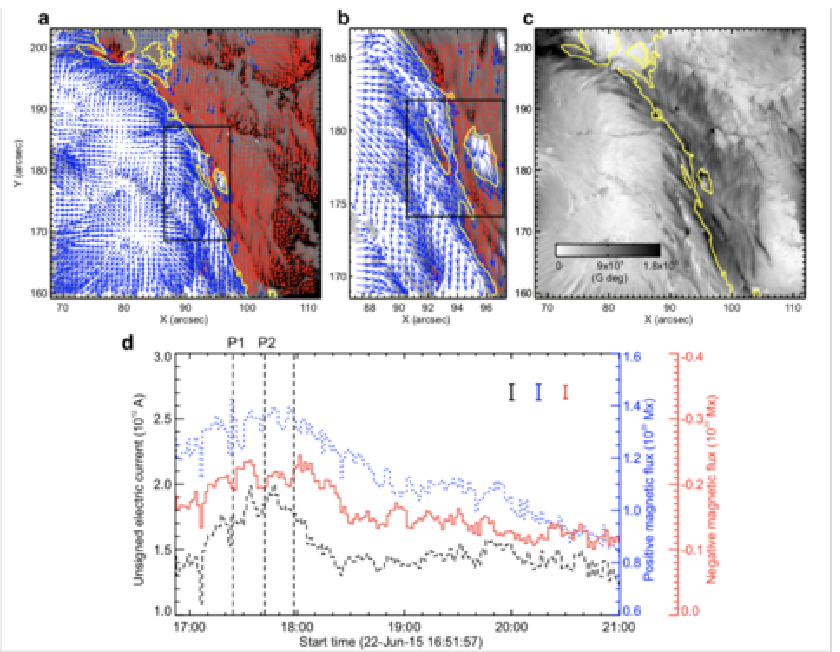}
  \caption{BBSO/GST observation of magnetic field in AR NOAA 12371
    before the M6.5-class flare at 18:23 UT on 2015 June 22.
    (a and b) GST/NIRIS photospheric vertical magnetic field
    (scaled between $\pm 1500\ {\rm G}$) at 17:35 UT,
    superimposed with arrows representing horizontal magnetic field vectors.
    The box in (a) denotes the FOV of (b),
    in which the magnetic channel structure can be obviously observed.
    (c) Distribution of magnetic shear
    in terms of a product of the field strength and shear angle.
    The overplotted yellow contour in (a)--(c) is the PIL.
    (d) Temporal evolution of total positive (blue dotted line)
    and negative (red solid line) magnetic fluxes
    and the unsigned electric current (black dashed line),
    calculated over the magnetic channel region enclosed by the box in (b).
    The first two vertical dashed lines indicate the times
    of two flare precursor episodes.
    {Image reproduced by permission from \citet{2017NatAs...1E..85W},
    copyright by ***.}
    }
\label{fig:wang2017}
\end{figure*}

The significance of the sheared PIL, magnetic channel, and small-scale trigger
was also verified by a super high-resolution observation by BBSO/GST.
Figure~\ref{fig:wang2017} shows the GST/NIRIS magnetogram of AR NOAA 12371.
Here, \citet{2017NatAs...1E..85W} found that the field is highly sheared with respect to the PIL,
especially in the precursor brightening region (panels (a) and (b)).
This signifies a high degree of non-potentiality,
as reflected by the concentration of magnetic shear along the PIL
(panel (c)).
In the region around the initial precursor brightening
enclosed by the box in panel (b),
they observed a miniature version of a magnetic channel with a scale of only 3,000 km,
which can also be recognized as the flare-triggering field.
Importantly, the evolutions of both polarities within the channel
are temporally associated with the occurrence of precursor episodes (panel (d)).

\subsubsection{Flow fields and spot rotations}
\label{subsubsec:spotrotation}

Given the high-$\beta$ condition in the photosphere,
it was speculated that
such flaring PILs are generated
by the sheared, converging flow fields around it.
In fact, \citet{1976SoPh...47..233H} observed
strong shear flows along the flaring PILs
and associated these flows with the occurrence of flares
\citep{2003A&A...412..541M,2004ApJ...617L.151Y,2006ApJ...644.1278D,2014PASJ...66S..14S}.
Also, \citet{1994ASPC...68..265K} showed that the flare kernels
correspond to the locations of convergence in the horizontal flows.
The converging flow and the sustained cancellation
of positive and negative polarities
on the two sides of the PIL
are thought to be the key process in building up a magnetic flux rope
\citep[][see also Sect.~\ref{subsubsec:sigmoids} of this article for detailed discussion]{1989ApJ...343..971V}.

The large-scale spot motions
drive the flow fields around the PILs
and, because of the
frozen-in state of the field,
the magnetic structures are reconfigured.
For instance, \citet{1982SoPh...79...59K} revealed that
the shear flow in the PIL is in association with rapid spot motions,
which enhances the magnetic shear at the PIL and leads to the series of flares.
\citet{1994SoPh..155..285W} observed that
magnetic shear development is intrinsically related to the newly emerging flux.

\begin{figure*}\sidecaption
  \centering
  \includegraphics[width=0.58\textwidth]{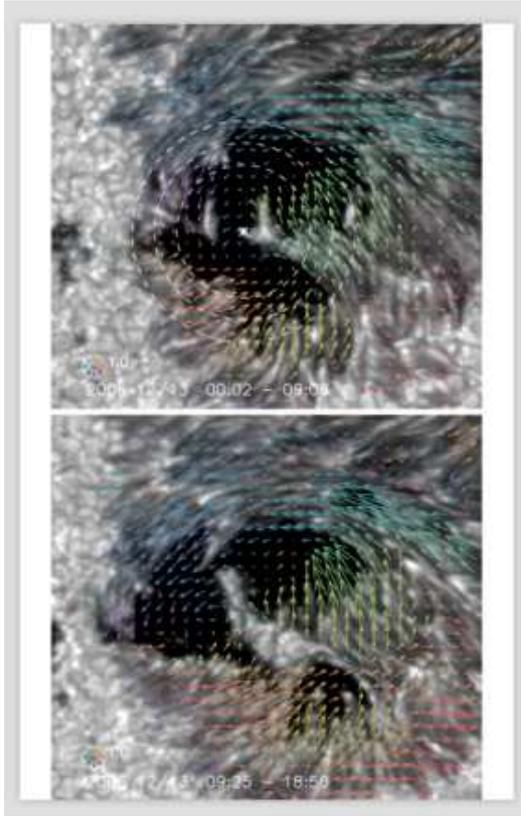}
  \caption{Velocity field of the southern sunspot in AR NOAA 10930
    over the FOV of $42''\times 38''$.
    The radius of the circle in the lower-left corner
    corresponds to a speed of $0.22\ {\rm km\ s}^{-1}$,
    and the color of an arrow corresponds to its direction.
    {Image reproduced by permission from \citet{2009SoPh..258..203M},
    copyright by ***.}
    }
\label{fig:min2009}
\end{figure*}

Strong spot rotations
(both the spot rotating around its center
and the spot rotating around its counterpart in the same AR)
are also often observed in the pre-flare state.
Figure~\ref{fig:min2009} is a clear example of rotating sunspots
in AR NOAA 10930 \citep{2009SoPh..258..203M}.
This figure highlights that the southern spot rotates
in the counter-clockwise direction before the X3.4-class flare occurs.
\citet{2003SoPh..216...79B} analyzed rotating sunspots in seven ARs
and found that the spots rotate around their umbral centers
up to 200$^{\circ}$ in 3--5 days.
The coronal loops are twisted as the spot rotates,
and six of them showed flares and/or CMEs
\citep{2006A&A...451..319R,2007ApJ...662L..35Z,2008SoPh..247...39Z,2012ApJ...761...60V,2014ApJ...784..165R,2016ApJ...829...24V}.
\citet{2003SoPh..216...79B} considered that
the spot rotation is caused by the flux tube emergence
(see Sect.~\ref{subsec:num_fe} for the discussion).
The observed association of spot rotations and eruptions
is consistent with the theoretical suggestion
by \citet{1969SoPh....8..115S} and \citet{1972ApJ...174..659B} that
such spot rotations accumulate flare energy in the atmosphere.
\citet{2008MNRAS.391.1887Y} surveyed 186 rotating sunspots in 153 ARs
and statistically investigated the relationship between the spot rotation and the flare productivity.
They found that ARs with sunspots of rotation direction opposite to
the global differential rotation are in favor of producing M- and X-class flares.

These flow fields and spot motions strongly suggest the possibility
that the flaring ARs, if not all,
are produced by the emergence of magnetic flux with a strong twist.
Through these processes, the magnetic flux
transports the energy and magnetic helicity (Sect.~\ref{subsubsec:helicity})
from the subsurface layer to the atmosphere.

\subsubsection{Injection of magnetic helicity}
\label{subsubsec:helicity}

Magnetic helicity is a measure
of magnetic structures
such as twists, kinks, and internal linkage \citep{1956RvMP...28..135E}
and is a useful tool to quantify and characterize
the complexity of flaring ARs.
The magnetic helicity of the magnetic field $\vec B$
fully contained in a volume $V$
(i.e., the normal component $B_{n}$ vanishes
at any point of the surface $S$)
is defined as
\begin{eqnarray}
  H=\int_{V} {\vec A}\cdot{\vec B}\, dV,
  \label{eq:helicity}
\end{eqnarray}
where ${\vec A}$ is the vector potential of ${\vec B}$,
i.e., ${\vec B}=\nabla\times{\vec A}$.
$H$ is invariant to gauge transformations
and, in ideal MHD, $H$ is a conserved quantity.
Even under resistive MHD where magnetic reconnection can occur,
it is shown that dissipation of $H$ is much slower
than dissipation of magnetic energy
\citep{1984GApFD..30...79B}.
However, in many practical situations,
the field lines cross the surface of the volume of interest $S$
(e.g., the photosphere)
and thus it is convenient to use the relative magnetic helicity
\citep{1984JFM...147..133B,finn1985}:
\begin{eqnarray}
  H_{\rm R}=\int_{V} ({\vec A}+{\vec A_{0}})\cdot({\vec B}-{\vec B_{0}})\, dV,
\end{eqnarray}
where ${\vec A}_{0}$ and ${\vec B}_{0}$ are
the reference vector potential and magnetic field, respectively
(${\vec B}_{0}$ has the same $B_{n}$ distribution on $S$).
$H_{\rm R}$ is also a gauge-invariant quantity,
and often the potential field ${\vec B}_{\rm p}$ $(=\nabla\times{\vec A}_{\rm p})$
is chosen as the reference field:
\begin{eqnarray}
  H_{\rm R}=\int_{V} ({\vec A}+{\vec A_{\rm p}})\cdot({\vec B}-{\vec B_{\rm p}})\, dV.
\end{eqnarray}

One way to calculate the relative helicity in the coronal volume
is to rely on 3D magnetic extrapolations
as it is not yet possible to fully measure the magnetic fields
in the atmosphere (Sect.~\ref{subsubsec:num_data_extrapolation}).
Alternatively,
it is also possible to monitor the helicity flux
(helicity injection rate)
through the photosphere over the AR,
\footnote{It is implicitly assumed here that
the net helicity flux through $S$ other than the photosphere is zero.}
\begin{eqnarray}
  \frac{dH_{\rm R}}{dt} = 2\int
  \left[
    ({\vec A}_{\rm p}\cdot{\vec B})v_{n}
    -({\vec A}_{\rm p}\cdot{\vec v})B_{n}
  \right]\, dS,
  \label{eq:relativeh}
\end{eqnarray}
where ${\vec v}$ is the velocity of the plasma
and $v_{n}$ is the component normal to the surface.
This parameter
has been used more commonly to investigate the accumulation of helicity
during the course of AR evolution
\citep{2001ApJ...560L..95C,2001ApJ...560..476C,2002SoPh..208...43G,2003ApJ...594.1033N,2004SoPh..223...39C}.
Note that in the last equation,
the first and second terms in the bracket
are called the ``emergence term'' and ``shear term,'' respectively.

\begin{figure*}
  \centering
  \includegraphics[width=0.8\textwidth]{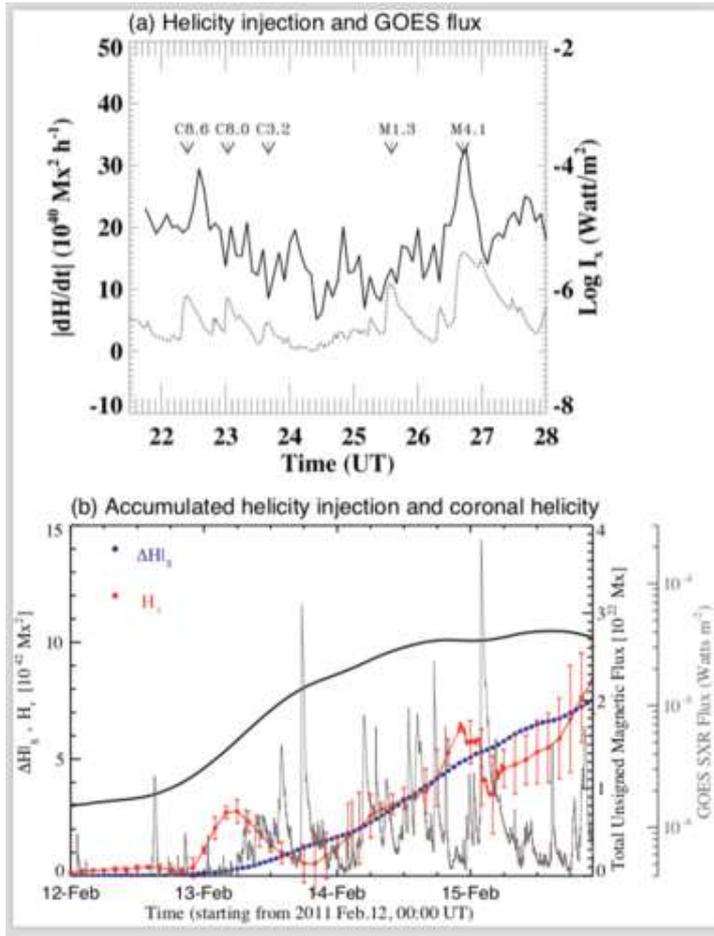}
  \caption{(a) Temporal evolution of the magnetic helicity injection rate (solid line)
    and the GOES soft X-ray flux (dotted line) over 6.5 hr.
    The arrows indicate the X-ray intensity peak of homologous flares
    in AR NOAA 8100.
    {Image reproduced by permission from \citet{2002ApJ...574.1066M},
    copyright by AAS.}
    (b) Temporal variation of magnetic helicity.
    Plotted are the coronal helicity
    derived from the NLFFF extrapolation $H_{\rm r}$ (red dots),
    the accumulated amount of helicity injection
    through the photosphere $\Delta H|_{S}$ (blue dots),
    total unsigned magnetic flux (black) and GOES flux (gray).
    The uncertainty in $H_{\rm r}$ is indicated by the error bars.
    The uncertainty in $\Delta H|_{S}$ is generally 0.5\%
    that is too small to be plotted.
    {Image reproduced by permission from \citet{2012ApJ...752L...9J},
    copyright by AAS.}
    }
\label{fig:moon2002}
\end{figure*}

Many observational studies have shown the temporal relationship
between the helicity injection and the occurrence of flares and CMEs
\citep{2002ApJ...574.1066M,2002ApJ...580..528M,2004SoPh..223...39C,2008PASJ...60.1181M,2008ApJ...686.1397P,2012ApJ...750...48P}.
For instance, \citet{2002ApJ...574.1066M,2002ApJ...580..528M} revealed that
the significant amount of helicity was impulsively injected
around the peak time of X-ray flux
of the flare events they studied,
especially for the strong ones
(Fig.~\ref{fig:moon2002}(a)).
The authors attributed the observed impulsive helicity injection
to the horizontal velocity anomalies near the PIL.
However, because the location of helicity injection
is near the flaring site (e.g., H$\alpha$ flare ribbons),
the possibility can not be ruled out
that the observation is affected
by an artifact of the magnetogram (SOHO/MDI)
due to emission caused by particle precipitation
that changes the spectral line's shape.

From long-term monitoring,
\citet{2008ApJ...686.1397P,2012ApJ...750...48P} found that
the helicity 
first increases monotonically
and then remains almost constant just before the flares.
Some events show the sign of injected helicity reverses and,
in such cases, the flares are more energetic and impulsive
and the accompanying CMEs are faster and more recurring.
\citet{2010ApJ...720.1102P} and \citet{2012ApJ...752L...9J}
compared the accumulated helicity injection
measured by integrating Eq. (\ref{eq:relativeh}) over time
and the coronal helicity
derived from the NLFFF extrapolation (Sect.~\ref{subsubsec:num_data_extrapolation})
and found close correlations between the two parameters
(see Fig.~\ref{fig:moon2002}(b)).

From the viewpoint of helicity budget,
the CME works as a carrier of helicity
that is taken away from a flaring AR
and leads the magnetic system of the AR to lower energy states
\citep[see illustration in Fig.~\ref{fig:cshkp}(b):][]{1994GeoRL..21..241R,2002A&A...382..650D,2002SoPh..208...43G}.
However, accumulated helicity may also be reduced by annihilation
of two magnetic systems of opposite helicity sign
(through magnetic reconnection).
Several observations show that
magnetic systems with oppositely singed helicity
commonly exist in a given AR
and the interaction of these systems play a key role in driving flares and CMEs
\citep{2002ApJ...577..501K,2004ApJ...615.1021W,2010SoPh..261..127C,2011A&A...525A..13R,2011A&A...530A..36Z}.
This scenario is further supported by MHD simulations
by \citet{2004ApJ...610..537K,2012ApJ...760...31K},
in which the emergence of reversed shear near the PIL triggers the eruption.

\begin{figure*}
  \centering
  \includegraphics[width=0.8\textwidth]{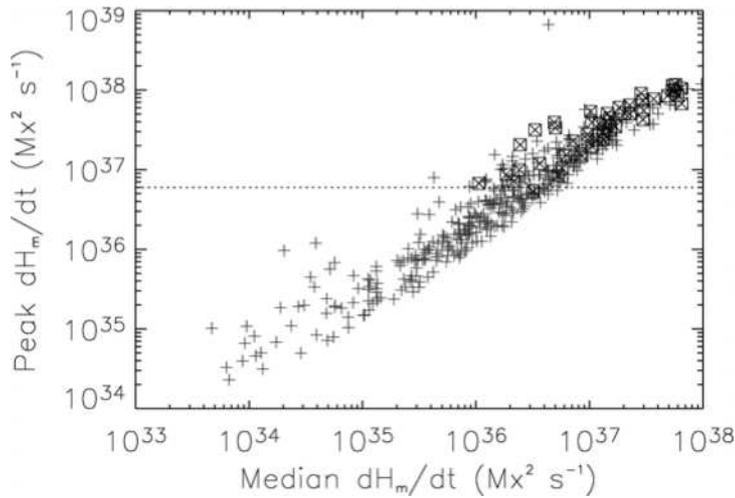}
  \caption{(a) Peak helicity injection rate
    during the observing interval vs. the median helicity flux over the interval.
    Non-X-flaring reference regions (345) are plotted as plus signs
    and X-flare regions (48) as boxed crosses.
    The necessary condition for the production of an X-flare
    is a peak helicity flux $>6\times 10^{36}\ {\rm Mx}^{2}\ {\rm s}^{-1}$.
    {Image reproduced by permission from \citet{2007ApJ...671..955L},
    copyright by AAS.}
    }
\label{fig:labonte2007}
\end{figure*}

Statistical investigations on a number of ARs clearly demonstrate
the tendency that
flare-productive ARs have a significantly higher amount of helicity
than flare-quiet ARs
\citep{2004ApJ...616L.175N,2010ApJ...718...43P}.
\citet{2007ApJ...671..955L} compared 48 X-flare-producing ARs
and 345 non-X-flaring regions
and derived an empirical threshold for the occurrence of an X-class flare
that the peak helicity flux exceeds
a magnitude of $6\times 10^{36}\ {\rm Mx}^{2}\ {\rm s}^{-1}$
(see Fig.~\ref{fig:labonte2007}).
\citet{2012ApJ...759L...4T,2014A&A...570L...1T} found a consistent monotonic scaling
between the relative helicity and the free magnetic energy
for both observational data sets and MHD simulations
\citep{2014SoPh..289.4453M}.
However, it should be noted that these results
do not take into account the area of ARs.
Because the magnetic helicity in a flux system scales
as the square of that system's magnetic flux,
we can compare,
by normalizing the magnetic helicity by the flux squared,
how much the magnetic configuration is stressed
in ARs of the same size
\citep{2009AdSpR..43.1013D}.

As mentioned above,
flaring ARs exhibit a fairly complicated distribution
of both positive and negative signs of magnetic helicity.
The helicity flux distribution can be measured
by computing and mapping the density of helicity flux
in Eq. (\ref{eq:relativeh}):
$G_{A}=2[({\vec A}_{\rm p}\cdot{\vec B})v_{n}-({\vec A}_{\rm p}\cdot{\vec v})B_{n}]$,
or simply $G_{A}=-2({\vec A}_{\rm p}\cdot{\vec v})B_{n}$.
However, \citet{2005A&A...439.1191P} showed that
$G_{A}$ is not a proper helicity flux density
as $G_{A}$ can be non zero ($G_{A}$ map can show variation)
even with simple translational motions
that do not inject any magnetic helicity.
Then, they proposed an alternative proxy
of the helicity flux density, $G_{\Phi}$,
which takes into account the magnetic field connectivity
and thus requires 3D magnetic extrapolations.
\citet{2013A&A...555L...6D,2014SoPh..289..107D}
developed a method to compute $G_{\Phi}$
and applied it to observational data
of the complex flaring AR NOAA 11158 (Fig.~\ref{fig:toriumi2014}),
showing that this proxy reliably and accurately maps
the distribution of photospheric helicity injection.


\subsubsection{Magnetic tongues and importance of structural complexity}
\label{subsubsec:tongue}

\begin{figure*}
  \centering
  \includegraphics[width=0.75\textwidth]{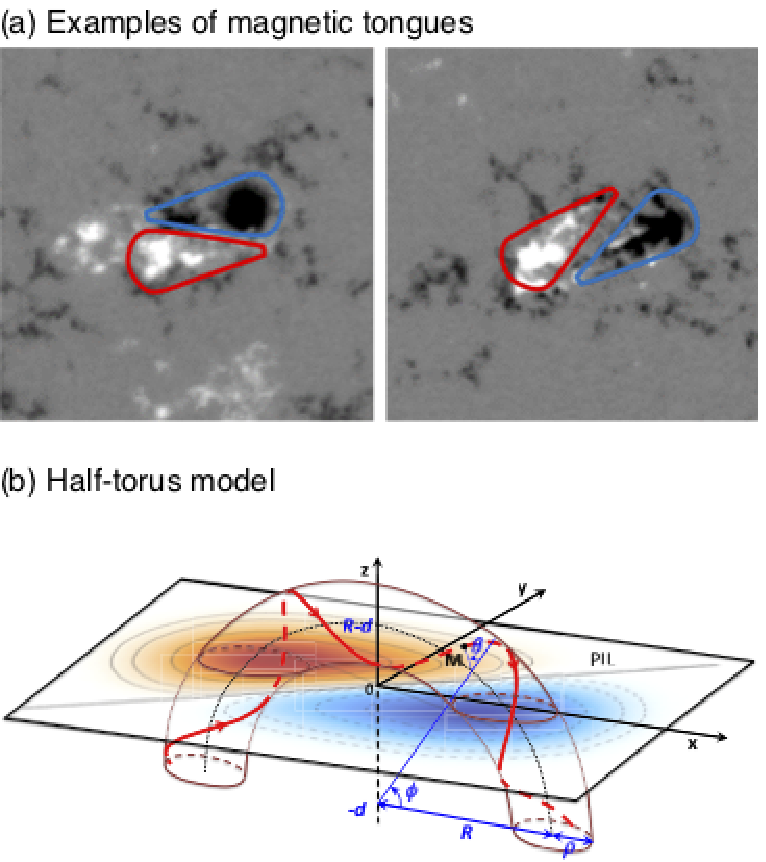}
  \caption{(a) Sample images of magnetic tongues
    resembling the yin-yang pattern.
    The left panel shows the tongue with negative helicity (left-handed twist),
    while the right panel is for positive helicity (right-handed twist).
    {Image reproduced by permission from \citet{2015SoPh..290.2093T},
    copyright by ***.}
    (b) Model of a twisted flux tube with a half-torus shape.
    The magnetic tongue (red-blue), separated by the PIL (straight line),
    is explained by the emergence of a twisted flux tube.
    In this case, the magnetic tongue has positive helicity
    due to the emergence of a flux tube with right-handed twist.
    {Image reproduced by permission from \citet{2016SoPh..291.1625P},
    copyright by ***.}
    }
\label{fig:takizawa2015}
\end{figure*}

In vertical (or LOS) magnetograms,
the newly emerging regions, especially of AR scales,
display ``magnetic tongue'' structures,
the extended magnetic polarities at both sides of the PIL
(Fig.~\ref{fig:takizawa2015}(a)),
first mentioned by \citet{2000ApJ...544..540L}.
The magnetic tongues that resemble the yin-yang pattern
are thought to be the vertical projection of the poloidal component
of the twisted emerging magnetic flux tube (Fig.~\ref{fig:takizawa2015}(b)),
and thus, the layout of tongues and the direction of PILs
are used as proxies of magnetic helicity sign of emerging fields
\citep[Sect.~\ref{subsubsec:helicity}:][]{2011SoPh..270...45L,2015SoPh..290.2093T,2015SoPh..290..727P,2016SoPh..291.1625P}.
Multiple observational studies showed that
such yin-yang tongues are seen in flaring ARs,
along with other observational characteristics
including sigmoids, sheared coronal loops, and {\sf J}-shaped flare ribbons
\citep{2007A&A...475.1081L,2007SoPh..246..365G,2009ApJ...693L..27C,2009SoPh..258...53C,2014SoPh..289.2041M}.
This may indicate that
the flaring ARs tend to
possess substantial magnetic helicity.

One of the important conclusions
from the series of statistical investigations
in Sect.~\ref{subsec:spots}
was that magnetic fields of flare-productive ARs exhibit
higher degrees of {\it complexity}.
While classical sunspot categorizations
(e.g., McIntosh and Mount Wilson schemes)
simply provide qualitative indices of the ARs' complexity,
one well-studied quantitative measure of the complexity is the fractal dimension,
an indication of self-similarity of structures
\citep{1983whf..book.....M}.
From the fractal dimension analysis
using full-disk magnetograms over 7.5 years,
\citet{2005ApJ...631..628M} found that the flare productivity,
in terms of both GOES magnitude and frequency,
has a good correlation with fractal dimension.
They showed a threshold fractal dimension of 1.2 and 1.25
as a necessary requirement for an AR to produce M- and X-class flares,
respectively, within next 24 hour period.
Interestingly,
\citet{2005ApJ...631..628M} also found that
the frequency distributions of the fractal dimension
for different Mount Wilson classes
($\alpha$, $\beta$, $\beta\gamma$, $\beta\gamma\delta$) are similar to each other
with a mean fractal dimension of 1.32.
Perhaps this result indicates that,
for the production of strong flares,
the complexity of mid-to-small scales
(smaller than the whole AR: detected by the fractal dimension analysis)
has to exist along with the large-scale complexity
(AR size: characterized by the Mount Wilson class).

\begin{figure*}
  \centering
  \includegraphics[width=0.8\textwidth]{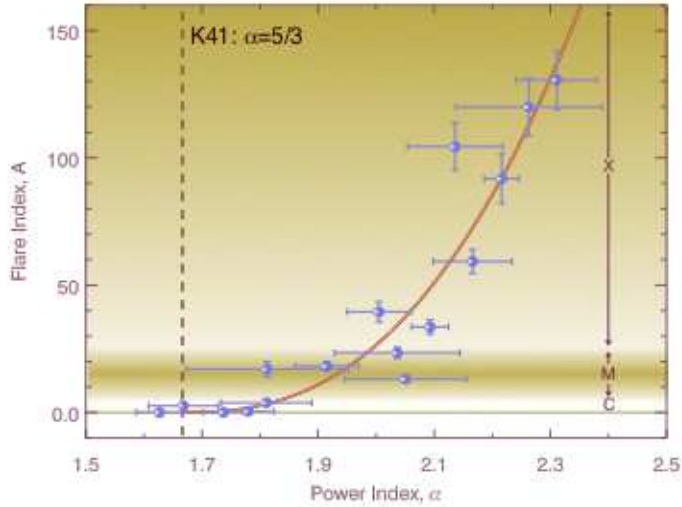}
  \caption{Power-law index $\alpha$ for 16 ARs
    of different flare index (denoted as $A$ in this panel).
    The dashed vertical line indicates $\alpha=5/3$
    for the Kolmogorov's turbulence theory.
    The positive relationship between the flare productivity
    and the power-law index is clearly illustrated.
    {Image reproduced by permission from \citet{2005ApJ...629.1141A},
    copyright by AAS.}
    }
\label{fig:abramenko2005}
\end{figure*}

Importance of small-scale fields in the flare production
is also demonstrated by plotting the power spectra of magnetograms.
\citet{2005ApJ...629.1141A} calculated the power-law index $\alpha$
of the magnetic power spectrum $E(k)\sim k^{-\alpha}$
of the magnetograms for 16 ARs,
where $k$ being the spatial wavenumber,
and compared $\alpha$ with the flare index $FI$,
which represents the flare productivity of a given AR:
\begin{eqnarray}
  FI=\frac{1}{\tau}
  \left[
    100\sum_{i}I_{X}+10\sum_{j}I_{M}+1.0\sum_{k}I_{C}+0.1\sum_{l}I_{B}
  \right],
  \label{eq:flareindex}
\end{eqnarray}
where $I_{X}$, $I_{M}$, $I_{C}$, and $I_{B}$ are the GOES magnitudes
of X-, M-, C-, and B-classes, respectively,
that occurred in a given AR in the period of $\tau$ days,
and indices $i$, $j$, $k$, and $l$ designate flares in each class.
As shown in Fig.~\ref{fig:abramenko2005},
it was revealed that higher flare productivity
is associated with steeper spectrum:
the power-law index is $\alpha >2.0$ for ARs producing X-class flares
and is $\alpha\approx 5/3$ for flare-quiet ARs
\citep[i.e., regime of classical Kolmogorov turbulence;][]{1941DoSSR..30..301K}.
Although not mentioned in \citet{2005ApJ...629.1141A},
the above result might also be explained by the observation
that larger ARs tend to produce stronger flares \citep[e.g.,][]{2000ApJ...540..583S}:
the spatial power spectrum of a large AR would have more power at low wavenumbers
but have the same power at higher wavenumbers,
which leads to a steeper power spectrum for a larger AR.

The works introduced in this subsubsection essentially show the fractal, multi-fractal,
and/or turbulent nature of flaring ARs
\citep{2002ApJ...577..487A,2003ApJ...597.1135A,2010ApJ...722..122A,2010AdSpR..45.1067M,2012SoPh..276..161G}.
Regarding the practical flare prediction,
\citet{2005SoPh..228....5G} revealed, however, that
the fractal dimension does not have significant predictability.
Rather, they suggested that the temporal evolution of the fractal diagnostics
may be practically useful in flare prediction.

\subsubsection{(Im)balance of electric currents}
\label{subsubsec:currents}

Magnetic energy that is released in solar flares
stems from the non-potential,
magnetic field associated with electrical currents.
An important and long-standing question about the electric current is
whether or not the current is neutralized in ARs, and, if not, to what extent and how
\citep[e.g.,][]{1991ApJ...381..306M,1995ApJ...451..391M,1996ApJ...471..497M,1996ApJ...471..489P}.

For the violation of current neutralization,
two basic mechanisms have been proposed,
which are (1) the magnetic field lines are stressed and twisted
by photospheric and sub-photospheric flow motions
\citep[e.g.,][]{1992ApJ...385..344K,2003A&A...406.1043T,2015ApJ...810...17D};
and (2) the current is provided by the emergence of twisted,
i.e., current-carrying flux tubes
\citep[e.g.,][]{1996ApJ...462..547L,2000ApJ...545.1089L,2001ApJ...554L.111F}.

The current neutralization is investigated by examining
whether the total electric current integrated
over a single magnetic polarity of an AR vanishes.
This is equivalent to whether the main (direct) current of a flux tube
is surrounded by the shielding (return) current
of equal strength and opposite direction.
A number of observers have tried to address this issue
by measuring the longitudinal (vertical) component of electric current density
from the vector magnetogram,
\begin{eqnarray}
  j_{z}=\frac{c}{4\pi} \left[ {\vec \nabla}\times {\vec B} \right]_{z}
  =\frac{c}{4\pi} \left(
  \frac{\partial B_{y}}{\partial x} - \frac{\partial B_{x}}{\partial y}
  \right),
  \label{eq:jz}
\end{eqnarray}
where $c$ is the speed of light.
Whereas \citet{1992ApJ...392L..39W} stated that
their data do not convincingly show a non-neutralized current system,
many observations have consistently suggested
the existence of twisted flux systems, in favor of the scenario (2)
(see a variety of observations introduced in previous sections).
To cite a case,
\citet{2000ApJ...532..616W} examined vector magnetograms for 21 ARs
and found that the electric currents in the positive and negative polarities
significantly deviated from zero in more than half of the ARs studied,
indicating that the AR currents are typically not neutralized.
Using vector magnetograms of the highest quality by Hinode/SOT/SP,
\citet{2012ApJ...761...61G} investigated the distribution of currents
in a flaring/eruptive AR (NOAA 10930) and a flare-quiet one (NOAA 10940).
They found that substantial non-neutralized currents are injected
along the photospheric PILs
and that more intense PILs yield stronger non-neutralized currents.
From statistical studies,
\citet{2017ApJ...846L...6L} and \citet{2017SoPh..292..159K} showed that
the flare- and CME-producing ARs are characterized
by strong non-neutralized currents.

However, because the measurement of electric currents is strongly hampered
by the limited resolution and ambiguities of magnetogram,
it has always been a challenging task to
accurately evaluate the distribution of currents
as in Eq. (\ref{eq:jz}).
Therefore, to figure out whether the ARs are born with net currents,
it is desirable to enlist the aid of numerical modeling
\citep[][see Sect.~\ref{subsec:num_fe}]{2014ApJ...782L..10T}.

\subsection{Atmospheric and subsurface evolutions}
\label{subsec:atm}

\subsubsection{Formation of flux ropes: sigmoids and filaments}
\label{subsubsec:sigmoids}

\begin{figure*}
  \centering
  \includegraphics[width=1.0\textwidth]{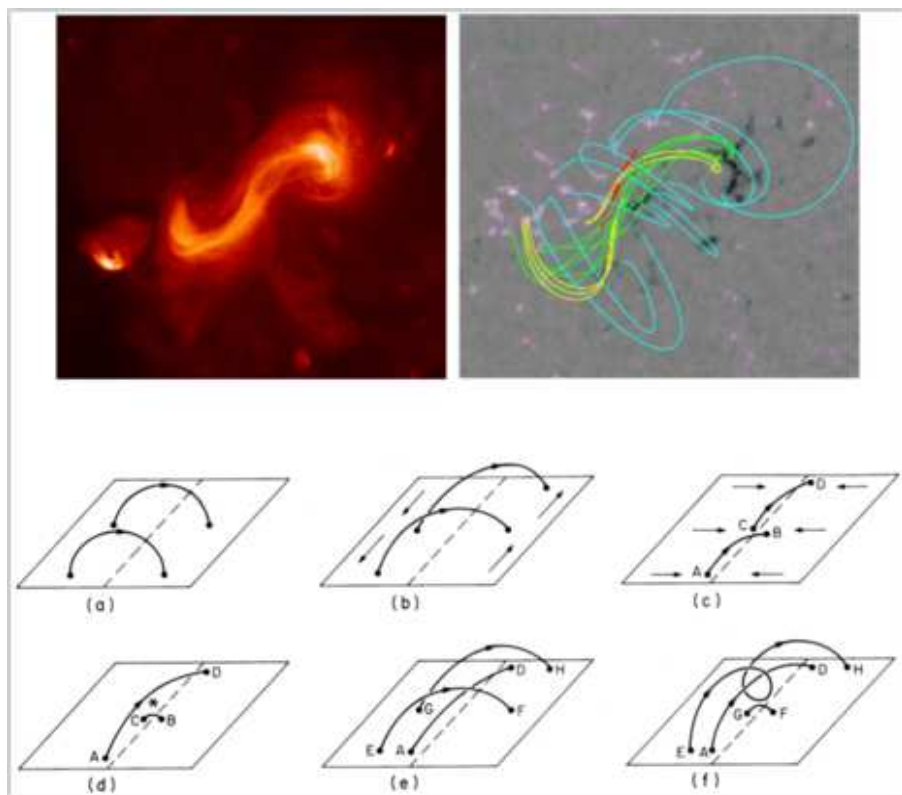}
  \caption{(Top left) Hinode/X-Ray Telescope \citep[XRT;][]{2007SoPh..243...63G} image
    of the sigmoid observed on February 12, 2007.
    (Top right) Field lines traced from the NLFFF extrapolation model.
    The cyan field lines belong to the potential arcade.
    The yellow {\sf J}-shaped and the green {\sf S}-shaped field lines
    are part of the flux rope,
    and the short red field lines lie under the flux rope.
    The background shows the LOS magnetogram.
    {Image reproduced by permission from \citet{2012ApJ...750...15S},
    copyright by AAS.}
    (Bottom) Filament formation model based on the flux cancellation scenario.
    Field lines above the PIL (dashed line) become sheared and converged
    due to the photospheric motions (panels a to c).
    Magnetic reconnection then produces
    a long overlying loop (A--D in panel d)
    and a short field line that submerges (B--C).
    Overlying arcades are further sheared and converged
    to produce a flux rope (panels e and f).
    {Image reproduced by permission from \citet{1989ApJ...343..971V},
    copyright by AAS.}
    }
\label{fig:savcheva2012a}
\end{figure*}

In flare-productive ARs,
free magnetic energy is stored in non-potential coronal fields
that harbor significant amount of shear and twist.
When observed in soft X-rays,
these coronal fields display forward or inverse {\sf S}-shaped structures,
which was first observed by \citet{1992Sci...258..618A}
and are called ``sigmoids'' \citep{1996ApJ...464L.199R}:
see review by \citet{2006SSRv..124..131G}.
Figure~\ref{fig:savcheva2012a}(top) shows a typical example of a sigmoid.
One may find that its structure is in good agreement
with the extrapolated coronal fields,
which shows the form of a magnetic flux rope.
From the statistical analysis of the data from Yohkoh's Soft X-ray Telescope
\citep[SXT;][]{1991SoPh..136...37T},
\citet{1999GeoRL..26..627C} revealed that
ARs are significantly more likely to be eruptive
if they are either sigmoid or large:
51\% of all ARs analyzed are sigmoid and they account for 65\% of the observed eruptions.
This result attracted interest in sigmoids as precursors of flare eruptions,
and the trend was confirmed later
by \citet{2007ApJ...671L..81C}, \citet{2014SoPh..289.3297S},
and \citet{2018ApJ...869...99K}.

Sigmoids are often accompanied by H$\alpha$ filaments
\citep[e.g.,][]{1996ApJ...473..533P,2002SoPh..207..111P},
and they form above and along the PILs in the evolving ARs.
It is therefore important to understand the formation mechanism of sigmoids
in relation to the large-scale/long-term evolution of the photospheric fields
(as we saw earlier in Sects.~\ref{subsec:deltaspots} and \ref{subsec:photo}).
In fact, the series of sigmoid observations indicate that
they are created in the manner anticipated in the filament formation model
by \citet{1989ApJ...343..971V} (see Fig.~\ref{fig:savcheva2012a}(bottom)),
in which the shearing and converging flow around the PIL
drives flux cancellation and twists up the arcade fields
to create a flux rope \citep[see also][]{2001ApJ...558..872M}.\footnote{It is also suggested that the flux ropes emerge bodily from below the surface \citep[e.g.,][]{1995ApJ...446..877L,2008ApJ...673L.215O}.}

\begin{figure*}
  \centering
  \includegraphics[width=0.75\textwidth]{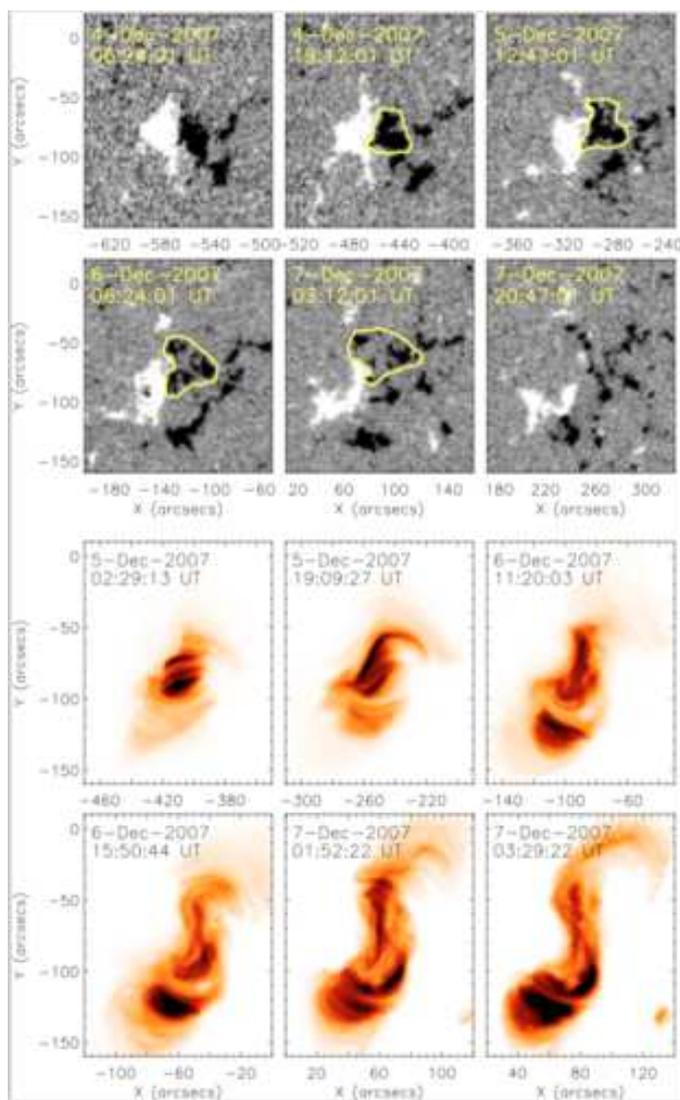}
  \caption{Day to day evolution of AR NOAA 10977.
    (Top) SOHO/MDI magnetogram saturating at $\pm 100\ {\rm G}$.
    (Bottom) Hinode/XRT C Poly filter images
    showing the transition from a sheared arcade to a sigmoid.
    {Images reproduced by permission from \citet{2011A&A...526A...2G},
    copyright by ESO.}
    }
\label{fig:green2011}
\end{figure*}

Figure~\ref{fig:green2011} is one of the most compelling examples
of the sigmoid formation through spot evolution \citep{2011A&A...526A...2G}.
At the central PIL of this AR,
about one third of the magnetic flux cancels in 2.5 days before the flare eruption
and the photospheric field shows an apparent shearing motion (top panels).
At the same time, the coronal structure transforms
first from a weakly to a highly sheared arcade
then to a sigmoid that lies over the PIL (bottom panels).
The sigmoid flux rope erupts eventually during the GOES B1.4-class flare,
leaving an arcade structure in soft X-ray images
\citep{1997ApJ...491L..55S,1998GeoRL..25.2481H,2000ApJ...532..628S}.
A similar long-term transition of coronal fields
from a sheared arcade or a pair of {\sf J}-shaped loops to the sigmoid
was also observed by \citet{2009ApJ...698L..27T},
\citet{2009ApJ...700L..83G}, and \citet{2012ApJ...759..105S}.
From these observations,
one can infer that the twisted flux rope in a flaring AR
is formed above the PIL
due to the photospheric driving before the eruption.

Then, it is natural to speculate that
magnetic helicity is the cause of the flux rope structure.
To this end, \citet{2005ApJ...624.1072Y} analyzed three sigmoid ARs
and found that in two regions,
the magnetic helicity injected through the sigmoid footpoints
is comparable to the helicity content of the sigmoid loops.
However, this is not true for the other AR,
which may be because the sigmoid consists of multiple loops.
They concluded that, excluding the latter complex AR,
the magnetic twist of sigmoids is consistent with
the helicity injected from the sigmoid footpoints.
Investigating various filament eruption events associated with sigmoids,
\citet{2007SoPh..246..365G} showed that the structure of a sigmoid
agrees with the helicity of a filament
(e.g., forward {\sf S}-shaped sigmoid for positive helicity filament)
and that the rotation of a filament apex during the eruption
is consistent with the helicity of the filament
(e.g., clockwise rotation for positive helicity filament).
The authors found that these behaviors agree with
the kink instability scenario
as numerically modeled by \citet{2005ApJ...630L..97T}.

\begin{figure*}\sidecaption
  \centering
  \includegraphics[width=0.65\textwidth]{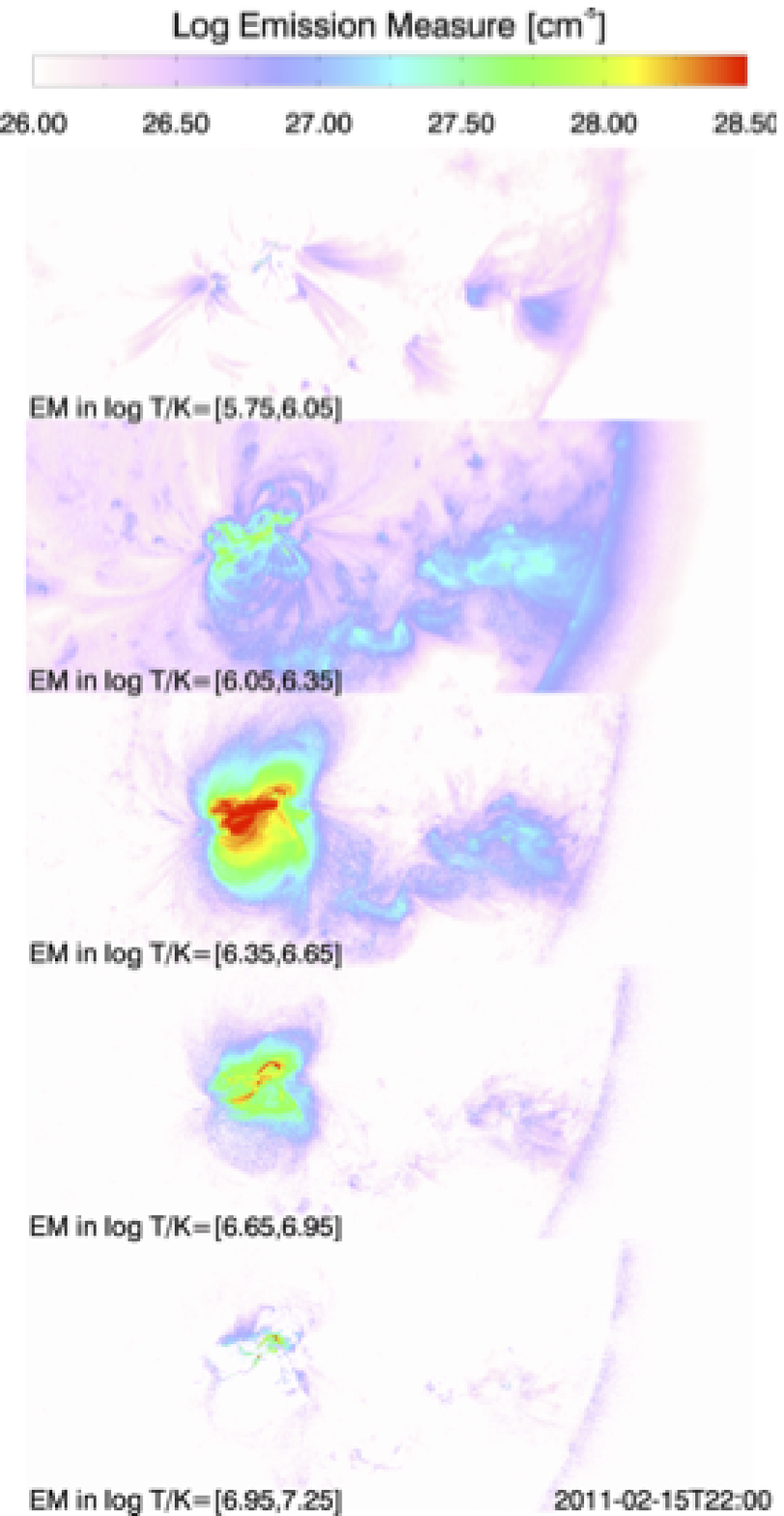}
  \caption{DEM maps of AR NOAA 11158 with the FOV of $1200''\times 480''$
    centered at $(600'', -268'')$.
    The color indicates the total EM contained within a $\log{(T\ {\rm [K]})}$ range
    indicated in the bottom left corner of each panel.
    {Image reproduced by permission from \citet{2015ApJ...807..143C},
    copyright by AAS.}
    }
\label{fig:cheung2015}
\end{figure*}

Thermal structures of sigmoid ARs have been investigated
by differential emission measure (DEM) analysis
\citep[for detailed account of this method, see Sects. 7 and 8 of][]{2018LRSP...15....5D}.
For instance,
the DEM maps of AR NOAA 11158 in Fig.~\ref{fig:cheung2015},
calculated from six EUV images of SDO/AIA by \citet{2015ApJ...807..143C},
clearly reveals that
a hot core structure is embedded in the center of AR
($\log{(T {\rm [K]})}>6.6$)
and covered by cooler overlying loops
($\log{(T {\rm [K]})}\lesssim 6.3$).
\citet{2016A&A...588A..16S} analyzed the pre-eruptive phase of NOAA 11429,
which is responsible for the two consecutive X-class flares with fast CMEs,
using data from both AIA and
Hinode's EUV Imaging Spectrometer \citep[EIS;][]{2007SoPh..243...19C}.
They found that the mean DEM of the flux ropes
in the temperature range of $\log{(T {\rm [K]})}=6.8$--7.1
gradually increased by an order of magnitude
about five hours before the CME eruption.
This increase was associated with the rising of the flux rope
and may be related to the observed heating in CME cores
\citep{2012ApJ...761...62C,2013A&A...553A..10H},
although the physical relationship with instabilities is not clear.

\subsubsection{Broadening of EUV spectral lines prior to flares}
\label{subsubsec:nonthermal}

\begin{figure*}
  \centering
  \includegraphics[width=0.75\textwidth]{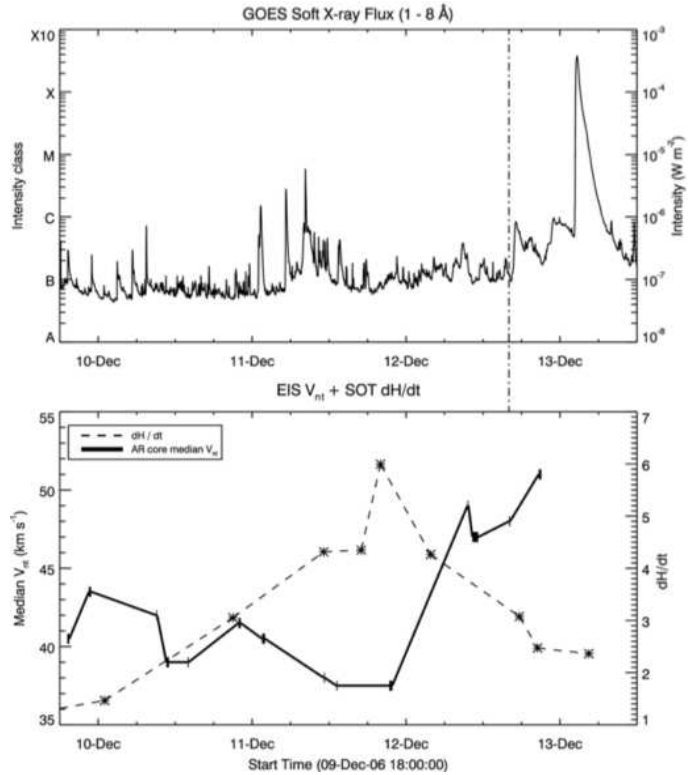}
  \caption{(Top) GOES soft X-ray light curve from December 9 to 13, 2006.
    The X3.4-class flare occurs at 02:14 UT on December 13.
    (Bottom) Helicity injection rate ($dH_{\rm R}/dt$)
    in the unit of $10^{36}\ {\rm Mx}^{2}\ {\rm s}^{-1}$,
    measured by Hinode/SOT/SP by \citet{2008PASJ...60.1181M} (asterisks with dashed line).
    The median of the top 95th percentile of non-thermal velocities
    observed in the AR core ($v_{\rm nt}$) for Hinode/EIS Fe\,{\sc xii} 195 {\AA} line
    is also plotted (solid line).
    The vertical dash-dotted line denotes the time of
    the third EIS measurement of December 12.
    {Image reproduced by permission from \citet{2009ApJ...691L..99H},
    copyright by AAS.}
    }
\label{fig:harra2009}
\end{figure*}

Another possible atmospheric response
to the photospheric evolution
is the pre-flare non-thermal broadening of coronal EUV spectral lines.
The observed line width consists of thermal width, instrumental width,
and non-thermal (excess) broadening, which are related via
\begin{eqnarray}
  W_{\rm obs}^{2}=W_{\rm inst}^{2}+4\ln{2}
  \left( \frac{\lambda}{c} \right)^{2}
  \left( v_{\rm t}^{2}+v_{\rm nt}^{2} \right),
\end{eqnarray}
where $W_{\rm obs}$ and $W_{\rm inst}$ are the observed and instrumental widths, respectively,
$\lambda$ the wavelength of the emission line, $c$ the speed of light,
$v_{\rm t}$ the thermal velocity, and $v_{\rm nt}$ the non-thermal velocity.

\citet{1998ApJ...494L.235A}, \citet{2000A&A...364..859R}, and \citet{2001ApJ...549L.245H}
showed that the non-thermal broadening peaks in the early phase of,
or even tens of minutes, before the flare occurrence,
and suggested that the broadening indicates turbulence
that is related to the flare triggering mechanism.
However, \citet{2009ApJ...691L..99H} revealed that
the pre-flare broadening starts much earlier.
They measured the non-thermal velocity of Fe\,{\sc xii} 195 {\AA} line
using Hinode/EIS
and found that,
as shown in Fig.~\ref{fig:harra2009},
the increase in the line width begins
up to one day before the X-class flare occurs
after the helicity injection saturates
\citep{2008PASJ...60.1181M}.
\citet{2014PASJ...66S..17I} revisited this event and showed that
this pre-flare broadening occurs in concurrence with
upflow of about 10 to $30\ {\rm km\ s}^{-1}$.
They speculated that the upflow indicates the expansion of outer coronal loops
and this rising motion (observed as the Doppler blueshift) causes the excess broadening.

\subsubsection{Helioseismic signatures in the interior}
\label{subsubsec:seismology}

\begin{figure*}
  \centering
  \includegraphics[width=1.\textwidth]{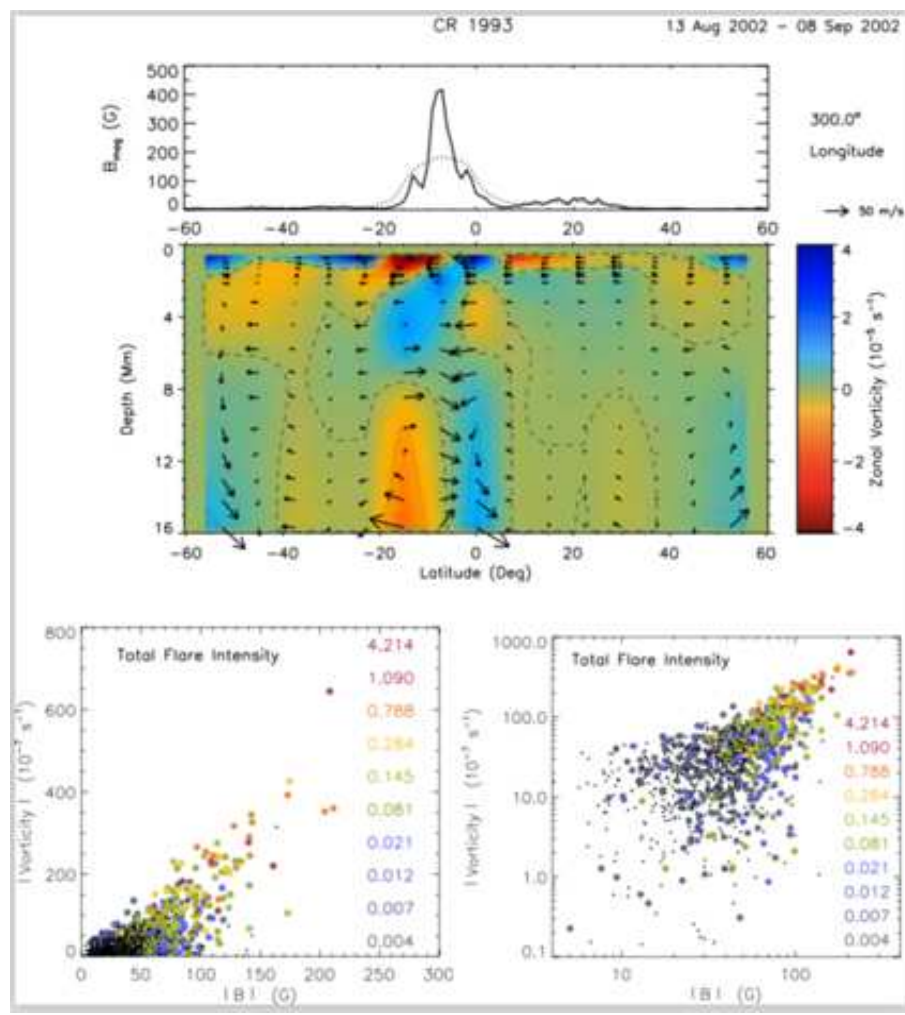}
  \caption{(Top) Vorticity distribution beneath a sample AR.
    The upper panel shows the latitudinal distribution of the unsigned magnetic flux
    across AR NOAA 10096 (solid) and that binned over 15$^{\circ}$ (dashed),
    whereas the lower panel displays
    the zonal vorticity component (the east-west component: $\omega_{x}$)
    as a function of latitude and depth,
    with arrows denoting the meridional flows.
    The strong zonal vorticity of opposite sign
    is concentrated at the location of the AR.
    {Image reproduced by permission from \citet{2006ApJ...645.1543M},
    copyright by AAS.}
    (Bottom) Total flare intensity of ARs during their disk passage
    (in the unit of $10^{-3}\ {\rm W\ m}^{-2}$,
    i.e. relative to an X10 flare)
    as a function of unsigned maximum magnetic flux density
    and unsigned subsurface vorticity at $-12\ {\rm Mm}$,
    plotted in linear scale to focus on large values (left)
    and logarithmic scale to focus on small ones (right).
    The colors indicate the maximum intensity of each subset.
    Black symbols are non-flaring ARs.
    {Image reproduced by permission from \citet{2009JGRA..114.6105K},
    copyright by ***.}
    }
\label{fig:mason2006}
\end{figure*}

Given the complex features of magnetic fields in flaring ARs,
it is natural to ask if there is any subsurface counterpart.
One of the earliest attempts to apply
the local helioseismology techniques
to search for the statistical relation between the subsurface flow field
and the flare occurrence
was done by \citet{2006ApJ...645.1543M}: Fig.~\ref{fig:mason2006} (top).
They applied the ring-diagram method to
408 ARs from the Global Oscillation Network Group (GONG) data
and 159 ARs from the SOHO/MDI data
to measure the vorticity of flows
(${\vec \omega}=\nabla\times {\vec v}$)
and compared it with the total flare intensity
(equivalent to the flare index $FI$: Eq. (\ref{eq:flareindex})).
It was found that the maximum unsigned vorticity components
at a depth of about 12 Mm,
calculated from a synoptic maps of global subsurface flows
that are generated by averaging the ring-diagram flow fields over 7 days
\citep{2002ApJ...570..855H},
are correlated well with
the flare intensity greater than $3.2\times 10^{-5}\ {\rm W\ m}^{-2}$.
For flare activity below this value,
the relation was not apparent.
\citet{2009JGRA..114.6105K} expanded the analysis to 1009 ARs
including non-flaring ones.
As shown in the bottom panels of Fig.~\ref{fig:mason2006},
they demonstrated a clear relation
between the magnetic flux
density (total magnetic flux averaged over area: in the unit of G)
and vorticity for flaring ARs
(correlation coefficient $CC=0.75$).
The non-flaring ARs show a similar trend
but the correlation is weaker ($CC=0.5$)
and the mean values of flux and vorticity are smaller.
The authors concluded that the inclusion of vorticity
helps to distinguish between flaring and non-flaring regions.

\citet{2010ApJ...710L.121R} put more focus
on the temporal evolution of subsurface flow fields.
By analyzing 1023 ARs with the ring-diagram method, they showed that
(1) at first, about 2--3 days before the flare occurrence,
the kinetic helicity density,
${\vec v}\cdot {\vec \omega}={\vec v}\cdot(\nabla\times {\vec v})$,
has a large spread in values with depth,
but the spread decreases
on the days of the flares,
and that (2) the degree of shrinking is greater for stronger flares.
The observed tendency lends support to the interpretation that
the subsurface rotational turbulent flows
twist the magnetic fields
into unstable configurations and drives the flare eruptions.
\citet{2011SoPh..268..389K} further applied discriminant analysis
to various magnetic and subsurface flow parameters
and found that the subsurface parameters improve the ability
to distinguish between the flaring and non-flaring ARs.
The most important parameter is the structure vorticity,
which estimates the horizontal gradient
of the horizontal vorticity components.

As an independent ring-diagram study,
\citet{2014Ap&SS.352..361L} compared the flare activity levels of 77 ARs
and the quantities that describe the subsurface structural disturbances.
According to the author,
there was no remarkable correlation between these parameters.

Another approach is to apply time-distance helioseismology.
Using the sequential SDO/HMI data of five flare-productive ARs,
\citet{2012ApJ...761L...9G,2014SoPh..289..493G} compared
the kinetic helicity density
measured from the subsurface velocity maps
and the current helicity density
calculated from the photospheric vector magnetograms,
${\vec B}\cdot(\nabla\times {\vec B})$,\footnote{Not to be
confused with the magnetic helicity density,
${\vec B}\cdot {\vec A}={\vec A}\cdot(\nabla\times {\vec A})$:
see Sect.~\ref{subsubsec:helicity}.}
and found a good correlation between the two values.
They found that eight out of a total of 11 events show a drastic amplitude change
of the kinetic helicity density,
and five of them are accompanied by flares stronger than M5.0 level within eight hours,
either before or after the amplitude change.
The spread of the kinetic helicity density in depth also showed strong variations,
which confirms the observational result of \citet{2010ApJ...710L.121R}.

\citet{2016ApJ...819..106B} used helioseismic holography
to more than 250 ARs observed between 2010 and 2014.
They found that individual ARs show mostly variations
associated with non-flare related evolution,
although correlations between the flare soft X-ray flux and subsurface flow indices
are in general similar to those found previously
by \citet{2009JGRA..114.6105K}.
Moreover, they detected no remarkable precursors or other temporal changes
that are specifically associated with the flare occurrences.

It should be pointed out that
whereas not a small number of results have been reported,
there is no clear physical model
that explains the statistical correlations found
between flaring and various properties of subsurface flows.
For instance, it is not clear why the subsurface vorticity is correlated with AR flux,
better for the flaring ARs
than for the non-flaring ARs (Fig.~\ref{fig:mason2006}).
Therefore, further investigation,
probably with the aid of numerical simulations, is required
to interpret the observational results.

The difficulty resides also in the observational techniques.
In many cases,
the existence of strong magnetic flux (i.e. ARs) is assumed as a small perturbation
when solving the linear inverse problem in seismology.
However, this may not be true \citep[see][Sect.~3.7]{2005LRSP....2....6G}.
Development of seismology techniques, again with the assistance of modeling,
may overcome this shortcoming
and deepen our understanding of subsurface evolutions.

\subsection{Summary of this section}
\label{subsec:longterm_summary}

In this section,
we have reviewed the important observational characteristics
that are created in the long-term and large-scale evolution of flare-productive ARs.
Many of these characteristics manifest the morphological and magnetic complexity
of such ARs
and prove the inherent high non-potentiality of the magnetic system.

The $\delta$-spots,
in which the umbrae of both polarities share a common penumbra
(Sect.~\ref{subsec:spots}),
are formed in three ways
(Sect.~\ref{subsec:deltaspots}):
Type 1 (Spot-spot), the tightly packed sunspot with multiple bipoles intertwined;
Type 2 (Spot-satellite), where a newly emerging flux appears
in close proximity to a pre-existing spot;
and Type 3 (Quadrupole), the head-on collision of two neighboring bipoles.
However, X-class flares also emanate from between two separated ARs,
albeit rarely (Inter-AR).
The $\delta$-spots develop
the strong-field, strong-gradient, highly-sheared PILs,
which sometimes show a magnetic channel,
a narrow lane structure consisting of elongated flux threads of opposite polarities
(Sect.~\ref{subsubsec:photo_pil}).
These magnetic evolutions are caused by the shearing and converging flows around the PIL,
where as remarkable sunspot rotations, both the self and mutual rotations,
are also observed
(Sect.~\ref{subsubsec:spotrotation}).

Injection of magnetic helicity is found to have
temporal correlation with flare productivity,
while X-class flares require a significantly higher amount of helicity injection
(Sect.~\ref{subsubsec:helicity}).
The magnetic tongue structure is thought to be
the manifestation of emergence of twisted magnetic flux
and is used as a proxy of magnetic helicity sign
(Sect.~\ref{subsubsec:tongue}).
In studies addressing the old question of whether AR currents are neutralized or not,
the preponderance of recent evidence supports
the view that electric currents are not neutralized,
particularly in regions prone to exhibit large flares
(Sect.~\ref{subsubsec:currents}).

Twisted flux ropes, observed as H$\alpha$ filaments and soft X-ray sigmoids,
can be produced in the atmosphere above the PILs
due to the shearing and converging flows and helicity injection,
which eventually erupt in the flares and evolves into CMEs
(Sect.~\ref{subsubsec:sigmoids}).

Though more extensive surveys are desired,
several works have shown that flaring ARs have
more smaller-scale features,
probably reflecting the morphological and magnetic complexity
(Sect.~\ref{subsubsec:tongue}),
coronal upflows with excess broadening of EUV emission lines
in response to the helicity injection
(Sect.\ref{subsubsec:nonthermal}),
and properties of vorticity in the convection zone
(Sect.~\ref{subsubsec:seismology}).

\begin{figure*}
  \centering
  \includegraphics[width=1.0\textwidth]{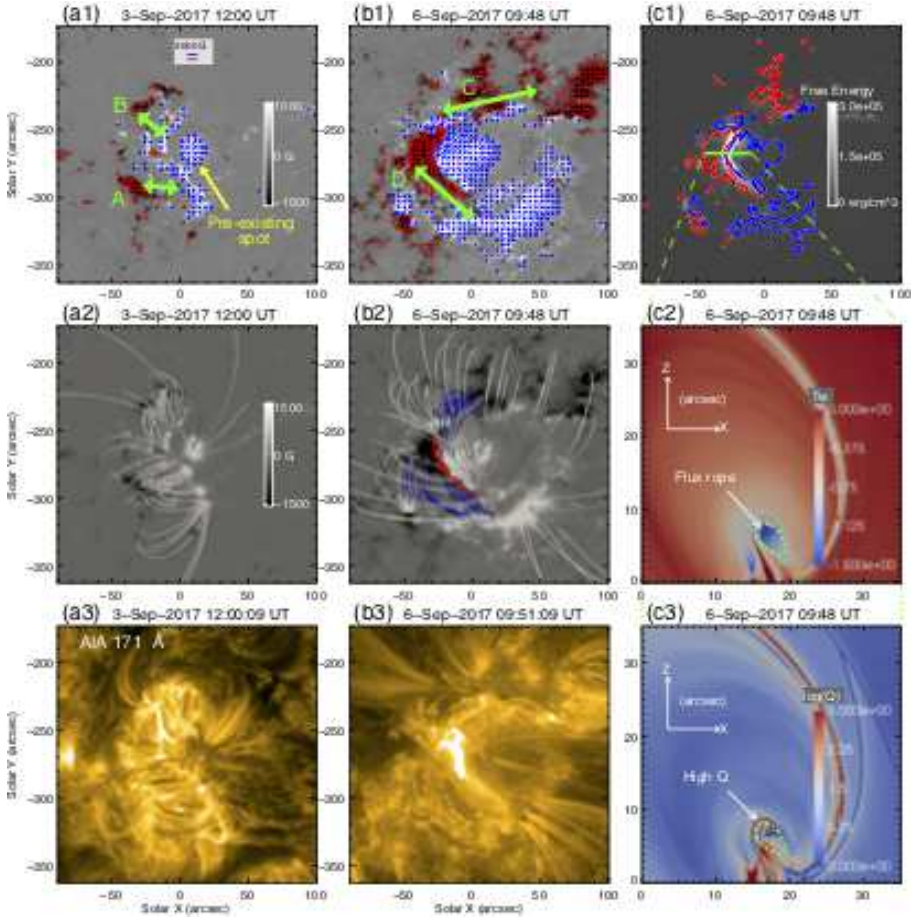}
  \caption{Evolution of AR NOAA 12673 and the formation of the flaring PIL.
    {Image reproduced by permission from \citet{2017ApJ...849L..21Y},
    copyright by AAS.}
    (a1--a3) SDO/HMI vector magnetograms at 12:00 UT on September 3,
    top view of the extrapolated field lines,
    and corresponding AIA 171 {\AA} image, respectively.
    (b1--b3) Similar to panels (a1)--(a3),
    but for the time at 09:48 UT on September 6.
    In panels (a1) and (b1), green arrows are overlaid to
    indicate bipoles A, B, C, and D,
    and yellow arrow shows the pre-existing sunspot.
    (c1) Free energy density corresponding to panel (b1)
    overlaid with the vertical magnetic field contours at $\pm 800\ {\rm G}$.
    Twist number $T_{w}$ \citep{2006JPhA...39.8321B}
    and squashing factor $Q$ \citep{1996A&A...308..643D,2002JGRA..107.1164T}
    distribution in the $x$-$z$ plane along the cut labeled in panel (c1).
    In panel (b2), the blue field lines connect the opposite patches of
    bipole C and bipole D, respectively,
    and the red field lines indicate a flux rope along the PIL.
    In panels (c2) and (c3), the green dotted curves outline
    the general shape of the flux rope.
    }
\label{fig:yang2017}
\end{figure*}

AR NOAA 12673,
which appeared in September 2017
and produced numerous flares including the X9.3-class event,
is characteristic of
the important features
introduced in this section.
Figure~\ref{fig:yang2017} by \citet{2017ApJ...849L..21Y}
shows the overall evolution and the formation of the flaring PIL.
This AR rotates on to the visible disk
as a simple $\alpha$-spot of positive polarity.
On September 3, two bipolar systems A and B suddenly emerge
to the east of the pre-existing central spot (panels a1--a3),
and two additional bipoles C and D emerge
more in the north-south direction within the first two pairs,
forming a highly complex $\delta$-spot (panels b1--b3).
This evolution reminds us of a Type-2 $\delta$-spot,
but at the same time the collision of the secondary bipoles C and D
is also reminiscent of the Type-3 structure.
\citet{2017RNAAS...1a..24S} pointed out that the emergence rate of this AR
is one of the fastest emergence events ever observed.

\begin{figure*}
  \centering
  \includegraphics[width=1.0\textwidth]{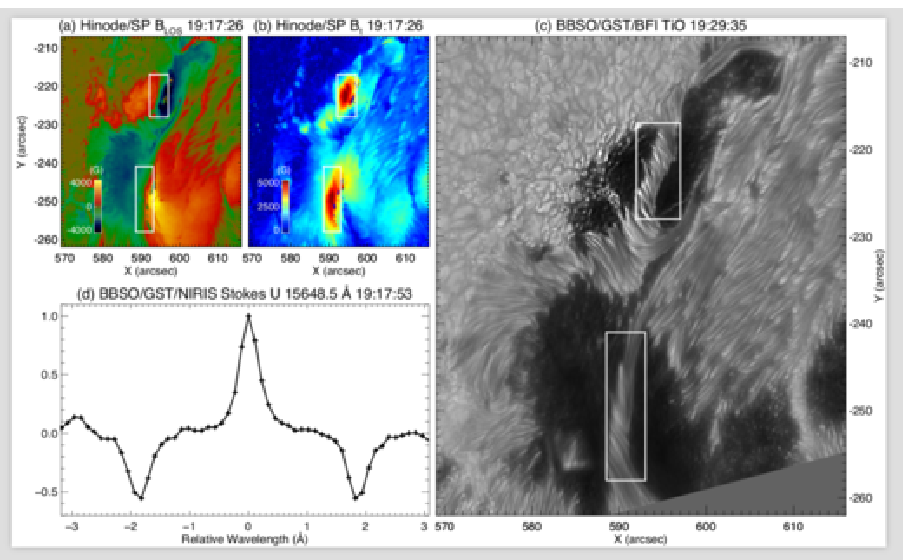}
  \caption{High-resolution observations of the flaring PIL of AR NOAA 12673.
    (a and b) Hinode/SOT/SP LOS and transverse magnetic field strength, respectively.
    Note that in many pixels near the PIL, transverse fields are saturated at 5000 G
    due to the limitation of inversion algorithm.
    (c) BBSO/GST TiO image. The two white boxes in (a)--(c) mark
    the two strong transverse field areas at the PIL,
    where twisted photospheric light-bridge structures
    of the $\delta$-configuration are present.
    (d) NIRIS Stokes-U profile of a selected strong transverse field pixel
    at the PIL within the northern box.
    The direct measurement of Zeeman splitting yields a field strength of 5570 G.
    {Image reproduced by permission from \citet{2018RNAAS...2a...8W},
    copyright by AAS.}
    }
\label{fig:wang2018}
\end{figure*}

As the negative polarity of D rapidly intrudes into the positive polarities,
it produces a strong-field, strong-gradient, highly-sheared PIL
(Fig.~\ref{fig:yang2017}: location where free energy is enhanced in panel c1).
According to \citet{2017ApJ...849L..21Y},
because the pre-existing central spot blocks the free development
of the newly-emerging fields,
the $B_{z}$ gradient at the PIL becomes much enhanced.
As Fig.~\ref{fig:wang2018} illustrates,
\citet{2018RNAAS...2a...8W} detected exceptionally strong transverse fields
of up to $5570\ {\rm G}$ around this PIL.
In the corona above this PIL,
a flux rope structure is clearly reproduced by the NLFFF modeling
(Fig.~\ref{fig:yang2017}: red field lines in panel b2),
which agrees well with the sigmoidal structure.
Moreover, \citet{2018A&A...612A.101V}, \citet{2018ApJ...853L..18Y},
and \citet{2019arXiv190109358V}
reported on the PIL shear flows, spot rotations,
and helicity injection, respectively,
which combined seems to activate the X9.3 flare.

\section{Long-term and large-scale evolution: theoretical aspects}
\label{sec:num}

As we saw in the preceding sections,
in its long history of solar observation,
a vast amount of key observational features
that differentiate the flare-productive ARs from the quiescent ones
have been discovered.
The essential questions we have are,
of course, how are they created
and what is the underlying physics?
The other side of solar physics,
the theoretical and numerical studies,
may provide answers to these questions.

Because there have already been substantial number of simulation models to date,
in order to offer the reader a guideline,
we introduce three genres of modeling,
following the discussion in \citet{2014LRSP...11....3C}.
The first group is the \emph{data-inspired} models,
which assume an ideal simulation setup
that is ``inspired'' by the observations.
Flux emergence and flux cancellation models fall into this group.
The second group is the \emph{data-constrained} models,
in which the models use observational data
at a single moment to drive computations.
The series of extrapolated magnetic fields,
computed from the sequential photospheric magnetogram,
is one representative model of this group.
However, it is less likely that
such static solutions are applicable to flare-producing,
i.e., dynamically evolving ARs.
So, another way of the data-constrained models
is to use the extrapolated field as the initial condition
and solve the time-dependent MHD equations to trace the temporal evolution.
The third group, the \emph{data-driven} models,
even utilizes a temporal sequence of observational data,
such as the series of magnetograms, to drive the models.

The flux emergence and flux cancellation models are introduced
in Sects.~\ref{subsec:num_fe} and \ref{subsec:num_fc}, respectively.
The data-constrained and data-driven models,
which are still rather the newcomers,
are jointly shown in Sect.~\ref{subsec:num_data}.

\subsection{Flux emergence models}
\label{subsec:num_fe}

The fundamental premise of the formation and evolution of flaring ARs
is that solar ARs are produced ultimately by emerging flux from the convection zone.
Therefore, it is not surprising that
many theoretical models have focused on
the evolution process of flaring ARs from below the surface of the Sun,
which we call the flux emergence models.
These models leverage the 3D flux emergence simulations,
such as those in Sect.~\ref{subsubsec:fe_photosphere_theory},
and try to capture some aspects of observed magnetic features of flaring ARs.
In fact, even classical models that configure a simple $\Omega$-loop
can explain some of the observed features.

\begin{description}
\item[Magnetic tongues:] As the series of observational studies predicted,
  magnetic tongues, the extended magnetic patches on the both sides of the PIL,
  are well reproduced by the emergence of a twisted flux tube
  (see, e.g., Fig.~\ref{fig:fan2001}(e)).
  \citet{2010A&A...514A..56A} compared the magnetogram of AR NOAA 10808
  and that produced in their numerical simulation
  and showed that the pattern of magnetic tongues
  depends on the azimuthal field of the emerging flux tube.

\item[Flux ropes and sigmoids:] It was \citet{2004ApJ...610..588M}
  who first reproduced the flux rope structures self-consistently
  in the 3D flux emergence simulation.
  In their model, where the buoyant segment of the flux tube is shorter
  than that of \citet{2001ApJ...554L.111F}'s model,
  the upper part of the emerged flux tube becomes detached
  from the main body and forms a coronal flux rope
  that erupts into the higher atmosphere
  as in a CME.
  \citet{2008A&A...492L..35A} explained the formation of a flux rope
  as magnetic reconnection between a set of emerging loops.
  Because the original flux tube is twisted,
  the emerged loops are sheared above the PIL
  and reconnect with each other, forming a flux rope structure.
  \citet{2009ApJ...691.1276A} revealed that the electric current sheets,
  which originally have a pair of {\sf J}-shaped configurations,
  are joined to form a sigmoid structure
  as observed in soft X-rays.
  Similar sigmoid structure was observed in the models by,
  e.g., \citet{2006ApJ...653.1499M}, \citet{2009ApJ...697.1529F},
  and \citet{2012A&A...537A..62A}.

\item[Shear flows:] The essential driver of the shear flows
  in the emergence simulations is the Lorentz force
  on the two sides of the PIL in opposite directions
  \citep{2001ApJ...547..503M}.
  When the twisted flux tube emerges into the atmosphere,
  the rapid expansion deforms the field lines of the flux tube
  and drives the shear flows around the PIL.
  \citet{2001ApJ...554L.111F} and \citet{2004ApJ...610..588M} explained
  the twisting up of the coronal field
  as a shear Alfv\'{e}n wave propagating upward,
  while \citet{2009ApJ...697.1529F} interpreted it as a torsional Alfv\'{e}n wave.
  The horizontal velocity vector of Fig.~\ref{fig:fan2001}(e) clearly displays
  the shear flows around the PIL.

\item[Helicity injection:] Injection of magnetic helicity flux
  through the photosphere was investigated by \citet{2003ApJ...586..630M},
  who revealed that in the earliest stage, the emergence term dominates,
  which then reduces and the shear term becomes
  the main source of the helicity injection for the rest of the period
  (see Sect.~\ref{subsubsec:helicity} for the definition of the terms).
  The helicity transport by the shear term is explained
  by the horizontal shearing and rotational motions
  at the footpoints of the emerged magnetic fields
  \citep{2000ApJ...545.1089L,2009ApJ...697.1529F}.

\item[Spot rotation:] This can be considered as the subtopic of the helicity injection.
  \citet{2000ApJ...545.1089L} proposed a theoretical model
  that treats both the expanded twisted flux tube in the corona
  and that remaining in the convection zone.
  In this model, as a twisted tube emerges,
  the torsional Alfv\'{e}n wave propagates downward into the convection zone
  due to the mismatch of twists between the two layers
  and causes the spot rotation.
  \citet{2003ApJ...586..630M} and \citet{2006ApJ...653.1499M} found that
  the rotational flows are formed in each of the spots
  soon after the rising flux tube becomes vertical,
  whereas \citet{2009ApJ...697.1529F} shows that significant vortical motions develop
  as a torsional Alfv\'{e}n wave propagates along the flux tube.
  \citet{2015A&A...582A..76S} used a toroidal tube model \citep{2009A&A...503..999H}
  and revealed that two sunspots do undergo rotation
  (not an apparent effect).
  They explained the rotation by unbalanced torque produced by magnetic tension.

\item[(Im)balance of electric currents:] \citet{2014ApJ...782L..10T} considered
  the emergence of a flux tube that contains neutralized electric currents
  (i.e., the situation where the direct current along the axis is
  balanced with the return current at the tube's periphery).
  As the significant emergence to the surface begins,
  the current rapidly deviates from the neutralized state
  and the total direct current remains several times larger
  than that of the return current throughout the whole evolution.
  They suggested that when the tube approaches the surface,
  the return current is pushed aside by the direct current.
  Also, most of the return currents remain beneath the surface
  because the tube does not undergo a bodily emergence.
  It was therefore concluded that ARs are born
  on the surface
  with substantial net electric currents.
\end{description}

The above features are formed as parts of relaxation processes
in which the twist of the flux tube
is released through the emergence from the convection zone to the corona.
However, in most of the these numerical models
that assume a simple buoyant emergence of flux tubes,
other important characteristics of flaring ARs,
such as tightly-packed $\delta$-spots with strong-field, strong-gradient, highly-sheared PILs,
are not reproduced.
The two photospheric footpoints of the emerging $\Omega$-loops
are prone to separation in a monotonous fashion
and never form a converged, $\delta$-shaped structure.
Therefore, to overcome this difficulty,
one needs to assume subsurface magnetic fields
with \emph{not-so-simple} configurations.

\subsubsection{Kinked tube model}
\label{subsubsec:kink}

The idea of the emergence of a kink-unstable magnetic flux tube is inspired by
the observations of flare-productive ARs, especially of Type 1 $\delta$-spots
(see Sect.~\ref{subsec:deltaspots}).
These regions have compact morphology and strong twists,
and the tilt often deviates so much from parallel to the equator
that sometimes it even violates Hale's polarity rule.
The 3D configurations inferred from the proper motion of the spots
strongly suggest the emergence of ``a knotted twisted flux tube''
\citep[][see Fig.~\ref{fig:tanaka1991}(a) of this article]{1991SoPh..136..133T}.

\begin{figure*}\sidecaption
  \centering
  \includegraphics[width=0.6\textwidth]{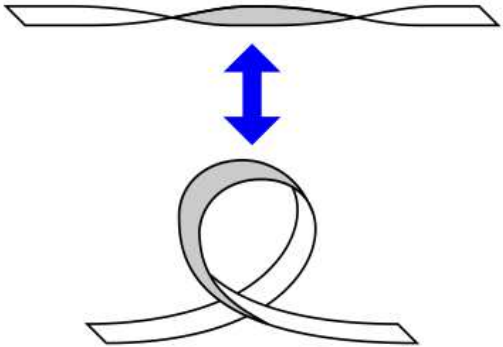}
  \caption{Conversion of twist and writhe.
    When a straight twisted ribbon (top) is loosened,
    the original twist converts into the writhe of the coiled ribbon (bottom).
    In an analogous way,
    a twisted flux tube deforms into a curled shape
    if the twist is sufficiently strong,
    which is the helical kink instability.
    }
\label{fig:ribbon}
\end{figure*}

According to \citet{1991LNP...387...39K},
it was \citet{1974SoPh...38..465P} who first proposed the concept
of emerging twisted flux tubes
for the energy source in the Alfv\'{e}n wave theory of solar flares.
In Appendix~\ref{app:kink}, we show the history about
who suggested the kink instability first
as the formation mechanism of the $\delta$-spots.

The helical kink instability is the instability of a highly-twisted flux tube,
in which the twist of the tube (turning of the field lines around the tube's axis)
is converted to writhe (turning of the axis itself)
due to the helicity conservation
\citep[see Fig.~\ref{fig:ribbon}:][]{1984JFM...147..133B,1992RSPSA.439..411M}.
It was applied to laboratory plasma
\citep[e.g.,][]{shafranov1957,1958PhFl....1..421K}
and to coronal plasma
\citep[e.g.,][]{1960MNRAS.120...89G,1968SoPh....3..298A,1972SoPh...22..425R,1980SoPh...66..113H,1981GApFD..17..297H},
before \citet{1996ApJ...469..954L} considered the kink instability
of flux tubes in a high-$\beta$ plasma.\footnote{Note that the kink instability is also suggested as one driving mechanism of CME eruption: see Sect.~\ref{subsec:flares}.}
For a uniformly twisted cylindrical flux tube
with the axial and azimuthal fields of $B_{x}(r)$ and $B_{\phi}(r)=qrB_{x}(r)$,
respectively,
where $r$ is the radial distance from the tube's axis
and the twist $q$ is constant,
the flux tube becomes unstable against the kink instability
when $q$ exceeds a critical value
\begin{eqnarray}
  q_{\rm cr}=a^{-1},
\end{eqnarray}
where $a^{-2}$ is the coefficient for the $r^{2}$ term
in the Taylor series expansion of the axial field $B_{x}$ about the flux tube:
$B_{x}(r)=B_{\rm tube}(1-a^{-2}r^{2}+\ldots)$.
In the case of commonly used Gaussian flux tubes, in which
\begin{eqnarray}
  B_{x}(r)=B_{\rm tube}\exp{\left(-\frac{r^{2}}{R_{\rm tube}^{2}}\right)}
\end{eqnarray}
and
\begin{eqnarray}
    B_{\phi}(r)=qrB_{x}(r),
\end{eqnarray}
with $R_{\rm tube}$ being the typical radius of the tube,
the critical twist is simply expressed as $q_{\rm cr}=R_{\rm tube}^{-1}$.
\citet{1996ApJ...469..954L} also argued that,
as the flux tube rises through the convection zone,
the originally stable tube may become unstable
because the tube expands ($R_{\rm tube}$ increases)
due to the decreasing surrounding pressure,
which lowers the critical twist ($q_{\rm cr}$ decreases).

\begin{figure*}
  \centering
  \includegraphics[width=0.95\textwidth]{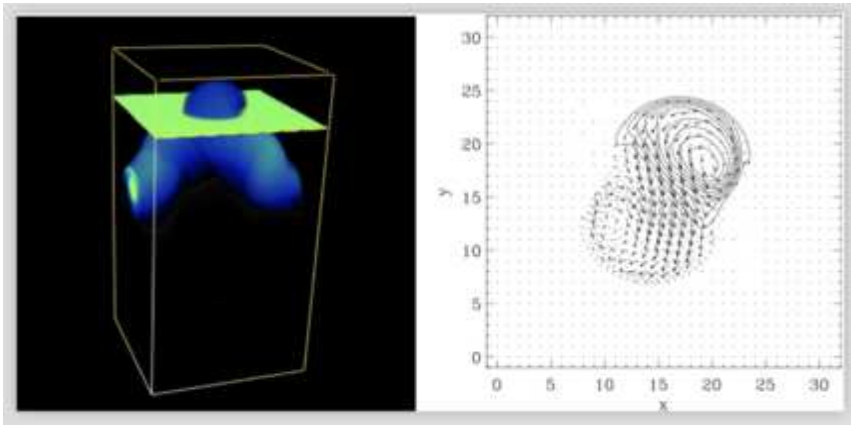}
  \caption{Emergence of a kink-unstable flux tube.
    {Image reproduced by permission from \citet{1998ApJ...505L..59F},
    copyright by AAS.}
    (Left) Snapshot of the flux tube during its rise
    as viewed from the side.
    The color shading indicates the absolute magnetic field strength.
    (Right) Horizontal cross-section of the upper portion of the flux tube
    (indicated by yellow plane in the left panel).
    The contours denote the vertical magnetic field $B_{z}$
    with solid line (dotted line) contours representing positive (negative) $B_{z}$.
    The arrows show the horizontal magnetic field.
    }
\label{fig:fan1998}
\end{figure*}

The first 3D non-linear simulation of the kink-unstable emergence
was done by \citet{1998ApJ...493L..43M}
for reproducing the sequence of sigmoid ARs
(top left panel of Fig.~\ref{fig:savcheva2012a}).
\citet{1998ApJ...507..404L,1999ApJ...522.1190L} performed linear and nonlinear calculations
of the kink instability in a uniform medium
without taking into account the effects
of gravity and stratification of external plasma.
Using the 3D anelastic MHD code,
\citet{1998ApJ...505L..59F,1999ApJ...521..460F} calculated the emergence
in an adiabatically stratified atmosphere
representing the solar convection zone (Fig.~\ref{fig:fan1998})
and found that, due to the kink instability,
the writhing of the tube increases the buoyancy at the apex
and accelerates the emergence.
The horizontal cross-section of the tube shows
a compact bipolar pair of $B_{z}$ with a highly sheared horizontal field along the PIL,
and the line connecting the two polarities
is deflected by more than 90$^{\circ}$ from its original orientation.
These structures are highly reminiscent of the $\delta$-spots.

\begin{figure*}
  \centering
  \includegraphics[width=0.9\textwidth]{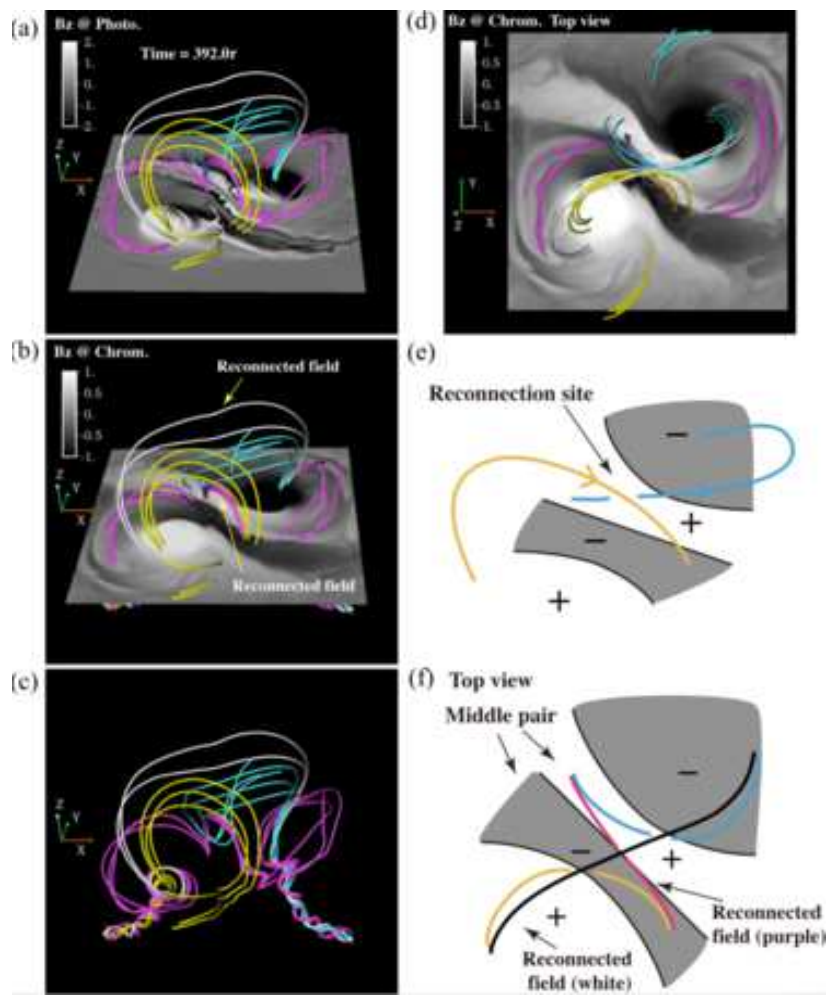}
  \caption{3D magnetic structure and photospheric and chromospheric fields $B_{z}$.
    Yellow and blue field lines denote the field lines
    passing by the current sheet between the two arcades.
    White field lines denote those enveloping the the arcade.
    Purple and white field lines denote those created by reconnection
    between the blue and yellow magnetic loops.
    (a)--(c) Bird's eye view. (d) Top view.
    (e) Schematic diagram of the magnetic field lines.
    (f) Schematic diagram of the magnetic field structure shown in panel (d).
    {Image reproduced by permission from \citet{2015ApJ...813..112T},
    copyright by AAS.}
    }
\label{fig:takasao2015}
\end{figure*}

However, because these emergence simulations were confined to the convection zone,
it remained unclear if the kinked tubes can really produce
observed characteristics when they emerge into the atmosphere.
To overcome this issue,
\citet{2015ApJ...813..112T} performed a fully compressible MHD simulation
in which a subsurface kink-unstable flux tube
rises from the convection zone seamlessly into the solar corona.
In their model,
the rising flux tube develops a knotted structure
as in the previous simulations \citep[e.g.,][]{1998ApJ...505L..59F,1999ApJ...522.1190L}
and, at the top-most convection zone,
it undergoes a strong horizontal expansion due to the strong stratification
and deforms into a pancake-like shape
(two-step emergence,
a commonly observed feature of large-scale flux emergence models:
see Sect.~\ref{subsubsec:fe_interior_theory} and Fig.~\ref{fig:toriumi2012}).
Interestingly,
as opposed to the simple bipolar structure
observed in the kinked tube simulations limited to the convection zone
(right panel of Fig.~\ref{fig:fan1998}),
the photospheric magnetogram in Fig.~\ref{fig:takasao2015}
shows a quadrupolar structure
consisting of the main bipolar pair of large roundish spots
that appears in the earlier phase
and the narrow, elongated middle pair formed later.
The middle pair is created due to the submergence of dipped fields,
which is a part of the emerged magnetic fields
(see also the accompanying movie).
The field lines in Fig.~\ref{fig:takasao2015} show that
magnetic reconnection takes place
between the two emerging loops (blue and yellow field lines)
and creates lower-lying and overlying post-reconnection field lines
(purple and white field lines, respectively).
Here, the lower-lying fields are almost parallel to the central PIL.
It is also found that,
as a consequence of Lorentz force exerted
by the two emerging loops (expanding arcades) on both sides of the central PIL,
a strong converging flow is excited around it
and the horizontal magnetic field becomes aligned more parallel to it.

Later, \citet{2018ApJ...864...89K} surveyed the evolution
of kink-unstable tubes with varying the twist intensity.
They revealed, for example, that the separation of both polarities on the surface
becomes smaller (i.e., more compact) with increasing the twist,
which underpins the kink instability
as a promising candidate for explaining $\delta$-spot formation.

It should be noted that the assumed twists
in these simulations may be too strong
compared to the twists of the actual ARs.
\citet{1994ApJ...425L.117P,1995ApJ...440L.109P} quantified the twist of ARs
by calculating the force-free parameter $\alpha$,
the constant of a force-free field $\nabla\times {\vec B}=\alpha {\vec B}$
(see Sect.~\ref{subsubsec:num_data_extrapolation})
measured from the vector field as
\begin{eqnarray}
  \alpha=\frac{\left[\nabla\times {\vec B}\right]_{z}}{B_{z}}
  =\frac{1}{B_{z}} \left(
  \frac{\partial B_{x}}{\partial y}-\frac{\partial B_{y}}{\partial x}
  \right),
\end{eqnarray}
and averaging it over the AR to obtain one global estimate of the twist.
The observed $\alpha$ is typically of the order of $0.01$ to $0.1\ {\rm Mm}^{-1}$
\citep[e.g.,][]{1995ApJ...440L.109P,1996ApJ...462..547L,1998ApJ...507..417L},
which yields $q\lesssim 0.1\ {\rm Mm}^{-1}$
under the simple relation of $\alpha\approx 2q$ \citep{1997ApJ...488..443L},
though there remains a possibility that
the observed ARs are inclined to regular, flare-quiet ones
due to selection bias.
On the other hand,
the threshold twist for the kink instability is, say,
$q_{\rm cr}=1\ {\rm Mm}^{-1}$ for the typical tube radius of 1 Mm
in the deeper convection zone.
Therefore, the twists of the flux tubes
assumed in the simulations, $q>q_{\rm cr}=1\ {\rm Mm}^{-1}$,
are at least one order of magnitude larger than the observed AR twists,
$q\lesssim 0.1\ {\rm Mm}^{-1}$,
even though each elementary bipole in ARs may satisfy the assumed condition
\citep{1999GMS...111...93L}.

\subsubsection{Multi-buoyant segment model}
\label{subsubsec:multibuoyant}

\begin{figure*}
  \centering
  \includegraphics[width=0.95\textwidth]{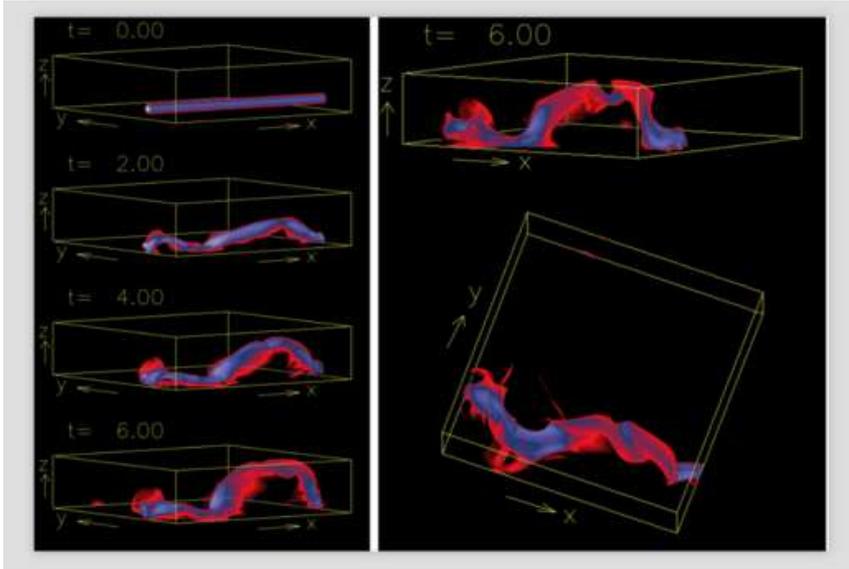}
  \caption{(Left) Evolution of the buoyant flux tube in the 3D convective flow
    for the case where the initial axial field is comparable to the equipartition field
    ($B_{\rm tube}=B_{\rm eq}$).
    The image shows the volume rendering
    of the absolute magnetic field strength of the flux tube.
    (Right) Two different views of the same tube at the final state,
    showing that the apex is pushed down by a local downflow.
    {Image reproduced by permission from \citet{2003ApJ...582.1206F},
    copyright by AAS.}
    }
  \label{fig:fan2003}
\end{figure*}

Type~3 $\delta$-spots like the quadrupolar AR NOAA 11158 (Fig.~\ref{fig:toriumi2014}),
in which two emerging bipoles collide against each other
to form a $\delta$-structure with a flaring PIL in between,
are redolent of a subsurface linkage of the two bipoles.
That is, the observed bipoles are the two emerging sections
of a single subsurface flux system,
distorted perhaps by convective buffeting during its rise
(Fig.~\ref{fig:tanaka1991}(c)).

An emerging flux tube can be affected by the convection
when the hydrodynamic force dominates
the restoring magnetic tension of the bent flux tube
\citep{2009LRSP....6....4F}:
\begin{eqnarray}
  \frac{B_{\rm tube}^{2}}{4\pi L} \lesssim C_{\rm D} \frac{\rho v^{2}}{\pi R_{\rm tube}},
\end{eqnarray}
which yields
\begin{eqnarray}
  B_{\rm tube} \lesssim \left(
    \frac{C_{\rm D}}{\pi} \frac{L}{R_{\rm tube}}
    \right)^{1/2} B_{\rm eq}
  \sim {\rm a\ few}\ B_{\rm eq},
\end{eqnarray}
where $B_{\rm eq}=(4\pi\rho)^{1/2}v$ is the equipartition field strength,
at which the magnetic energy density is comparable to
the kinetic energy density of convective flows,
$B_{\rm eq}^{2}/(8\pi)=\rho v^{2}/2$,
$L$ and $v$ are the size scale and speed of the convection, respectively,
and $C_{\rm D}$ is the aerodynamic drag coefficient,
which is of order unity.
At the bottom of the convection zone,
$(L/R_{\rm tube})^{1/2}=3$--$5$ and $B_{\rm eq}\sim 10\ {\rm kG}$
\citep{2009LRSP....6....4F}.
In fact,
\citet{2003ApJ...582.1206F} numerically demonstrated that
flux tubes of $B_{\rm tube}\sim B_{\rm eq}$ are
significantly influenced by turbulent convection.
As Fig.~\ref{fig:fan2003} shows,
the section of the emerging flux tube
within convective upflows is strongly pushed up
while the downdraft sections are pinned down.
To make things intriguing,
the apex of the rising $\Omega$-tube encounters another local downdraft
and takes an {\sf M}-shaped structure.

\begin{figure*}
  \centering
  \includegraphics[width=1.0\textwidth]{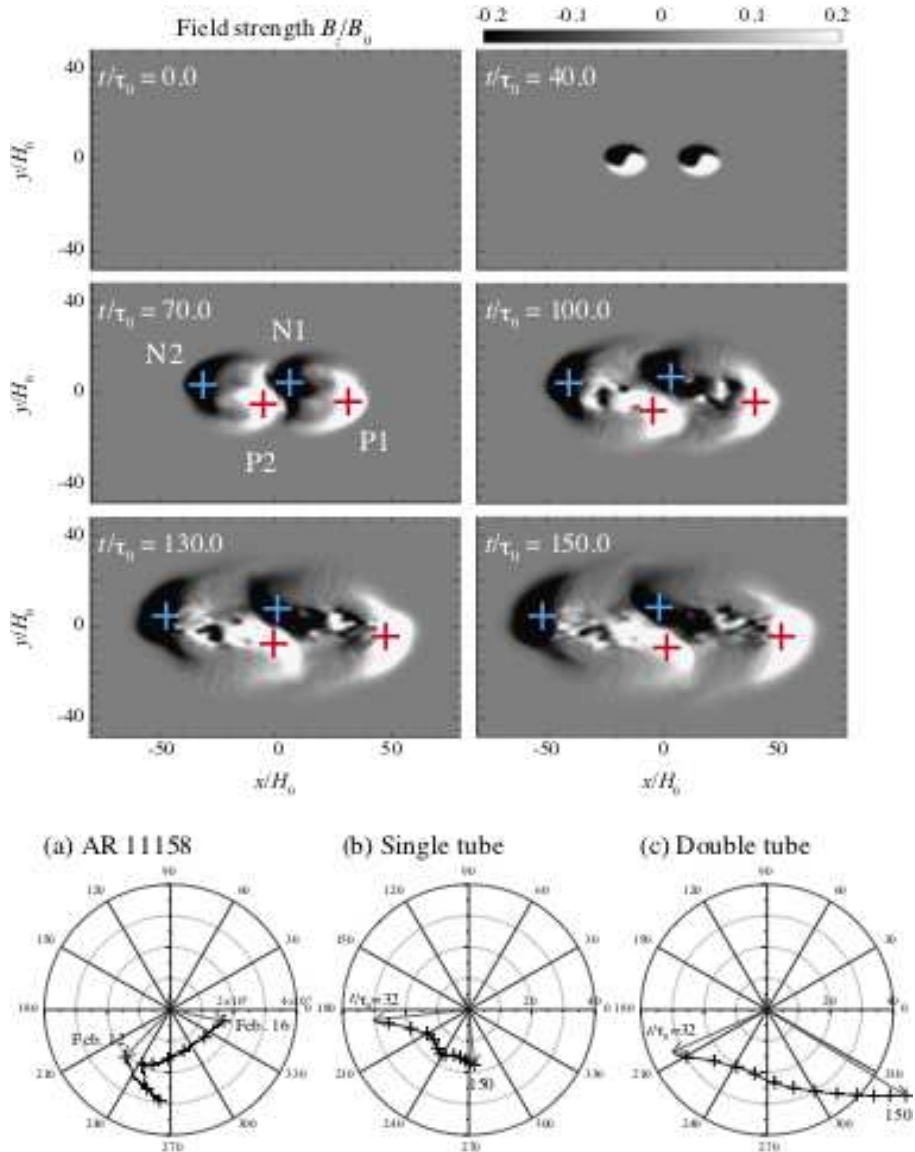}
  \caption{(Top) Emergence of a double-buoyant segment flux tube.
    The shown are the temporal evolution of vertical fields at the surface
    (photospheric magnetogram).
    Two emerging bipoles P1--N1 and P2--N2 collide at the center
    and form a sheared PIL with a compact $\delta$-spot structure.
    (Bottom) Relative motion of the photospheric polarities N1 and P2
    for (a) AR NOAA 11158 (Fig.~\ref{fig:toriumi2014}),
    (b) the simulation with a single double-buoyant-segment tube
    (i.e., top panels),
    and (c) another simulation with two parallel tubes.
    The center of each diagram indicates the position of N1
    and the horizontal axis is parallel to the $x$-axis.
    Approaching of the two polarities in NOAA 11158
    is reproduced only in the single tube model.
    {Image reproduced by permission from \citet{2014SoPh..289.3351T},
    copyright by ***.}
    }
  \label{fig:toriumi2014_2}
\end{figure*}

Such a situation was modeled by \citet{2014SoPh..289.3351T},
who reproduced NOAA 11158 (Fig.~\ref{fig:toriumi2014})
by simulating the emergence of a single horizontal flux tube
that rises at two sections along the tube.
As the photospheric magnetogram of Fig.~\ref{fig:toriumi2014_2}(top) displays,
the two buoyant segments produce a pair of emerging bipoles P1--N1 and P2--N2,
and the inner polarities (N1 and P2) become tightly packed
to create a $\delta$-spot.
The strong confinement of the central polarities happens
because the two emerging loops (P1--N1 and P2--N2) are joined
by a dipped field beneath the photosphere.

These authors also modeled the emergence of two buoyant flux tubes
that are placed closely in parallel (but not connected).
In this case, the inner polarities of the two emerging bipoles
move closer but just fly-by and never form a compact $\delta$-spot.
Bottom of Fig.~\ref{fig:toriumi2014_2} compares the relative motion
of the two inner polarities (time evolution of the vector from N1 to P2)
for NOAA 11158, the single tube case, and the double tube case.
In the actual AR (see also Fig.~\ref{fig:toriumi2014}),
P2 continuously drifts along the southern edge of N1 from east to west
in a counter-clockwise direction and becomes closer to N1,
producing a highly-sheared, strong-gradient PIL.
Between the two simulation cases,
only the single tube case shows the monotonic decrease of the distance.
Therefore,
they concluded that this Type 3 quadrupolar AR is,
between the two scenarios,
more likely to be created from a single multi-buoyant-segment flux tube.

\begin{figure*}\sidecaption
  \centering
  \includegraphics[width=0.6\textwidth]{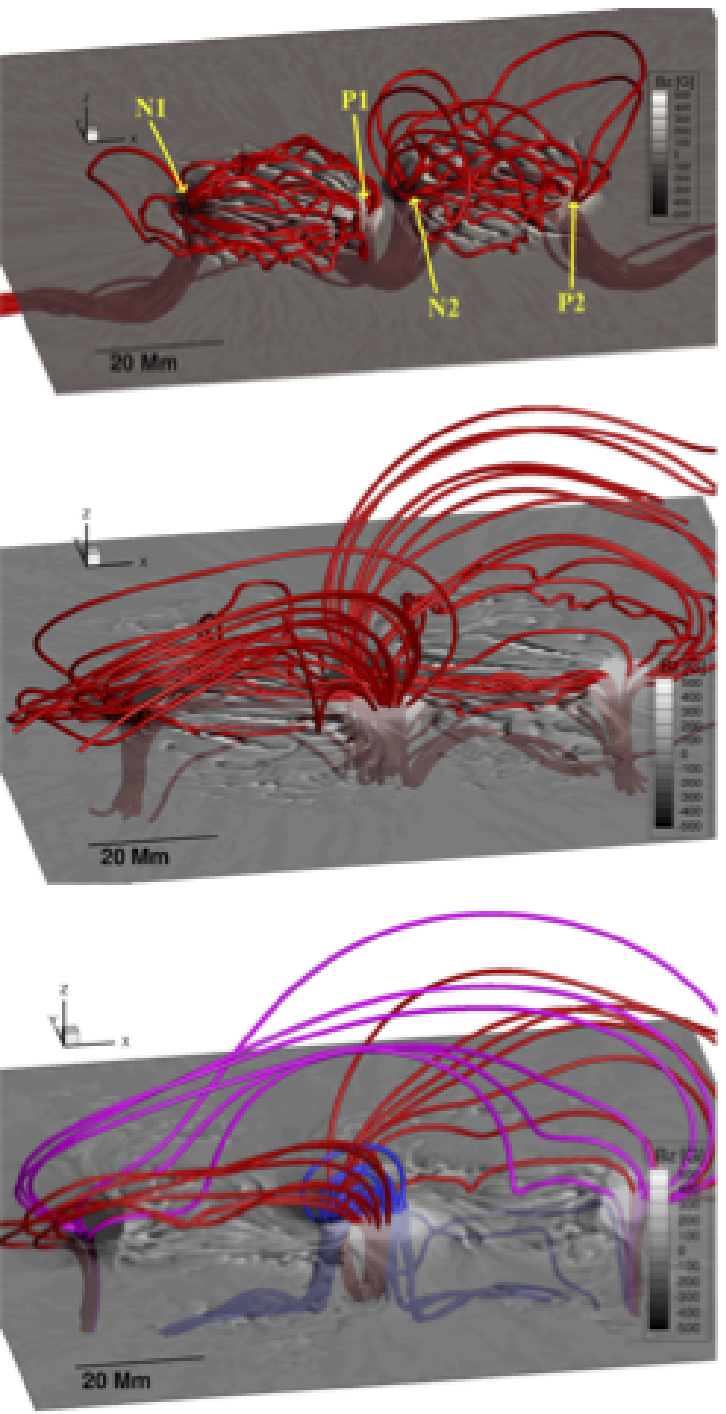}
  \caption{Simulation results by \citet{2015ApJ...806...79F},
    showing 3D structure of the {\sf M}-shaped emerging loops (red lines)
    at three different time steps.
    The plane shows the photospheric magnetogram.
    Note that the notation of the four polarities is different
    from that in Figs.~\ref{fig:toriumi2014} and \ref{fig:toriumi2014_2}.
    In the final state, magnetic reconnection between the two loops (red)
    produces overlying (magenta) and low-lying (blue) field lines.
    {Image reproduced by permission,
    copyright by AAS.}
    }
  \label{fig:fang2015}
\end{figure*}

Exactly the same situation was investigated later by \citet{2015ApJ...806...79F},
but in a much larger computational domain of a realistic AR size
with an adaptive mesh refinement code to resolve fine-scale structures.
Fig.~\ref{fig:fang2015} shows three snapshots from their simulation,
which clearly shows that the {\sf M}-shaped emerging loop
produces two arcades in the corona and,
through magnetic reconnection,
overlying and lower-lying field lines,
which is expected from the coronal observation of NOAA 11158
(see bottom panels of Fig.~\ref{fig:toriumi2014}).
The striking consistency between the more realistic simulation and the observation
further supports the scenario of multi-buoyant-segment flux tubes
for the Type 3 $\delta$-spots.


\subsubsection{Interacting tube model}
\label{subsubsec:interactingtube}

Another possible origin of the complexity of ARs
is the subsurface interaction of multiple rising flux systems.
Based on the study of potential flow around circular cylinders,
\citet{1978ApJ...222..357P,1979ApJ...231..270P} predicted that
when two cylindrical flux tubes are rising in a fluid one above the other,
the lower tube is attracted toward the other
because of the wake of the tube ahead and,
when rising side by side,
the tubes attract each other due to the Bernoulli effect.
However, from 2D simulations on the cross-sectional evolution,
\citet{1998ApJ...493..480F} found that the interaction
of the two tubes is much more complicated.
When the tubes rise side by side,
because the wake behind each tube interacts with that of the other,
each tube sheds a succession of eddies of alternating signs
and gains Magnus force in the lateral direction,
leading to the repeated attractive and repulsive motions during their ascents.
On the other hand,
when the tubes do not have the same initial height,
the tube behind is drawn into the wake of the tube ahead
and eventually merges with it.
At the interface between the two tubes,
dissipation of oppositely directed field components (twists) occurs.

\begin{figure*}
  \centering
  \includegraphics[width=1.0\textwidth]{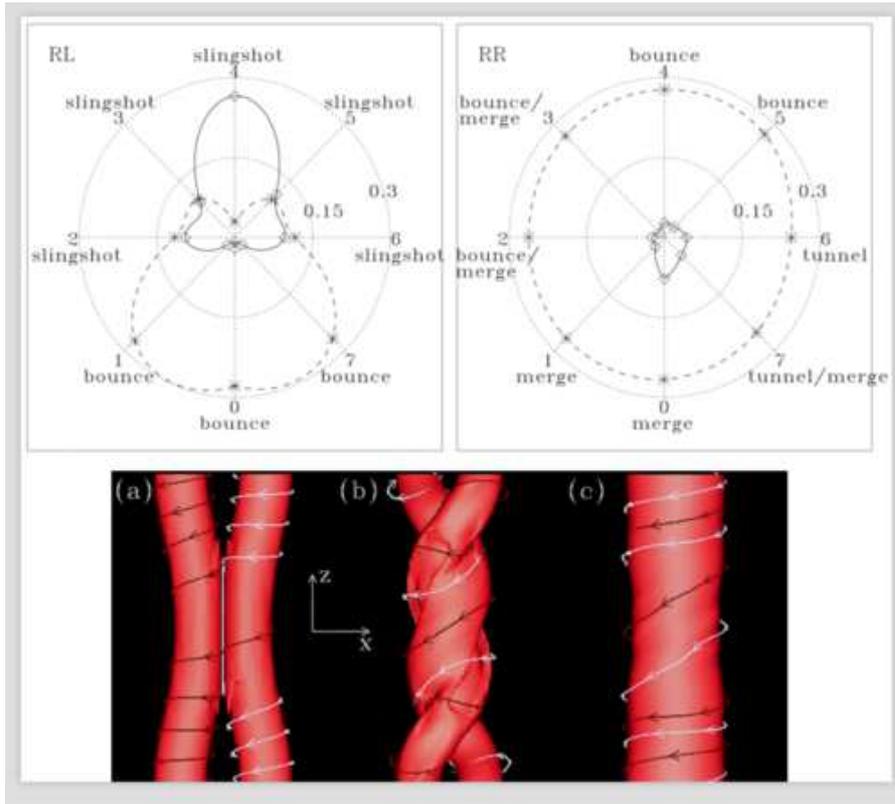}
  \caption{(Top) Polar plots showing the types of interaction
    of right-handed (R) and left-handed (L) twist tubes.
    Each radial spoke corresponds to a simulation RL$i$,
    where one R tube is in the reference position and another tube is in front of it,
    rotated by an angle $i\pi/4$ clockwise to it
    in such a way that RL0/RR0 is parallel and RL4/RR4 is anti-parallel.
    The solid curves show $2({\rm KE}_{\rm peak}-{\rm KE}_{0})/{\rm ME}_{0}$,
    where KE$_{\rm peak}$ is the peak global kinetic energy during the simulation,
    KE$_{0}$ is the initial global kinetic energy,
    and ME$_{0}$ is the initial global magnetic energy.
    The dashed curves show the global magnetic energy near the end of the simulations
    normalized by ME$_{0}$.
    The dotted circles are the normalized energy levels of 0.15 and 0.3.
    (Bottom)
    Merge interaction of RR0.
    Isosurface of $|B|_{\rm max}/3$ and field lines for three time steps
    are shown.
    {Image reproduced by permission from \citet{2001ApJ...553..905L},
    copyright by AAS.}
    }
  \label{fig:linton2001}
\end{figure*}

\citet{2001ApJ...553..905L} focused more on magnetic reconnection
between two strongly-twisted flux tubes
in the 3D low-$\beta$ volume (i.e., the solar corona)
to study the triggering of flares and eruptions.
They found that,
depending on the helicity (twist handedness) and the relative angle of the tube axes,
the interaction can be classified into four distinct classes
(see Fig.~\ref{fig:linton2001}):
(1) bounce, in which the two tubes bounce off each other with very little reconnection,
occurring for example between parallel counter-helicity tubes (RL0);
(2) merge, in which the tubes merge due to reconnection of azimuthal components,
e.g., between parallel co-helicity tubes
(RR0: bottom of Fig.~\ref{fig:linton2001});
(3) slingshot, in which the tubes reconnect and ``slingshot'' away
in a manner analogous to the classical 2D reconnection,
e.g., between anti-parallel counter-helicity tubes (RL4);
and (4) tunnel, in which field lines of the tubes undergo reconnection twice
and the tubes pass through each other,
occurring when the co-helicity tubes are placed in the orthogonal direction like RR6.
These interactions were also investigated by \citet{1992SoPh..142..399S}.
\citet{2005ApJ...625..506L} and \citet{2006JGRA..11112S09L} demonstrated that
the situations may differ depending on the level of twist
and the balance of magnetic flux contained in the two tubes.

\begin{figure*}
  \centering
  \includegraphics[width=1.0\textwidth]{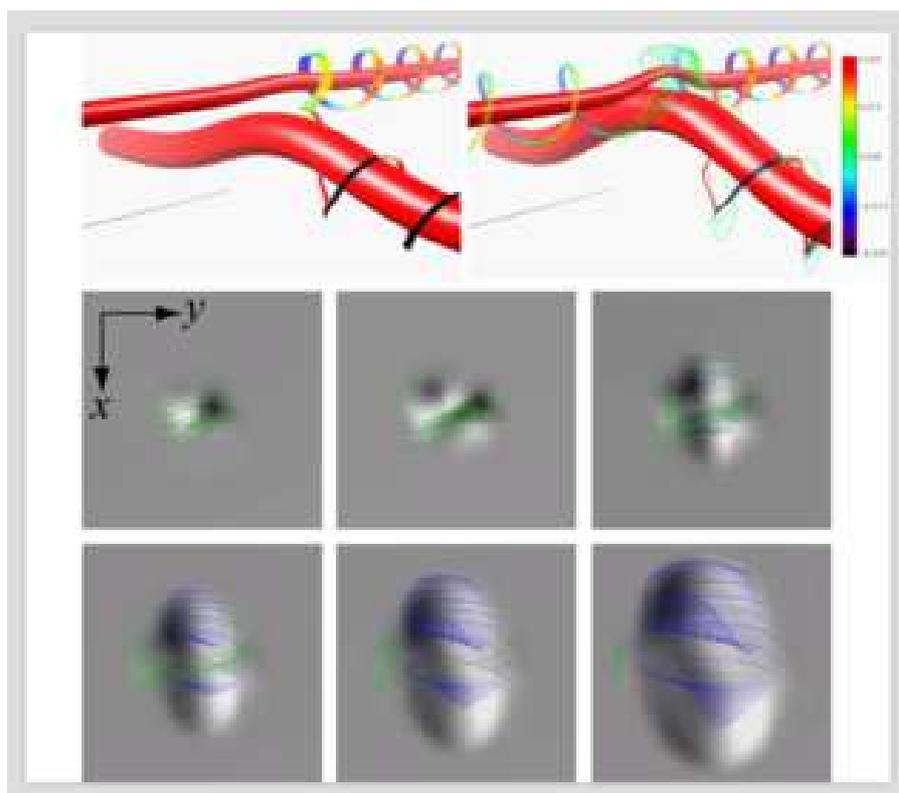}
  \caption{(Top) Two snapshots from the simulation
    of interacting orthogonal flux tubes.
    The field lines are colored according to local $B_{z}$,
    while the red isosurface gives a constant-$|B|$ layer.
    (Bottom) Synthesized magnetogram at the photospheric height,
    in which darker and lighter colors represent
    $B_{z}<0$ and $B_{z}>0$, respectively.
    The green and blue lines are selected field lines,
    traced from the upper and lower tubes, respectively.
    {Image reproduced by permission from \citet{2007A&A...470..709M},
    copyright by ESO.}
    }
  \label{fig:murray2007}
\end{figure*}

\citet{2007A&A...470..709M} simulated the interaction of emerging flux tubes
in the stratified high-$\beta$ medium representing the solar interior.
They examined the cases where two horizontal tubes are placed
in such a way that the lower one is buoyant
whereas the upper one remains stable.
For the case of parallel tubes,
or LL0 (the mirror symmetry of RR0)
following the notation by \citet{2001ApJ...553..905L},
they found that the tubes gradually merge, though not totally,
and the photospheric magnetogram shows
a simple ying-yang pattern similar to that of the single tube case
(like in Fig.~\ref{fig:fan2001}).
Of more interest is the case with orthogonal tubes in Fig.~\ref{fig:murray2007},
or LL2 (corresponding to RR6),
where the two tubes are expected to perform a slingshot reconnection
due to their lower degrees of twist \citep{2006JGRA..11112S09L}.
The authors found that, as opposed to the expectation,
the two tubes do not undergo a complete slingshot
because the tubes differ much in strength.
The resultant magnetogram becomes much more complicated.
As Fig.~\ref{fig:murray2007} illustrates,
the polarity layout is at first
positive negative from left to right when the upper tube emerges.
However, as the lower tube reaches the photosphere,
the layout reveals a quadrupolar structure
and transits to negative positive,
eventually recovering the classical ying-yang pattern.

\begin{figure*}
  \centering
  \includegraphics[width=1.0\textwidth]{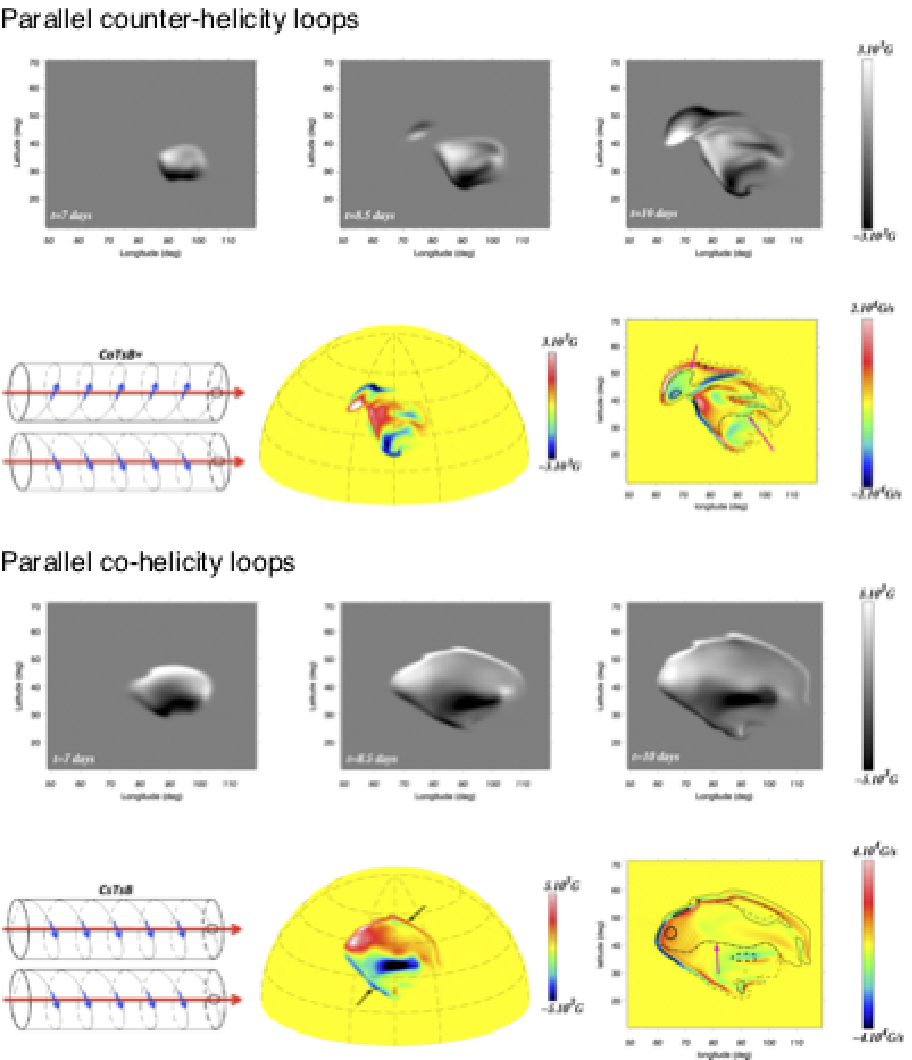}
  \caption{(Top) Simulation results of global-scale toroidal loops
    for the case with the same axial field but opposite handedness (RL0),
    which is illustrated as the cartoon.
    The panels in the first row and on the second middle indicate
    the radial magnetic field at the near-top layer at $0.93R_{\odot}$.
    The panel on the second right shows the radial current,
    on which the contours of the radial field at 80\% (thick) and 20\% (thin)
    of its maximum (solid) and minimum (dashed) are overplotted.
    The magenta arrows point to the PILs.
    Due to the bounce interaction of the emerging tubes,
    the surface magnetogram shows
    two emerging bipoles with different helicity signs.
    (Bottom) The same as the top panels but for the case with
    the same handedness and axial field (RR0).
    In this merging case,
    the emerging region consists of a large single bipole
    but shows a higher degree of non-neutralized currents.
    {Image reproduced by permission from \citet{2018ApJ...857...83J},
    copyright by AAS.}
    }
  \label{fig:jouve2018}
\end{figure*}

The interaction of two emerging flux tubes inside the solar interior
was also examined by \citet{2018ApJ...857...83J} in a global scale.
By extending their anelastic MHD models
of the flux emergence in a spherical convective shell
with large-scale mean flows \citep[e.g.,][]{2013ApJ...762....4J},
they conducted simulations on the pairs of emerging toroidal loops
that have different combinations of the twist handedness and axial direction.
They found that
if the two loops are given opposite handedness and the same axial direction
or the same handedness but opposite axial direction,
they bounce against each other through rising,
which is in good agreement with RL0 and RR4 of \citet{2001ApJ...553..905L}.
Consequently, as in the top panels of Fig.~\ref{fig:jouve2018},
the map of the radial magnetic field
near the top boundary (substituting the solar surface)
shows a quadrupolar region constituted of two emerging bipoles.
On the other hand,
the case with parallel co-helicity loops (corresponding to RR0)
yields a simple bipolar pattern due to the merging of the loops
(Fig.~\ref{fig:jouve2018} (bottom)),
just like the first model of \citet{2007A&A...470..709M}.
However, in such a case,
the non-neutralized currents,
suggested to be the origin of eruptive events
(Sect.~\ref{subsubsec:currents}),
are much more pronounced than the other cases
because the return currents
contained in the periphery of each loop
are annihilated at the current sheet between the merging loops.
From the series of simulation runs in \citet{2018ApJ...857...83J},
a variety of AR structures are formed by interaction of two rising flux tubes,
from simple bipolar to complex quadrupolar ones.
Since the magnetograms investigated in this study are at $0.93R_{\odot}$
(i.e., about 50 Mm below the actual surface of the Sun)
due to the limitation of anelastic models,
further investigations with the fully compressible calculations
that enable the direct access to the surface are needed
to elaborate how much of the emerging flux does reach the photosphere
and what the possible AR configurations at the surface are.

ARs with much higher degree of complexity
were modeled by \citet{2016GApFD.110..432P},
who simulated the buoyant emergence of braided magnetic fields
from the convection zone to the corona.
For instance, their ``pigtail'' field,
in which three flux tubes are entangled with each other, develops
a magnetogram with a number of positive and negative polarities intertwined:
see Fig. 13 of \citet{2016GApFD.110..432P}.

\subsubsection{Effect of turbulent convection}
\label{subsubsec:turbulentconv}

As we have discussed in Sect.~\ref{subsec:fe} and above,
thermal convection exerts a diverse range of impacts on the emerging flux,
and the series of realistic simulations have revealed the dynamic interactions
between the magnetic fields and convective flows,
such as boost-up and pin-down of large-scale emerging fields
\citep{2003ApJ...582.1206F,2009ApJ...701.1300J},
elongation of the surface granular cells
\citep{2008ApJ...679..871M,2008ApJ...687.1373C},
and the local undulation of emerging fields
\citep{2009A&A...507..949T,2010ApJ...714.1649F,2010ApJ...720..233C}.

\begin{figure*}
  \centering
  \includegraphics[width=0.9\textwidth]{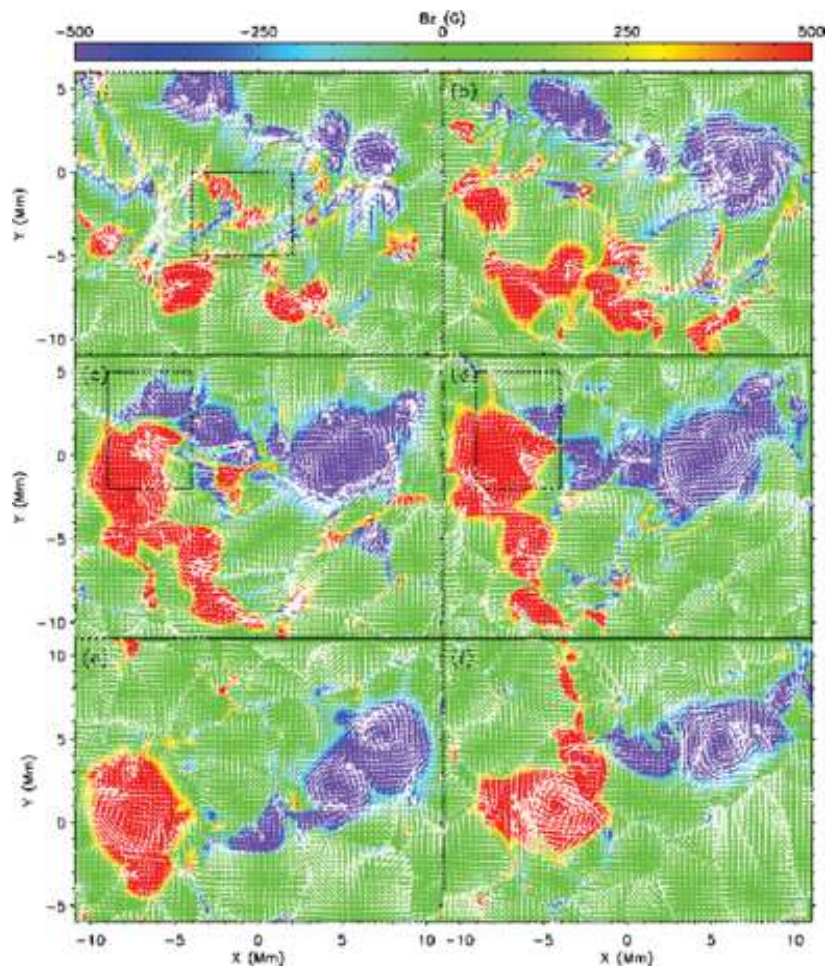}
  \caption{Temporal evolution of vertical magnetic field at the solar surface
    at (a) 3:45:00, (b) 4:15:00, (c) 5:10:00, (d) 5:35:00, (e) 6:23:00, and (f) 7:41:00
    from the start of the simulation.
    Arrows show the horizontal velocity field.
    Noticeable shearing/converging flows are highlighted with the boxes.
    {Image reproduced by permission from \citet{2012ApJ...745...37F},
    copyright by AAS.}
    }
  \label{fig:fang2012a}
\end{figure*}

\begin{figure*}
  \centering
  \includegraphics[width=0.9\textwidth]{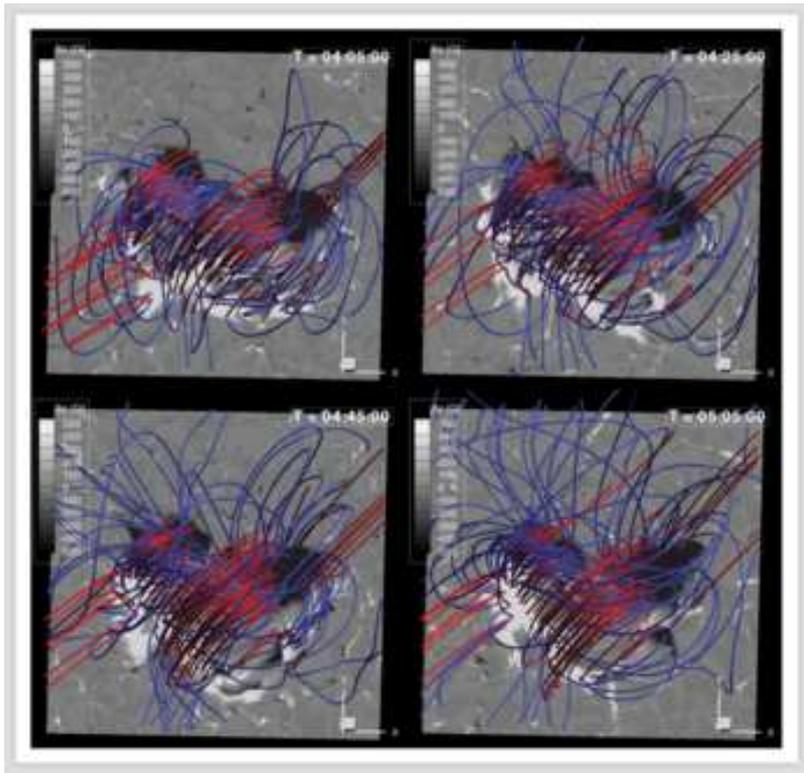}
  \caption{Comparison of the model field (blue)
    with the extrapolated potential field (red)
    at the times of 04:05:00, 04:25:00, 04:45:00, and 05:05:00
    plotted on the photospheric magnetogram.
    The formation of non-potential sigmoidal field is clearly seen.
    {Image reproduced by permission from \citet{2012ApJ...754...15F},
    copyright by AAS.}
    }
  \label{fig:fang2012b}
\end{figure*}

\citet{2012ApJ...745...37F,2012ApJ...754...15F} simulated
the buoyant rise of a twisted flux tube
from the convection zone in which turbulent convection resides.
Fig.~\ref{fig:fang2012a} shows the evolution of photospheric magnetograms,
which reveals the rapid growth of magnetic concentrations (spots)
with the unsigned total flux of up to $1.37\times 10^{21}\ {\rm Mx}$ (at $t=5$ hr),
the strong spot rotations (see the large negative spot at $x=6\ {\rm Mm}$),
and the shearing and converging motions around the PIL.
Here, both the shearing and rotational motions are driven by the Lorentz force
and these motions transfer the magnetic energy and helicity into the corona
\citep[consistent with, e.g.,][]{2001ApJ...547..503M,2001ApJ...554L.111F}.
The authors found that the convection-driven convergence flow
produces a strong magnetic gradient and flux cancellation at the PIL.
Together with the shear flow,
the field lines above the PIL undergo a tether-cutting reconnection
and produce long overlying sheared arcades
and short submerging loops \citep{2001ApJ...552..833M}.
Comparison of the model and extrapolated field lines
in Fig.~\ref{fig:fang2012b} clearly illustrates
the development of non-potential, sigmoidal structure above the PIL
that is covered by the more potential coronal loops.

Similar convective emergence simulation was
also performed by \citet{2016PhRvL.116j1101C},
who employed a horizontal magnetic flux sheet instead of a tube
at the start of the simulation.
The flux sheet breaks up into several flux bundles
due to the undular mode instability \citep{2001ApJ...546..509F}
and develops into a large-scale {\sf U}-shaped loop,
which appears in the photosphere
as a pair of colliding flux concentrations (i.e., a $\delta$-spot).
The strong cancellation between the two spots manifests
as a series of flare eruptions
with magnitudes comparable to GOES C- and B-class events
\citep{2018ApJ...857..103K}.
Through the creation of a $\delta$-spot and the flaring activity,
they observed the repeated formation of cool dense filaments above the PIL
and the ejection of helical flux ropes.

Another intriguing possibility of $\delta$-spot formation
was suggested by \citet{2014MNRAS.445..761M},
who conducted the direct numerical simulation
of the strong stratified dynamo with forced turbulence.
Their 3D computation box holds two-layered turbulence,
the helical and large-scale dynamo in the lower layer
and the non-helical turbulence in the upper layer.
As a result,
they observed the formation of strong bipolar flux concentrations
with super-equipartition fields,
which sometimes move closer to take a $\delta$-spot configuration.
While the large-scale magnetic field in the deeper layer
is created through a large-scale dynamo ($\alpha$ effect),
the spontaneous spot formation in the upper layer may be due to
the so-called negative effective magnetic pressure instability (NEMPI),
which is caused by suppression of the turbulent hydromagnetic pressure and tension
due to the mean magnetic field
\citep{2011ApJ...740L..50B}.

\subsubsection{Toward the general picture}
\label{subsubsec:unified}

The numerical simulations introduced above
have suggested the possibility
that different types of flare-productive ARs
have different subsurface origins and evolution histories
\citep{1987SoPh..113..267Z,2017ApJ...834...56T}.
For example,
the $\delta$-spots of Types~1 (Spot-spot) and 3 (Quadrupole)
may be produced from the kinked and multi-buoyant-segment flux systems,
respectively
\citep{1999ApJ...522.1190L,1999ApJ...521..460F,2015ApJ...813..112T,2014SoPh..289.3351T,2015ApJ...806...79F}.

\begin{figure*}
  \centering
  \includegraphics[width=1.\textwidth]{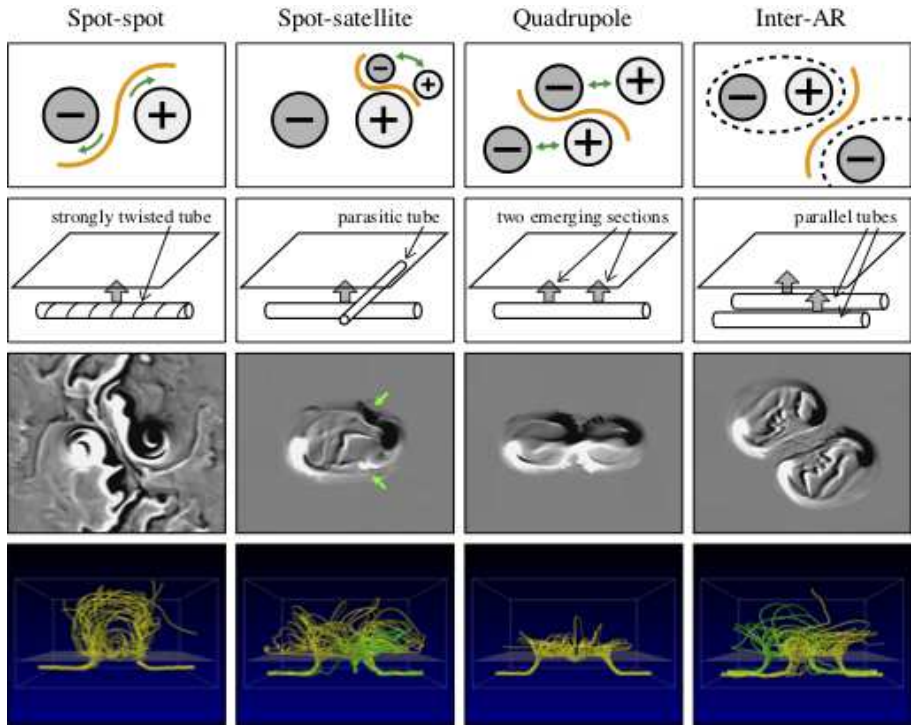}
  \caption{3D numerical simulations of the four representative types
    of flare-productive ARs, as introduced in Fig.~\ref{fig:toriumi2017a}.
    {Images and movie reproduced by permission from \citet{2017ApJ...850...39T},
    copyright by AAS.}
    (Top) Polarity distributions.
    (Second) Schematic diagrams showing the numerical setup.
    (Third) Surface vertical magnetic fields (magnetogram).
    The green arrows for the Spot-satellite case point to
    the satellite spots, which originate from the parasitic flux tube.
    (Bottom) Magnetic field lines.
    The green field lines are for the parasitic tube and the parallel tube. 
    (For movie see Electronic Supplementary Material.) 
    The accompanying movie shows the temporal evolutions
    for the four cases.
    }
  \label{fig:toriumi2017b}
\end{figure*}

In order to scrutinize the differences between the above three cases
plus another type of X-flaring ARs, the Inter-AR case,
created by two independent but closely neighboring episodes of flux emergence,
\citet{2017ApJ...850...39T} conducted a systematic survey of flux emergence simulations
by using similar numerical conditions with as little difference as possible,
and explored the formation of $\delta$-spots,
flaring PILs, and their evolution processes.
Figure~\ref{fig:toriumi2017b} summarizes the numerical conditions and results.
For the Spot-spot case, the initial twist strength is intensified so as to exceed
the critical value for the kink instability
\citep[][see also Sect.~\ref{subsubsec:kink}]{1996ApJ...469..954L}.
The Spot-satellite is modeled by introducing a parasitic flux tube
above the main tube in a direction perpendicular to it,
the situation similar to the interacting tube models
in Sect.~\ref{subsubsec:interactingtube}.
The Spot-satellite may also be produced from a single bifurcating tube,
which, however, was not considered for the sake of simplicity.
The Quadrupole flux tube has two buoyant sections along the axis,
resembling the simulations in Sect.~\ref{subsubsec:multibuoyant}.
Finally, for the Inter-AR case, two flux tubes are placed in parallel.

As the movie of Fig.~\ref{fig:toriumi2017b} demonstrates,
all cases except for Inter-AR produce $\delta$-shaped polarities
with strongly-sheared, strong-gradient PILs in their cores
that are coupled with flow motions,
but the most drastic evolution appears for the Spot-spot case.
As discussed in Sect.~\ref{subsubsec:kink},
the knotted apex enhances the buoyancy
that leads to the fastest emergence among the four cases.
The total unsigned magnetic flux in the photosphere
\begin{eqnarray}
  \Phi=\int_{z=0} |B_{z}|\, dS
  \label{eq:usflux}
\end{eqnarray}
and the free magnetic energy stored in the atmosphere
\begin{eqnarray}
  \Delta E_{\rm mag}\equiv E_{\rm mag}- E_{\rm pot}
  =\int_{z>0} \frac{{\vec B}^{2}}{8\pi}\, dV
  -\int_{z>0} \frac{{\vec B}_{\rm pot}^{2}}{8\pi}\, dV,
  \label{eq:efre}
\end{eqnarray}
where ${\vec B}_{\rm pot}$ is the potential field,
are also largest for the Spot-spot case.

\begin{figure*}
  \centering
  \includegraphics[width=0.9\textwidth]{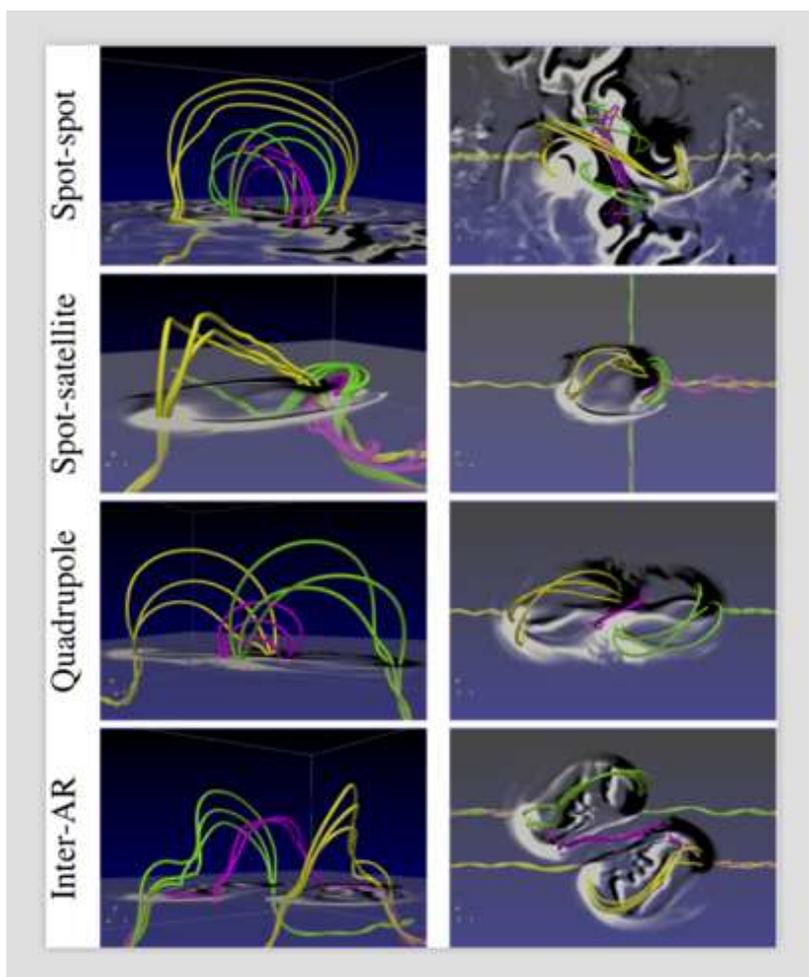}
  \caption{Modeled 3D magnetic structures for the four types of flare-producing ARs
    in \citet{2017ApJ...850...39T}.
    The purple field lines are the newly formed flux rope structure,
    created through magnetic reconnection of emerged loops
    indicated with yellow and green lines.
    Except for the Spot-spot case,
    the flux ropes are exposed and have an access to the outer space.
    On the contrary, the Spot-spot flux rope is covered by the overlying arcade.
    {Image reproduced by permission, copyright by AAS.}
    }
  \label{fig:toriumi2017b_3}
\end{figure*}

It is also suggested from these models
that the difference in initial simulation setup
may determine the fate of a CME eruption.
As shown in Fig.~\ref{fig:toriumi2017b_3},
in the case of Spot-satellite, Quadrupole, and Inter-AR,
the newly formed flux rope above the sheared PIL
is exposed to outer space,
an ideal situation for successful CME eruption.
However, in the Spot-spot case,
the flux rope is trapped and confined by the overlying loops.
Very strong confinement
may explain the flare-rich but CME-poor nature
of the Spot-spot AR NOAA 12192
(see Fig.~\ref{fig:20141024} and discussion
on successful and failed eruptions in Sect.~\ref{subsec:flares}).

In addition, this model is able to account for the formation
of ``magnetic channels,''
another important feature of the flaring PILs
\citep[][see Sect.~\ref{subsubsec:photo_pil}]{1993Natur.363..426Z}.
In the magnetogram of the Spot-spot case (Fig.~\ref{fig:toriumi2017b}),
one may find that the central PIL has an elongated alternating pattern
of positive and negative polarities, resembling the magnetic channel.
This structure is produced because the photospheric fields are highly
inclined to horizontal
and almost parallel to the PIL with slight undulations.

The series of simulations above provides
a unified, general view of the birth of flare-productive ARs.
Within the solar interior, probably due to convective evolution,
the emerging flux systems
that form $\delta$-spots
are severely twisted to take on tortuous structures,
partially pinned down to bear multiple rising segments,
bifurcated into entangled branches,
or hit against other flux systems to undergo mutual interactions.
All of these processes are prone to enhancement of free magnetic energy.
As the fluxes reach the photosphere,
complex magnetic structures,
prominently manifested by $\delta$-spots, sheared PILs,
sheared coronal arcades, and flux ropes,
develop.
The $\delta$-spots are likely generated
by multiple emerging loops instead of a single $\Omega$-loop,
and the different patterns of polarity layouts,
such as Types 1, 2, and 3,
stem from the difference in the subsurface evolution.
Even two separated, seemingly independent ARs may intensify the free energy
if located in the closer proximity (Inter-AR case).
The stored free energy is, if accumulated enough,
released in the form of flares and CMEs.

One possibility that was not considered in \citet{2017ApJ...850...39T}
is the situation where a new, delayed flux emerges into a pre-existing flux system
(i.e. the concepts of successive emergence, complexes of activities,
and sunspot nests in Sect.~\ref{subsec:deltaspots}).
\citet{2007ApJ...655L.117S} interpreted the formations of flaring PILs with this idea,
and \citet{2008ASPC..383..429W} overall agreed.
This situation is qualitatively similar to the Spot-satellite case,
in which a minor bipole appears in the close proximity to the major sunspot,
but the scale is much larger.
Therefore, toward a more complete view,
we may need to take into account this successive emergence case.

\subsection{Flux cancellation models}
\label{subsec:num_fc}

It is thought that coronal flux ropes
can also form post-emergence as a coronal response
to photospheric driving.
\citet{1994ApJ...420L..41A} and \citet{2000ApJ...539..954D}
demonstrated that a sheared arcade lying above a PIL,
produced by shearing motion in the photosphere (without convergence),
contains a dipped structure that supports the prominence material.
In the theory of \citet{1989ApJ...343..971V}
(see Fig.~\ref{fig:savcheva2012a}),
coronal loops above the PIL become sheared and converged
due to photospheric motions
and eventually reconnect against each other to form a flux rope.
Most of the simulations based on this theory,
often referred to as the ``flux cancellation'' models,
deal with the evolution of coronal field lines
within the computational box above the photospheric surface,
i.e., the situation after the magnetic flux is emerged.

\begin{figure*}
  \centering
  \includegraphics[width=0.7\textwidth]{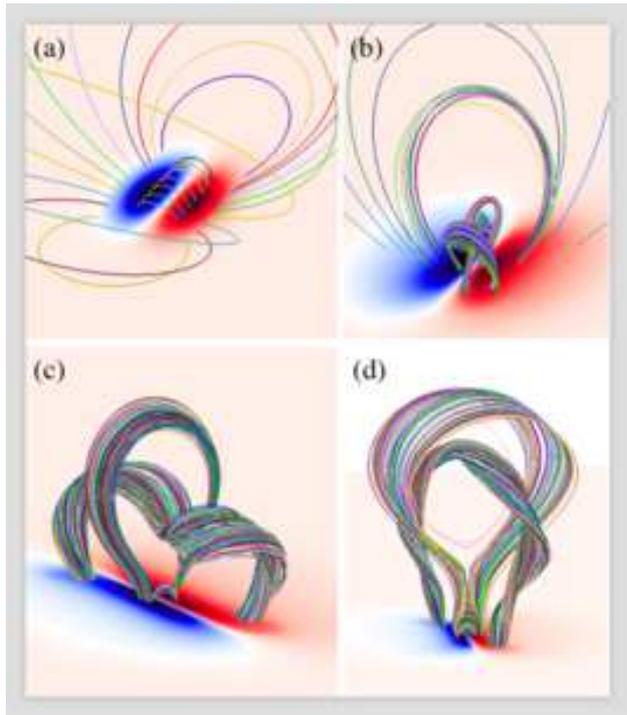}
  \caption{Flux cancellation model by \citet{2003ApJ...585.1073A}.
    {Image reproduced by permission, copyright by AAS.}
    (a) Initial bipolar potential fields (i.e., $t=0$).
    A pair of counter-clockwise twisting motions is imposed
    at the bottom boundary from $t=0$ to $t_{\rm s}$,
    followed by a viscous relaxation from $t=t_{\rm s}$ to $t_{0}$.
    (b) Field lines of the magnetic configuration
    after the converging flow is applied
    from $t_{0}=400\tau_{\rm A}$ to $450\tau_{\rm A}$,
    where the unit $\tau_{\rm A}$ denotes the Alfv\'{e}n transit time.
    Shown is the case for $t_{\rm s}=200\tau_{\rm A}$,
    in which the sheared loops are obvious around the PIL.
    (c) The state after the convergence is applied to $t=498\tau_{\rm A}$.
    A helical flux rope, low-lying arcade, and overlying arcade are now formed
    through magnetic reconnection between the sheared loops.
    (d) The convergence is further applied to $t=530\tau_{\rm A}$.
    The flux rope erupts upward
    with entraining the overlying arcades successively.
    }
  \label{fig:amari2003a}
\end{figure*}

Figure~\ref{fig:amari2003a} shows the representative 3D calculation
by \citet{2003ApJ...585.1073A}.
Here, the original potential field (panel a) is twisted
by two co-rotating vortices imposed at the photospheric boundary.
After the system is relaxed (panel b), converging motion is applied
and magnetic reconnection between the sheared loops leads to
the formation of a twisted flux rope, with a small low-lying arcade below,
and an overlying arcade above (panel c).
As the reconnection goes on, the unstable flux rope is ejected (panel d).

For instigating the flux cancellation of sheared loops,
several types of mechanisms have been considered
\citep[see, e.g.,][]{2010SSRv..151..333M,2014IAUS..300..184A}.
Other than the convergence flow \citep{2003ApJ...585.1073A,2010ApJ...708..314A},
proposed mechanisms include
decrease of photospheric flux through shearing motion
\citep{2000ApJ...529L..49A,2010ApJ...717L..26A},
turbulent diffusion
\citep{2003ApJ...595.1231A,2006ApJ...641..577M,2009ApJ...699.1024Y,2010ApJ...708..314A},
and reversal of magnetic shear \citep{2004ApJ...610..537K}.

\begin{figure*}
  \centering
  \includegraphics[width=0.75\textwidth]{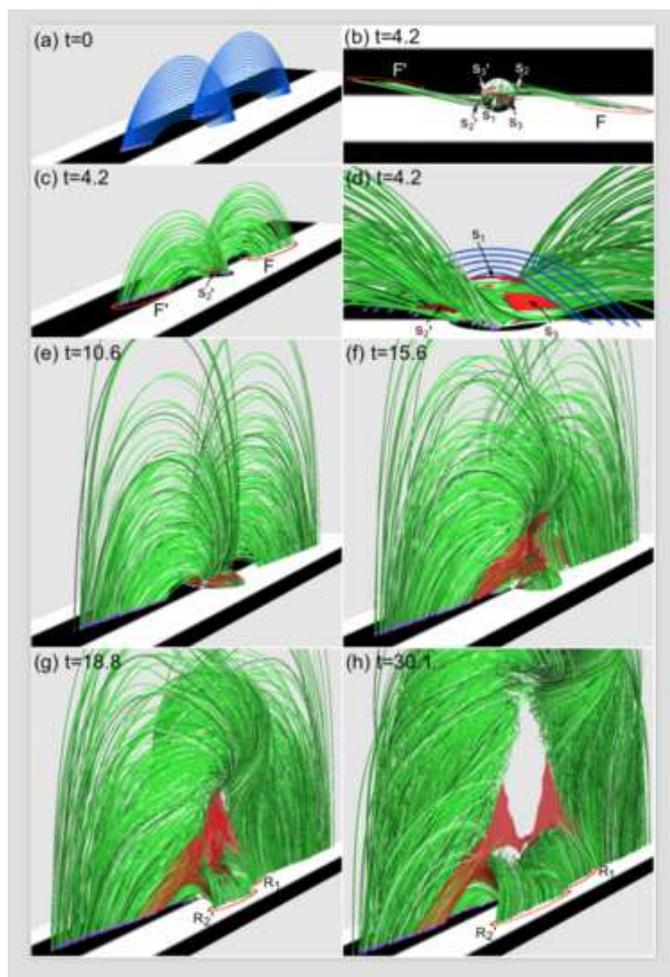}
  \caption{Flux rope formation and eruption by opposite-polarity type emerging flux.
    {Image reproduced by permission from \citet{2012ApJ...760...31K},
    copyright by AAS.}
    Green tubes show the field lines with connectivity
    that differs from the initial state,
    while the blue tubes in panels (a) and (d)
    are the original sheared arcades.
    Gray scale at the bottom indicates the vertical field $B_{z}$
    (white, positive; black, negative)
    and red contours denote the strong current layer.
    The initial sheared arcades (blue lines in panel a)
    go through reconnection triggered
    by the emerging flux at the bottom boundary
    and a helical flux rope is created (panels b to d).
    The flux rope is ejected
    leaving a current sheet underneath (panels e to h).
    }
  \label{fig:kusano2012}
\end{figure*}

\citet{2012ApJ...760...31K} investigated the process
where the sheared arcade field above the PIL
reconnects to create a flux rope and erupts,
triggered by emerging flux from the photospheric surface
(rather than the convergence flow or diffusion).
This model sheds light on the importance of small-scale magnetic structures,
which are often observed around flaring PILs,
in the destabilization of the entire system
\citep{2013ApJ...773..128T,2013ApJ...778...48B,2017NatAs...1E..85W}.
In the particular simulation case of Fig.~\ref{fig:kusano2012},
emerging flux with the field direction opposite to that of the arcades
triggers the reconnection and produces an erupting flux rope.
From a systematic survey on the orientations of arcade and emerging flux,
it was found that there exist two kinds of emerging flux
capable of initiating the cancellation:
the opposite-polarity type (shown as Fig.~\ref{fig:kusano2012})
and the reversed-shear type \citep[comparable to][]{2004ApJ...610..537K}.

As a more recent attempt,
\citet{2014ApJ...780..130X}, \citet{2016ApJ...823...22X}, and \citet{2017ApJ...845...12K}
performed 3D flux cancellation simulations
that take into account the effect of thermodynamical processes.
Due to the strong radiative cooling,
coronal plasma within the helical field lines of the flux rope
becomes condensed and piles up on the dipped part at the bottom.
In this way, these authors successfully reproduced filaments (prominences)
in a more realistic manner
than those lacking in the thermodynamical processes.

\subsection{Data-constrained and data-driven models}
\label{subsec:num_data}

\subsubsection{Field extrapolation methods}
\label{subsubsec:num_data_extrapolation}

One way to trace the development of coronal magnetic field
is to sequentially compute the field lines
from the routinely measured photospheric magnetograms
by using extrapolation methods
which neglect non-magnetic forces (such as pressure gradient)
and assume that the Lorentz force vanishes,
i.e., the force-free condition,
\begin{eqnarray}
  {\vec j}\times {\vec B}=0,
  \label{eq:forcefree}
\end{eqnarray}
where ${\vec j}$ is the current density
\begin{eqnarray}
  {\vec j}=\frac{c}{4\pi}{\vec \nabla}\times {\vec B}.
\end{eqnarray}

The potential (current-free) field is the simplest approximation,
under which $\nabla\times {\vec B}=0$.
This can be replaced by
\begin{eqnarray}
  {\vec B}=-\nabla\psi,
\end{eqnarray}
where $\psi$ is the scalar potential,
and combined with the solenoidal condition ($\nabla\cdot{\vec B}=0$),
further rewritten as
\begin{eqnarray}
  \nabla^{2}\psi=0.
\end{eqnarray}
The potential coronal field is calculated by solving this equation
with using the normal component of the photospheric field $B_{z}$
as the boundary condition.
\citet{2005ApJ...628..501S} and \citet{2016ApJ...820..103S}
assessed the non-potentiality of coronal fields of 95 and 41 ARs
by comparing potential field extrapolations to the corresponding coronal images
from the Transition Region and Coronal Explorer \citep[TRACE;][]{1999SoPh..187..229H}
and SDO/AIA, respectively.
They concluded that, in most cases,
significant non-potentiality exists
in ARs with newly emerging flux within $\sim 30$ hours
or when opposite-polarity concentrations are evolving and in close contact.

The force-free condition, Eq. (\ref{eq:forcefree}), is also expressed as
\begin{eqnarray}
  \nabla\times {\vec B}=\alpha {\vec B},
\end{eqnarray}
where $\alpha$ is called the force-free parameter.
If $\alpha$ is constant everywhere in the coronal volume under consideration,
the magnetic field is called a linear force-free field (LFFF);
otherwise, a non-linear force-free field (NLFFF).
In these models, all components of the vector magnetogram are used
as the bottom boundary condition.
As Figs.~\ref{fig:savcheva2012a} and \ref{fig:yang2017} show,
the NLFFF extrapolations provide realistic coronal fields
comparable to the actual observations.
By applying NLFFF methods to the complex quadrupolar AR NOAA 11967,
\citet{2016NatSR...634021L} and \citet{2017ApJ...842..106K}
investigated the topology of coronal fields
and elucidated the homologous occurrence of X-shaped flares.
However, it has been shown that
the NLFFF models are sensitive to
the quality of photospheric boundary conditions,
and thus do not faithfully reproduce observed coronal loop structures
\citep[e.g.,][]{2009ApJ...696.1780D,2015ApJ...811..107D}.
Moreover, the input vector magnetograms are subject
to the intrinsic ambiguity in the direction of the transverse magnetic field
and this hampers fundamentally
any magnetogram-driven coronal field reconstructions.

Representative NLFFF techniques include
the optimization method, MHD relaxation method,
and flux-rope insertion method.
For the basis and comparison of various extrapolation methods,
we refer the reader to \citet{2009ApJ...696.1780D,2015ApJ...811..107D},
\citet{2012LRSP....9....5W}, and \citet{2016PEPS....3...19I}.

\subsubsection{Data-constrained models}
\label{subsubsec:num_data_constrained}

\begin{figure*}
  \centering
  \includegraphics[width=1.0\textwidth]{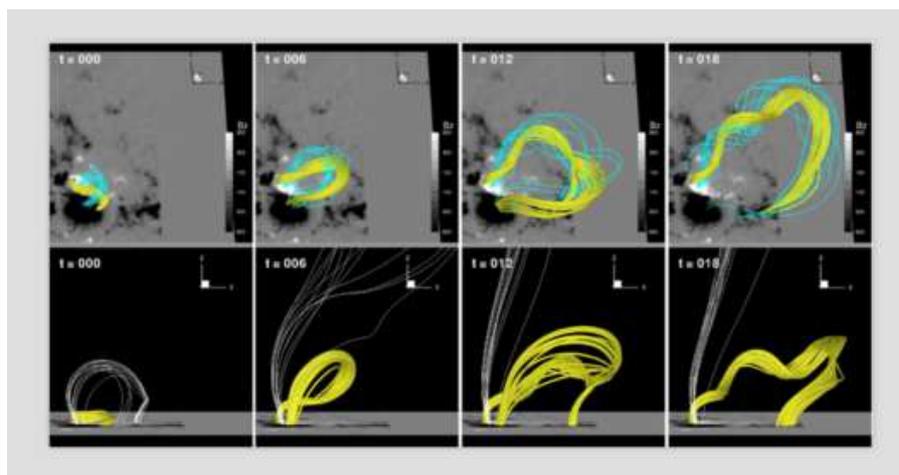}
  \caption{Data-constrained MHD simulation
    of the flux rope eruption in AR NOAA 11283.
    Yellow and cyan lines are the magnetic field lines traced
    from the same positive polarity.
    Another set of field lines (white)
    are those that pass through the null point, and reconnect and open.
    Bottom boundary is the photospheric magnetogram.
    The sigmoidal flux rope (yellow field lines at $t=0$, reproduced with NLFFF)
    becomes unstable and launched.
    {Image reproduced by permission from \citet{2013ApJ...771L..30J},
    copyright by AAS.}
    }
\label{fig:jiang2013}
\end{figure*}

Even if one applies the most sophisticated technique of the NLFFF extrapolations
to the accurate sequential magnetograms by Hinode/SOT and SDO/HMI,
the obtained temporal evolution is still far from the real one
because these models unavoidably assume a static state.
One approach to overcome this issue is to use time-evolving data-constrained modeling.
In this more physics-based method,
the temporal evolution is obtained by solving the MHD equations
with setting the reconstructed coronal field for the initial condition.
\citet{2013ApJ...771L..30J} were the first to apply this method to the actual AR.
As in Fig.~\ref{fig:jiang2013},
they reconstructed the initial coronal field of AR NOAA 11283 with the NLFFF model
and demonstrated the CME eruption from this AR.
According to the authors,
due to small numerical errors in the extrapolation
(i.e., their NLFFF was not perfectly force free),
the system became unstable and the flux rope was erupted via the torus instability.

\begin{figure*}
  \centering
  \includegraphics[width=1.\textwidth]{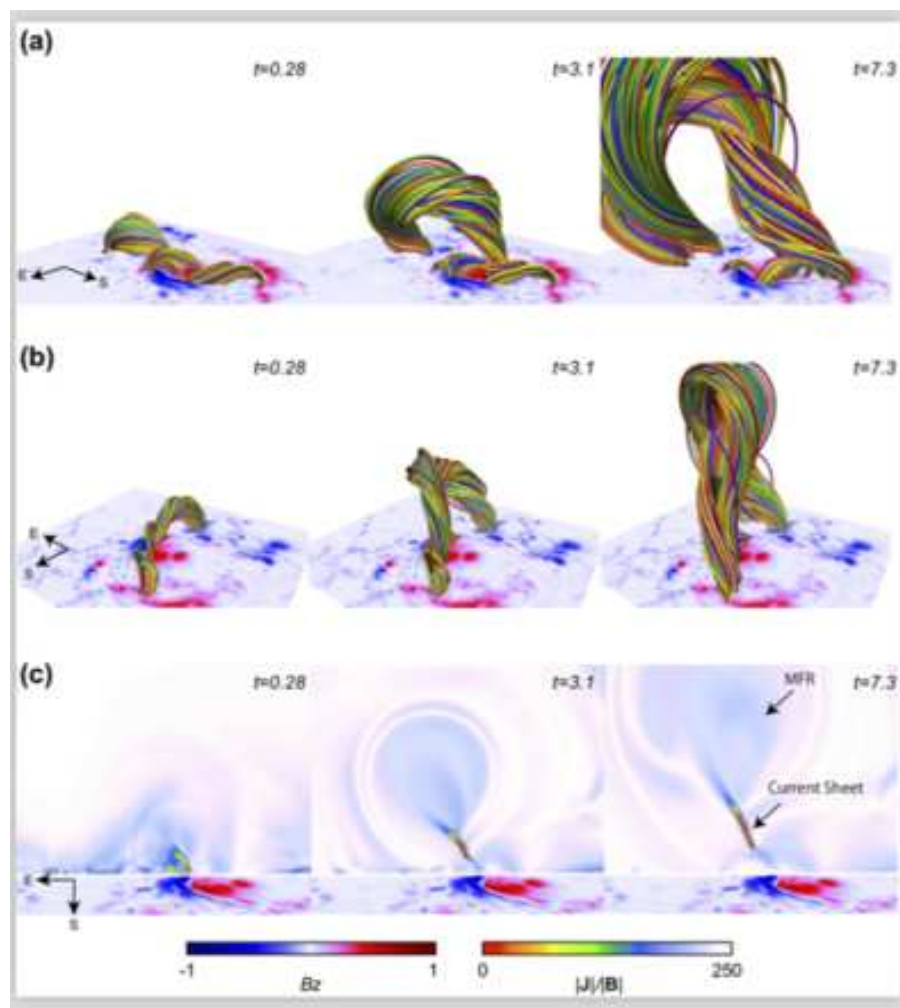}
  \caption{The formation and evolution of an eruptive flux rope
    in the X9.3-class flare in AR NOAA 12673.
    The top and second rows provide the field lines and magnetogram ($B_{z}$)
    that are viewed from two different angles
    and the bottom row shows the distribution of electric current
    in a vertical cross-section.
    In this model, multiple flux ropes along the PIL at the initial stage ($t=0.28$)
    reconnect and merge into a single flux rope ($t=3.1$),
    which eventually erupts into the higher atmosphere ($t=7.3$).
    {Image reproduced by permission from \citet{2018ApJ...867...83I},
    copyright by AAS.}
    }
\label{fig:inoue2018}
\end{figure*}

Since then, the data-constrained approach has become the hot topic
\citep[][]{2013ApJ...779..129K,2014Natur.514..465A}.
\citet{2014ApJ...788..182I,2015ApJ...803...73I} modeled the X2.2-class event in NOAA 11158
(Fig.~\ref{fig:toriumi2014}) and found that, interestingly,
the flux rope at the core of this AR does not erupt directly
but rather reconnects with ambient weakly twisted fields.
Then, the ambient field transforms into a flux rope,
which eventually exceeds the critical height of the torus instability.
\citet{2017ApJ...842...86M} applied this method to NOAA 10930
(e.g., Figs.~\ref{fig:20061213} and \ref{fig:bamba2013})
and, by inserting emerging flux at the PIL from the bottom boundary,
they succeeded in triggering the flux rope eruption,
which is in line with the flare-triggering scenario by \citet{2012ApJ...760...31K}.
The dramatic eruption in the X9.3 flare in NOAA 12673,
which we introduced in Sect.~\ref{subsec:longterm_summary},
was modeled by \citet{2018ApJ...867...83I}.
They found that, as in Fig.~\ref{fig:inoue2018},
multiple compact flux ropes lying along the sheared PIL
reconnect with each other and merge into a large, highly twisted flux rope
that eventually erupts.


\subsubsection{Data-driven models}
\label{subsubsec:num_data_driven}

Even more realistic reconstruction of the evolving coronal field
is to sequentially update the photospheric boundary condition,
which is called the data-driven model.
The first approach of the data-driven models we show here
is the magneto-frictional method \citep{1986ApJ...309..383Y},
in which the magnetic field evolves due to the Lorentz force,
\begin{eqnarray}
  {\vec v}=\frac{1}{\nu c}\, {\vec j}\times {\vec B},
\end{eqnarray}
where $\nu$ is the frictional coefficient.
In this formulation,
the (pseudo) velocity is simply proportional to the Lorentz force.
\citet{2012ApJ...757..147C} applied this method to the sequential magnetogram
of NOAA 11158 and reproduced flux ropes that were ejected
in the series of M- and X-class flares in this AR.

\begin{figure*}
  \centering
  \includegraphics[width=1.0\textwidth]{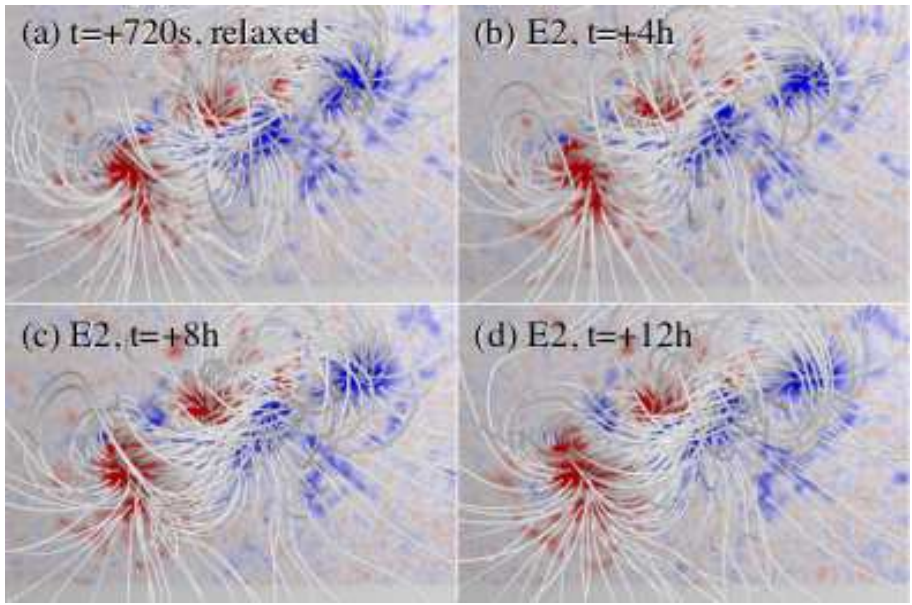}
  \caption{Data-driven model of NOAA 11158,
    performed with a time-evolving photospheric electric field.
    The initial relaxed coronal field (a) is stretched and sheared over time
    especially above the central PIL.
    {Image reproduced by permission from \citet{2018ApJ...855...11H},
    copyright by AAS.}
    }
\label{fig:hayashi2018}
\end{figure*}

Another recent,
yet nascent attempt is to directly solve the MHD equations
with sequentially replacing the magnetogram
to self-consistently reconstruct the coronal evolution
\citep{2006ApJ...652..800W}.
This was demonstrated by \citet{2016NatCo...711522J,2016ApJ...828...62J}
for ARs NOAA 11283 and 12192, respectively.
\citet{2018ApJ...855...11H} calculated the photospheric electric field ${\vec E}$
from the sequential magnetogram ${\vec B}$
and drove the model of NOAA 11158 through Faraday's law
\begin{eqnarray}
  \frac{\partial {\vec B}}{\partial t}=-c\nabla\times {\vec E},
\end{eqnarray}
instead of solving the induction equation
\begin{eqnarray}
  \frac{\partial {\vec B}}{\partial t}=\nabla\times({\vec v}\times {\vec B}).
\end{eqnarray}
Here, ${\vec E}$ is determined, for instance, by solving Ohm's law
(${\vec E}=-{\vec V}\times{\vec B}/c$)
by using the velocity ${\vec V}$ obtained
with flow tracking techniques
\citep[see][and references therein]{2007ApJ...670.1434W}.
As Fig.~\ref{fig:hayashi2018} displays,
the initial coronal field,
obtained by matching the potential field
to the observed vector magnetogram and relaxing it,
undergoes substantial elongation and twisting, especially above the central PIL,
in response to the shear motion in the photosphere.

A data-driven, dynamic model is supposed to calculate
the coronal field that matches the changing photospheric magnetogram.
An accurate model would, in principle, produce a flare or eruption
at the same time that the actual Sun does.
Inevitable simplifications of the model and inaccuracies in its initial state,
however, suggest that it may be difficult to reproduce flares or eruptions.
This is because the observed, gradual photospheric change
(before and around the flare onset) might be insufficient
to cause any drastic change in the (inaccurate) model's coronal field.

Another caveat is that the model is limited
by the temporal frequency of the driving data.
Using the flux emergence simulation as the ground-truth data set,
\citet{2017ApJ...838..113L} performed a data-driven simulation
with the assumption that the photospheric information
is provided every 12 minutes
(the default cadence of the SDO/HMI vector magnetogram).
They showed that the data-driven models can reproduce
the slowly emerging ARs over 25 hour
with only $\sim 1$\% error in the free magnetic energy.
However, the modeling was largely affected by rapidly evolving features.
Even if one applies interpolation to the driving data,
the coarse sampling generates a strobe effect,
in which smoothly evolving features appear to jump across the photosphere.
For an emerging bipole with a spatial extent of $L=1\ {\rm Mm}$
with an apparent horizontal velocity
of $v_{\rm h}=20\ {\rm km\ s}^{-1}$,
the sampling interval needs to be less than $L/v_{\rm h}=50\ {\rm s}$.
Note that this may be partly overcome
by using faster-cadence LOS magnetograms.

\subsection{Summary of this section}

In this section, we presented theoretical investigations
that try to address the subsurface origin and physical mechanisms
behind the large-scale/long-term evolution of flare-producing ARs.
We first showed
in the beginning of Sect.~\ref{subsec:num_fe}
that classical flux emergence simulations
of the $\Omega$-loop emergence can explain several characteristics,
such as magnetic tongues,
formation of flux ropes and sigmoids,
generation of shear flows and spot rotation,
helicity injection,
and non-neutralized currents.
However, most of these models do not reproduce
other important features of flaring ARs
such as the highly-sheared PIL between closely neighboring
opposite-polarity sunspots.

From the observational evidence of emergence of top-curled flux tube,
the helical kink instability was invoked as the possible production mechanism
of the $\delta$-sunspots
(Sect.~\ref{subsubsec:kink}).
3D models demonstrate that
(1) a tightly twisted tube develops a kink instability;
(2) the rise speed of the kinked tube is accelerated due to the enhanced buoyancy;
and (3) the tube reproduces a quadrupolar polarity pattern with a sheared PIL
on the photospheric surface.
These models can reproduce
the observed characteristics of Type 1 (Spot-spot) $\delta$-spots.

Type 3 (Quadrupole) $\delta$-spots may be produced
by the emergence of a flux tube with multiple buoyant segments
(Sect.~\ref{subsubsec:multibuoyant}).
Such a top-dent configuration is in fact created
in a large-scale convective emergence model.
Inspired by the observation of the quadrupolar AR NOAA 11158,
the emergence of a flux tube
that rises at two sections along the axis
was investigated.
It was found that the time evolution of the photospheric polarities,
i.e., the collision, shearing, and converging motions of the central bipole,
is fairly consistent with that of the actual AR.
Such evolutions were not achieved
by a pair of emerging flux tubes that are placed in parallel.
Together with the follow-up study,
the multi-buoyant segment model is considered as a likely candidate
for quadrupolar $\delta$-spots.

Interaction of emerging flux systems is also
recognized as a source of complexity
(Sect.~\ref{subsubsec:interactingtube}).
In fact, 3D simulations showed that
complex-shaped ARs can be
created by interaction of multiple tubes
in the solar interior.
One interesting consequence of the interaction,
both aerodynamic and bodily,
is that even simple bipolar ARs may originate from multiple flux systems through merging.
In this case, non-neutralized currents can be significant
because the return currents are annihilated.

Turbulent convection results in
a multitude of effects on the rising flux
(Sect.~\ref{subsubsec:turbulentconv}).
The convective emergence simulation
revealed that
the two polarities on the photosphere
undergo shearing and rotational motion
due to the Lorentz force
and that the converging motion at the PIL causes flux cancellation,
which leads to the production of a flux rope in the atmosphere.
It was also found that the strong collision of opposite polarities
results in a series of flare eruptions.

With the aim to obtain a unified perspective of production of flaring ARs,
a comparison of different modeling setups
was performed (Sect.~\ref{subsubsec:unified}).
It was assumed
that the production of
Spot-spot, Spot-satellite, Quadrupole, and Inter-AR types
are due to the emergence of a kink-unstable tube, two interacting tubes,
a multi-buoyant-segment tube, and two independent tubes, respectively.
Although all models except for the Inter-AR case
successfully reproduced $\delta$-spots with flaring PILs,
the Spot-spot case showed a by far fastest rising with the largest free magnetic energy.
Therefore,
the difference in the observed evolution on the solar surface
likely stems from the subsurface history, probably caused by turbulent convection,
such as a strong twisting, downward pinning, and collision with other flux systems.

Flux rope formation and the consequent eruption
have been extensively surveyed
in the sheared arcade and flux cancellation models
(Sect.~\ref{subsec:num_fc}).
Many of these simulation models are based on the filament formation theory
by \citet{1989ApJ...343..971V}:
the coronal fields are tied to the photospheric bottom boundary
and the photospheric motion, such as shearing, converging, and/or diffusion,
drives the overall evolution.
However, the reversed-shear and small-scale emerging field at the PIL
are also suggested as the trigger of magnetic reconnection between coronal arcades.
Flux cancellation models
that take into account the effect of thermodynamics
now reproduce the condensation of filament plasma due to radiative cooling.

Along with the extrapolation methods
(Sect.~\ref{subsubsec:num_data_extrapolation}),
recent progress in the more physics-based modeling of the coronal field
is facilitated by the development in magnetographs,
especially by the advanced
vector magnetograms of Hinode/SOT and SDO/HMI.
There are two methods in this category, 
which are data-constrained models,
where a single snapshot is used for creating the initial coronal field
(Sect.~\ref{subsubsec:num_data_constrained}),
and data-driven models,
where the bottom boundary is sequentially updated to drive the calculation
(Sect.~\ref{subsubsec:num_data_driven}).
These methods, although still in the stage of development,
provide the means to trace the evolution of coronal fields
in a more realistic manner,
such as the formation of flux ropes in response to the photospheric motion
and the resultant eruptions,
and may open the door to real-time space weather forecasting.

\section{Rapid changes of magnetic fields associated with flares}
\label{sec:change}

As we saw in the previous sections,
the gradual magnetic field evolution (in the time scale of hours to days)
is the key factor for the energy build up
of solar eruptions.
Then, can solar eruptions in the corona cause
rapid (within minutes) magnetic field changes in the photosphere?
The changes in the photosphere in response to the coronal eruptions
have been expected to be small
because the photospheric plasma density is much larger than that of the corona.
\citet{2016NatPh..12..998A} gave a review of this topic
from both observational and modeling perspectives
and provided a physical analysis of this issue
called the ``tail wags the dog'' problem.
Under certain circumstances,
the coronal eruption can cause rapid changes
in the photospheric magnetic topology.

Earlier, \citet{2008ASPC..383..221H} and \citet{2012SoPh..277...59F}
quantitatively assessed the back reaction on the solar surface and interior
resulting from the coronal field evolution required to release energy
and made the prediction that after flares,
the photospheric magnetic field
would become more horizontal at the flaring PILs.
Their analysis is based on the principle of energy and momentum conservation
and builds upon the proposal by \citet{2000ApJ...531L..75H}
that the coronal field should, in an overall sense,
contract or implode
if there is a net decrease in magnetic energy
(coronal implosion).
This is one of the very few models that specifically predict that
magnetic destabilization associated with flares can be
accompanied by rapid and permanent changes
of photospheric magnetic fields and the pattern of the field changes.
One special case related to this scenario is the tether-cutting reconnection model
for sigmoids \citep{2001ApJ...552..833M,2006GMS...165...43M},
which involves a two-stage reconnection process.
At the eruption onset,
the near-surface reconnection between the two sigmoid elbows
produces a low-lying shorter loop across the PIL
and a larger twisted flux rope connecting the two far ends of the sigmoid.
The second stage reconnection occurs
when the large-scale loop cuts through the arcade fields,
which causes the erupting flux rope to evolve into a CME
and precipitation of electrons to produce flare ribbons
(see Fig.~\ref{fig:cshkp}(a) for illustration).
If scrutinizing the magnetic topology close to the surface,
one would find a permanent change of magnetic fields
that conforms to the scenario as described above:
the magnetic fields turn more horizontal near the flaring PIL
due to the newly formed short loops there.

Whereas an earlier review by \citet{2015RAA....15..145W} summarizes
certain aspects of research up to that time,
focusing primarily on the results obtained before the SDO era,
this section summarizes more
recent observational findings
of rapid magnetic field and sunspot structure changes associated with flares
and briefly discusses the related theoretical insights.

\subsection{Magnetic transients}
\label{subsec:transients}

\begin{figure*}
  \centering
  \includegraphics[width=1.0\textwidth]{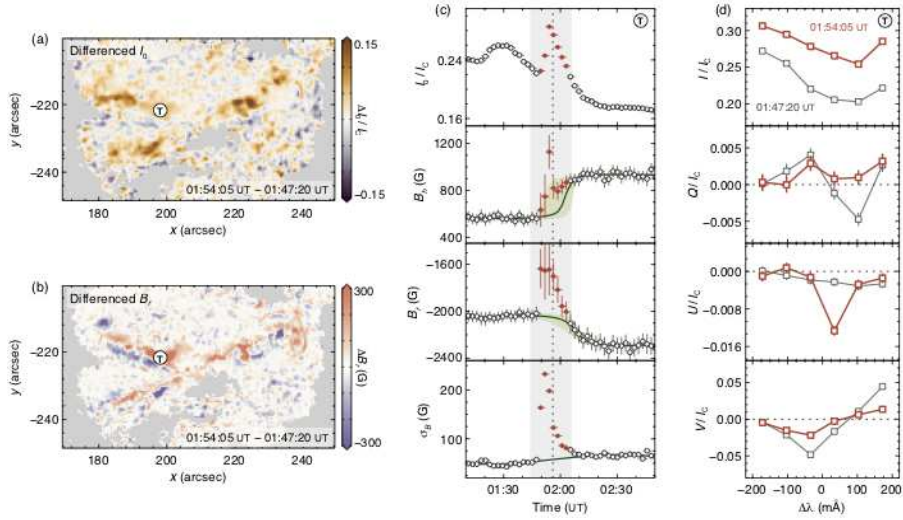}
  \caption{Flare-induced artifact as ``magnetic transient.''
    (a) Differenced map of intensitygram.
    Symbol ``T'' marks the sample pixel.
    (b) Differenced map of magnetic field in the radial direction $B_{r}$.
    (c) Temporal evolution of the sample pixel.
    Red symbols show the frames affected by flare emission.
    Green curves show the fitted step-like function for the horizontal field $B_{h}$
    and the radial field $B_{r}$ and a fitted third-order polynomial
    for the formal uncertainty of field strength ${\sigma}_{B}$;
    green bands show the $1\sigma $ fitting confidence interval.
    (d) Stokes profiles of the sample pixel at two instances,
    near (red) and before (gray) the flare peak.
    {Image reproduced by permission from \citet{2017ApJ...839...67S},
    copyright by AAS.}
    }
\label{fig:sun2017}
\end{figure*}

Before the discovery of the persistent photospheric magnetic field changes
associated with flares,
some studies showed observations of
the so-called ``magnetic transients''--the rapid, but short-lived change
in the LOS magnetic fields.
In the earlier studies \citep[e.g.,][]{1978SoPh...58..149T,1984ApJ...280..884P},
these apparent transient reversals of magnetic polarity
associated with flare footpoint emissions
were interpreted as real physical effects
of change in magnetic topology.
Some later studies demonstrated that the short-lived magnetic transients are
the observational effect due to changes in profiles of observing spectral lines
caused by the flare emissions \citep{2001ApJ...550L.105K,2003ApJ...599..615Q,2009RAA.....9..812Z},
so they are sometimes called magnetic anomalies.
The most comprehensive study in this topic is
a recent paper by \citet{2017ApJ...839...67S},
who analyzed the 135-s cadence HMI data
and demonstrated the line profile changes and associated field signatures of transients
(Fig.~\ref{fig:sun2017}).
Non-LTE\footnote{In local thermodynamical equilibrium (LTE), it is assumed that the state of plasma is described simply by the Saha-Boltzmann equations, i.e., as a function of the local kinetic temperature and electron density alone. Non-LTE indicates that this assumption is not valid.} modeling by \citet{2018ApJ...857L...2H} explained the profile changes
of Fe I 6173 {\AA} line that the HMI uses
and provided a quantitative assessment of magnetic transients.
\citet{2018ApJ...854...64S} suggested that magnetic transients and white-light flares
are closely related spatially and temporally.

\begin{figure*}
  \centering
  \includegraphics[width=0.8\textwidth]{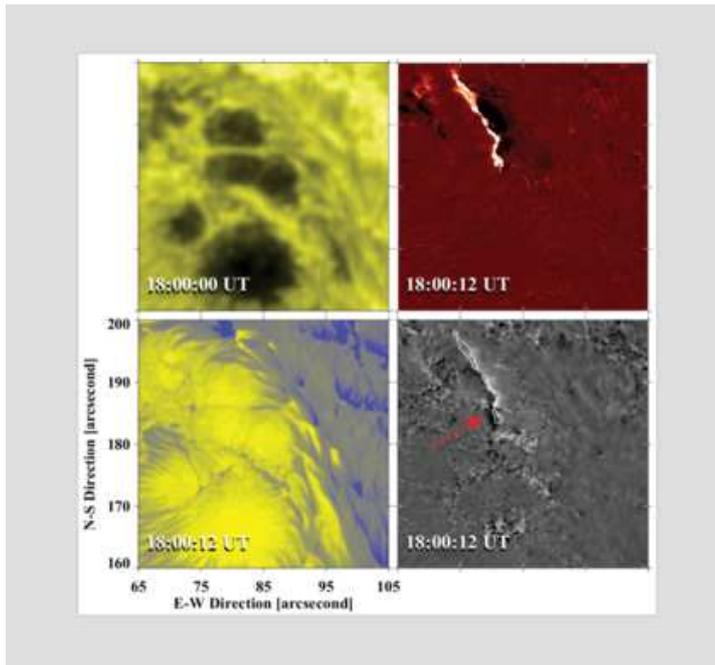}
  \caption{Azimuth angle changes in association with flare emission of 2015 June 22.
    The FOV is 40'' $\times$ 40''.
    (Top left) SDO/HMI white-light map.
    (Top right) Running difference image in H$\alpha$ blue wing (line core $-1.0$ {\AA}),
    showing the eastern flare ribbon.
    The bright part is the leading front and the dark part is the following component.
    (Bottom left) The GST/NIRIS LOS magnetogram,
    scaled in a range of $-2500\ {\rm G}$ (blue) to $2500\ {\rm G}$ (yellow).
    (Bottom right) Running difference map of azimuth angle generated
    by subtracting the map taken at 17:58:45 UT from the one taken at 18:00:12 UT.
    The dark signal pointed by the pink arrow represents the sudden,
    transient increase of azimuth angle at 18:00:12 UT.
    {Image reproduced by permission from \citet{2018NatCo...9...46X},
    copyright by ***.}
    }
\label{fig:xu2018}
\end{figure*}

All the above magnetic transients are for the LOS component of the magnetic fields.
Taking advantage of the unprecedented resolution provided
by the 1.6-m GST at BBSO,
\citet{2018NatCo...9...46X} showed a sudden rotation of the magnetic field vector
by about 12$^{\circ}$--20$^{\circ}$ counterclockwise,
in association with the M6.5-class flare on June 22, 2015.
Such changes of the azimuth angles of the transverse magnetic field
are well pronounced within a ribbon-like structure ($\sim 600\ {\rm km}$ in width),
moving co-spatially and co-temporally with the flare emission
as seen in the H$\alpha$ line  (see Fig.~\ref{fig:xu2018}).
However, they are not related to the magnetic transients as shown above.
A strong spatial correlation between the azimuth transient
and the ribbon front indicates that the energetic electron beams
are very likely the cause of the rotation.
During the rotation, the measured azimuth becomes closer to that of the potential field,
which indicates the process of energy release (untwisting motion) in the associated flare loop.
The magnetic fields restored their original direction
after the flare ribbons swept through over the area.
This was the first time that a transient field rotation was observed.
Possible explanations of this phenomenon include
(1) effect of induced magnetic fields;
(2) effect of downward-drafting plasma;
(3) polarization of emission lines due to return current
and/or filamentary chromospheric evaporation
(different from the original concept of magnetic transient);
and (4) effect of Alfv\'{e}n waves.
The authors claimed that the observed field change cannot be explained by existing models.
This new, transient magnetic signature in the photosphere may offer
a new diagnostic tool for future modeling
of magnetic reconnection and the resulting energy release.

\subsection{Rapid, persistent magnetic field changes}
\label{subsec:vector}

In the early 1990s,
the Caltech solar group discovered obvious rapid and permanent changes
of vector magnetic fields associated with the flares
using the BBSO data
\citep{1992SoPh..140...85W,1994ApJ...424..436W}.
They found that the transverse field shows much more prominent changes
compared to the LOS component.
Some of the results appeared to be puzzling:
the magnetic shear angle (an indicator of non-potentiality),
defined as the angular difference between the potential magnetic field
and the measured field (see Sect.~\ref{subsubsec:photo_pil}),
increases following flares.
It is well known that, in order to release the energy for a flare to occur,
the coronal magnetic field has to evolve to a more relaxed state to release energy.
For this reason, there have been some doubts to these earlier measurements,
especially because the data were obtained from ground-based observatories
that may suffer from certain effects such as atmospheric seeing
and lack of continuous observing coverage.

\citet{2001ApJ...550L.105K} studied high-resolution SOHO/MDI magnetogram data
for the ``Bastille Day Flare'' on 2000 July 14,
and found regions with a permanent decrease of magnetic flux,
which are related to the release of magnetic energy.
Using high cadence GONG data, 
\citet{2005ApJ...635..647S} found solid evidence
of step-wise field changes associated with a number of flares.
The time scale of the changes is as fast as 10 minutes
(GONG cadence is 1 minute),
and magnitude of change is in the order of 100 G.
\citet{2010ApJ...724.1218P}, \citet{2012ApJ...760...29J},
\citet{2012ApJ...756..144C}, and \citet{2013SoPh..283..429B}
also surveyed more comprehensively the rapid and permanent changes
of LOS magnetic fields with GONG data,
which were indeed associated with almost all the X-class flares studied by them.

The above studies using the LOS field data demonstrated
the step-wise property of flare-related photospheric magnetic field change.
However, the underlying cause of those changes was not clearly revealed.
The work by \citet{1999ApJ...525L..61C} was the first to use
near-limb magnetograph observations
to characterize flare-related changes of magnetic fields,
taking advantage of the projection effect.
In a number of papers, it was found that, for the LOS magnetic field,
the limb-ward flux increases in general, while the disk-ward flux in the flaring ARs decreases
\citep{2002ApJ...576..497W,2006ApJ...649..490W,2004ApJ...605..546Y,2002ApJ...572.1072S,2010ApJ...716L.195W}.
Such a behavior suggests that after flares,
the overall magnetic field structure of ARs may change
from a more vertical to a more horizontal configuration,
which is consistent with the scenario
that the Lorentz force change pushes down the field lines.
Note that most of the observations listed in \citet{2010ApJ...716L.195W}
are made by SOHO/MDI, which has a cadence of up to one minute.
The drastic change in inclination angle of magnetic fields in sunspots
associated with the flare eruption
was also detected by \citet{2016RAA....16...95Y}
by using vector magnetograms from the SDO/HMI,
and the observational result was consistent with
the expectation of the coronal implosion scenario.

\begin{figure*}
  \centering
  \includegraphics[width=1.0\textwidth]{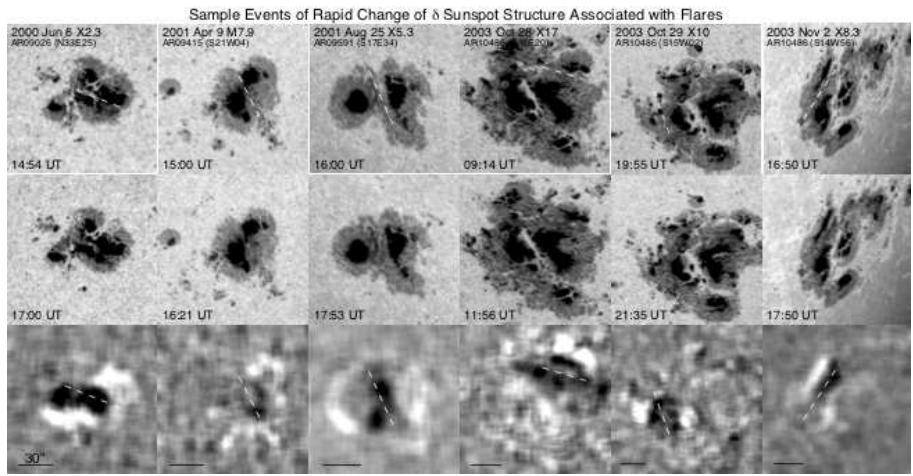}
  \caption{TRACE white-light images covering associated with six major flares.
    The rapid changes of $\delta$-sunspot structures are observed.
    The top, middle, and bottom rows show the pre-flare, the post-flare,
    and the difference images between them after some smoothing, respectively.
    The white pattern in the difference image indicates the region of penumbral decay,
    while the dark pattern indicates the region of darkening of penumbra.
    The white dashed line denotes the flaring PIL
    and the black line represents a spatial scale of 30''.
    {Image reproduced by permission from \citet{2005ApJ...622..722L},
    copyright by AAS.}
    }
\label{fig:liu2005}
\end{figure*}

\begin{figure*}
  \centering
  \includegraphics[width=1.0\textwidth]{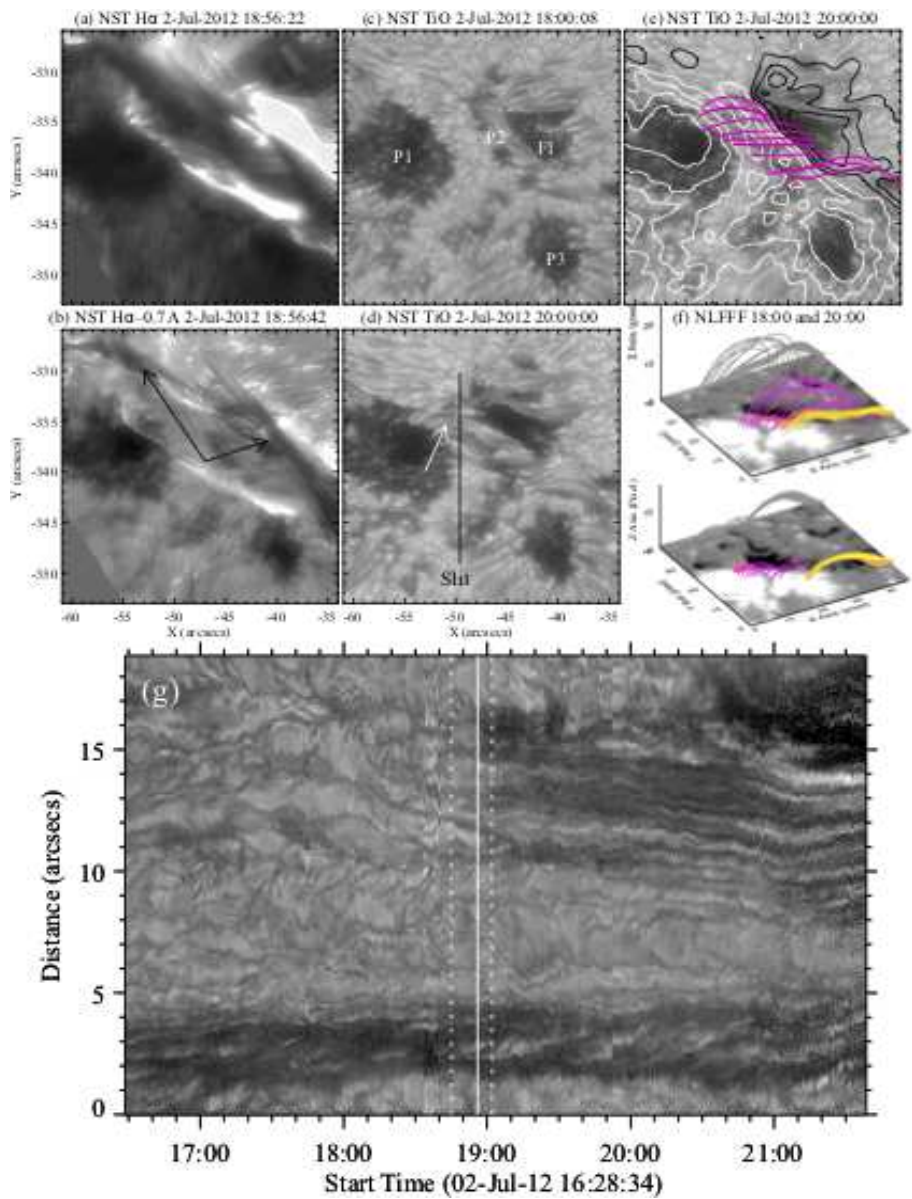}
  \caption{BBSO/GST H$\alpha$ center (a) and blue-wing (b) images
    at the peak of the 2011 July 2 C7.4 flare,
    showing the flare ribbons and possible signatures of a flux rope eruption
    (the arrows in panel (b)).
    The GST TiO images about 1 hour before (c) and 1 hour after (d) the flare
    clearly show the formation of penumbra (pointed to by the arrow in panel (d)).
    The same post-flare TiO image in panel (e) is superimposed
    with positive (white) and negative (black) HMI LOS field contours,
    and NLFFF lines (pink).
    (f) Perspective views of the pre- and post-flare 3D magnetic structures
    including the core field (a flux rope) and the arcade field from NLFFF extrapolations.
    The collapse of arcade fields is obvious.
    (g) TiO time slices for a slit (black line in panel (d))
    across the newly formed penumbra area.
    The dashed and solid lines denote the time of the start, peak, and end
    of the flare in GOES 1--8 {\AA}.
    The sudden turning off of the convection associated with the flare is obviously shown.
    {Images reproduced by permission from \citet{2013ApJ...774L..24W} and \citet{2014ApJ...784L..13J},
    copyright by AAS.}
    }
\label{fig:jing2014}
\end{figure*}

As more and more evidence indicates the irreversible photospheric magnetic field changes following flares,
it is natural to find whether these changes are detectable in white-light structures of ARs.
The white-light signatures of topological changes are indeed discovered in a number of papers
\citep[e.g.][]{2004ApJ...601L.195W,2005ApJ...622..722L,2005ApJ...623.1195D,2009ScChG..52.1702L,2009ApJ...690..862W,2013ApJ...774L..24W,2018ApJ...853..143W}.
The most prominent changes are the enhancement (i.e., darkening) of penumbral structure
near the flaring PILs and the decay of penumbral structure
in the peripheral sides (outer edges) of $\delta$-spots.
Fig.~\ref{fig:liu2005} clearly demonstrates some examples of such spot structure changes.
The difference image between pre- and post-flare states always shows a dark patch
at the flaring PIL that is surrounded by a bright ring.
They correspond to the enhancement of the central sunspot penumbrae
and the decay of the peripheral penumbrae, respectively.
These examples were discussed in detail by \citet{2005ApJ...622..722L},
in which they showed that
(1) these rapid changes are associated with flares and are permanent,
and (2) the decay of sunspot penumbrae is related
to the magnetic field
in the outer edge of AR that turns to a more vertical direction,
while the darkening of sunspot structure near the central PIL is
related to the magnetic field that turns to a more horizontal direction.
\citet{2007ChJAA...7..733C} statistically studied over 400 events
using TRACE white-light data
and found that the significance of sunspot structure change
is positively correlated
with the magnitude of flares.
Using Hinode/SOT G-band data,
\citet{2012ApJ...748...76W} further studied
the intrinsic linkage of penumbral decay to magnetic field changes.
They took advantage of the high spatio-temporal resolution Hinode/SOT data
and observed that in sections of peripheral penumbrae swept by flare ribbons,
the dark fibrils completely disappear
while the bright grains evolve into faculae
where the magnetic flux becomes even more vertical.
These results again suggest that
the component of horizontal magnetic field of the penumbra
is straightened upward (i.e., turning from horizontal to vertical)
due to magnetic field restructuring associated with flares.
Also notably, the flare-related enhancement of penumbral structure
near central flaring PILs has also been unambiguously observed with BBSO/GST.
Using GST TiO images with unprecedented spatial (0.1'') and temporal (15 s) resolution,
\citet{2013ApJ...774L..24W} reported on a rapid formation of sunspot penumbra
at the PIL associated with the 2012 July 2 C7.4 flare
(see Fig.~\ref{fig:jing2014} and the corresponding movie).
The most striking observation is that the solar granulation evolves
to the typical pattern of penumbra consisting of alternating dark and bright fibrils.
Interestingly, a new $\delta$-sunspot is created by the appearance of
such a penumbral feature,
and this penumbral formation also corresponds to the enhancement of the horizontal field.
Similar pattern of penumbral formation is shown by \citet{2018ApJ...853..143W}.

\begin{figure*}
  \centering
  \includegraphics[width=1.0\textwidth]{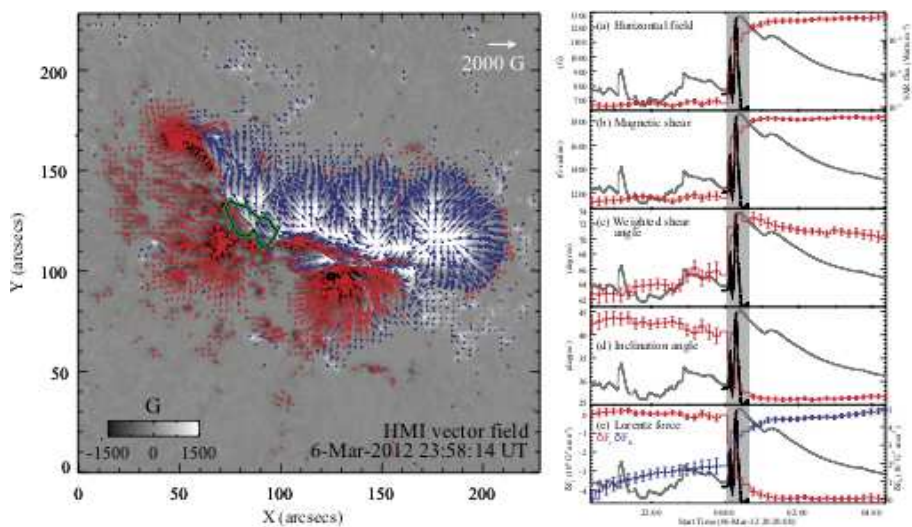}
  \caption{(Left) HMI vector magnetogram on 2012 March 7
    showing the flare-productive AR NOAA 11429 right before the X5.4 flare.
    (Right) Temporal evolution of various magnetic properties
    of a compact region (green contour in the left panel) at the central PIL,
    in comparison with the light curves of GOES 1--8~\AA\ soft X-ray flux (gray)
    and its derivative (black).
    Note that in panel (d), the inclination is measured from horizontal direction.
    The shaded interval denotes the flare period in the GOES flux.
    {Image reproduced by permission from \citet{2012ApJ...757L...5W},
    copyright by AAS.}
    }
\label{fig:wang2012}
\end{figure*}

A very clear demonstration of flare related changes in vector magnetic fields
came from the analysis of SDO/HMI vector data by \citet{2012ApJ...745L..17W}.
The analysis of the X2.2 flare in AR NOAA 11158 on 2011 February 15
clearly demonstrated a rapid/irreversible increase
of the horizontal magnetic field at the flaring PIL. 
The mean horizontal fields increased by about 500 G within 30 minutes after the flare.
The authors also found that the photospheric field near the flaring PIL became
more sheared and
more inclined towards horizontal,
consistent with the earlier results
\citep[e.g.,][]{1992SoPh..140...85W,1994ApJ...424..436W,2005ApJ...622..722L}.
Following that initial study, a number of papers using HMI data demonstrated
the consistent changes of magnetic fields
\citep{2012ApJ...745L...4L,2012ApJ...748...77S,2012ApJ...757L...5W,2012ApJ...759...50P,2013SoPh..287..415P,2014ApJ...786...72Y,2018ApJ...852...25C}.
The found patterns of the changes are consistent in the sense
that the transverse field enhances in a region across the central flaring PIL.
Figure~\ref{fig:wang2012} shows the typical time profiles of such field changes.

\begin{figure*}
  \centering
  \includegraphics[width=0.8\textwidth]{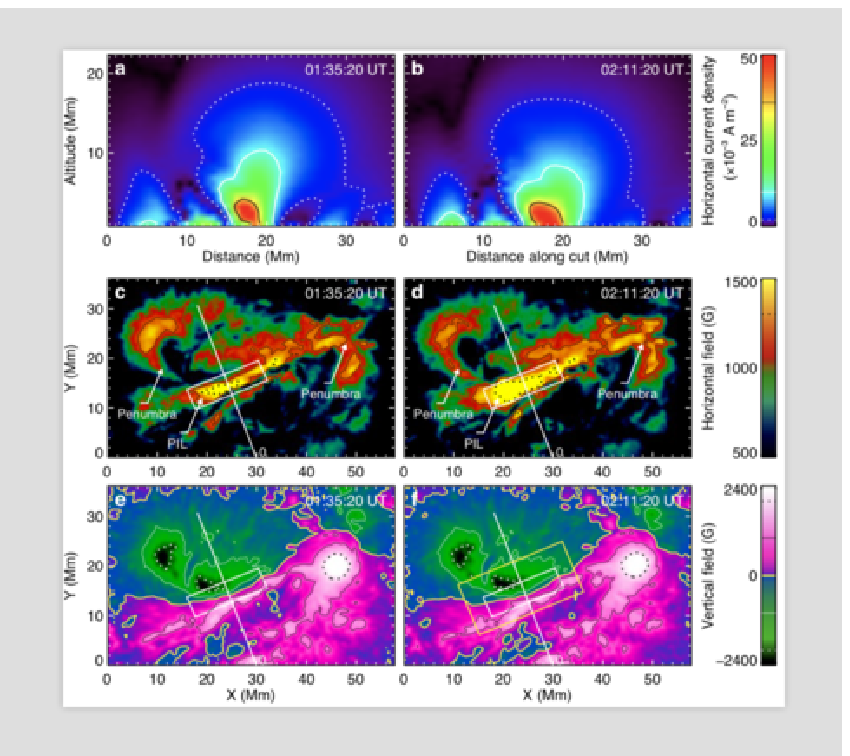}
  \caption{Modeled and observed field changes
    from before (01:00 UT; (a), (c), and (e))
    to after (04:00 UT; (b), (d), and (f))
    the 2011 February 15 X2.2 flare.
    (a--b) Current density distribution on a vertical cross section
    indicated in (c)--(f). (c--d) HMI horizontal field strength.
    Contour levels are 1200 G and 1500 G. (e--f) HMI vertical field.
    Contour levels are $\pm 1000\ {\rm G}$ and $\pm 2000\ {\rm G}$.
    {Image reproduced by permission from \citet{2012ApJ...748...77S},
    copyright by AAS.}
    }
\label{fig:sun2012}
\end{figure*}

Associated with the above findings in the 2D photospheric magnetic fields,
there must be a corresponding magnetic field evolution in 3D above the photosphere.
The NLFFF extrapolation works as a powerful tool to reconstruct
the 3D magnetic topology of the solar corona
(see Sect.~\ref{subsubsec:num_data_extrapolation} for the extrapolation methods).
Using Hinode/SOT magnetic field data,
\citet{2008ApJ...676L..81J} showed that the magnetic shear (indicating non-potentiality)
only increases at lower altitude
while it still largely relaxes in the higher corona,
therefore the total free magnetic energy in 3D volume
should still decrease after energy release of a flare.
Using HMI data, \citet{2012ApJ...748...77S} clearly showed
that the electric current density indeed increases
at the flaring PIL near the surface while it decreases higher up,
which may explain the overall decrease of free magnetic energy
together with a local enhancement at low altitude (see Fig.~\ref{fig:sun2012}).
The above results may also imply that magnetic fields collapse toward the surface.
Such a collapse was even detected in a C7.4 flare on 2012 July 2
as reported by \citet{2014ApJ...784L..13J} and shown in Fig.~\ref{fig:jing2014}.
The collapse (or contraction) of magnetic arcades
as reflected by NLFFF models across the C7.4 flare
is spatially and temporally correlated with the formation of sunspot penumbra
on the surface \citep{2013ApJ...774L..24W},
as observed in high resolution observations of GST.
The physics of this phenomenon is not fully understood:
this could be due to newly reconnected magnetic fields above the PIL,
or perhaps the reduction of local magnetic pressure
due to a removal/weakening of the magnetic flux rope instigates the collapse.

\begin{figure*}
  \centering
  \includegraphics[width=1.0\textwidth]{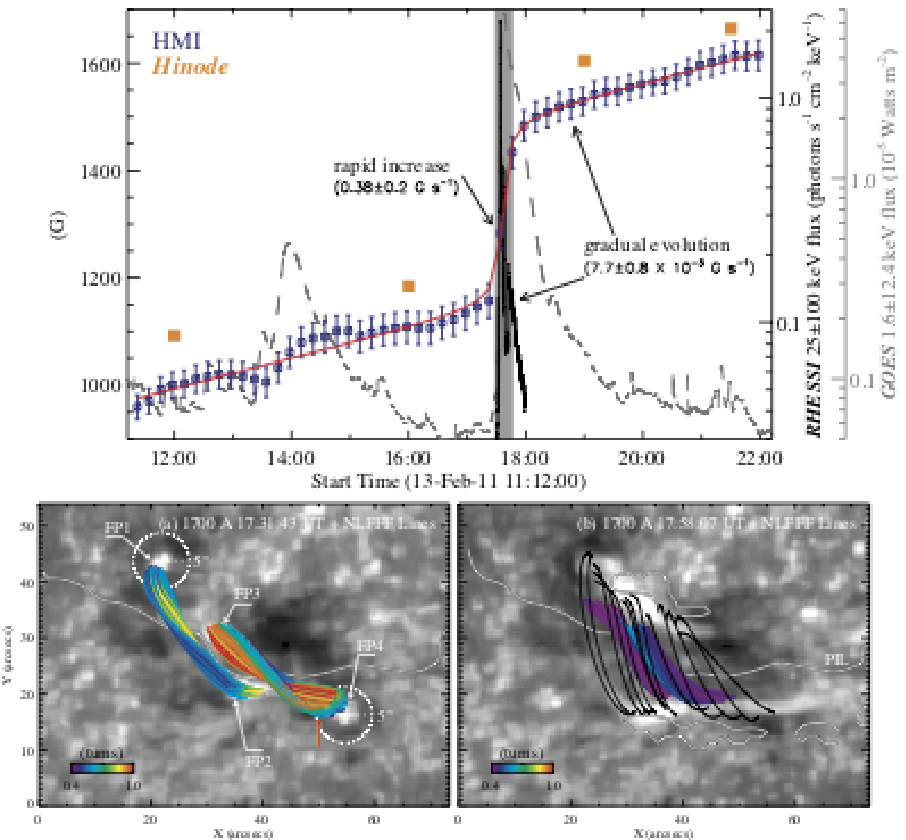}
  \caption{(Top) Temporal evolution of horizontal magnetic field
    measured by HMI and Hinode/SOT in a compact region around the PIL,
    in comparison with X-ray light curves
    for the M6.6 flare on 2011 February 13.
    The red curve is the fitting of HMI data with a step function.
    (Bottom) Extrapolated NLFFF lines before and after the event,
    demonstrating the process of magnetic reconnection
    consistent with the tether-cutting reconnection model.
    {Images reproduced by permission from \citet{2012ApJ...745L...4L} and \citet{2013ApJ...778L..36L},
    copyright by AAS.}
    }
\label{fig:liu2012}
\end{figure*}

Using vector magnetograms from HMI together with those from Hinode/SOT
with high polarization accuracy and spatial resolution,
\citet{2012ApJ...745L...4L} revealed similar rapid and persistent increase
of the transverse field associated with the M6.6 flare on 2011 February 13,
together with the collapse of coronal currents
toward the surface at the sigmoid core region.
\citet{2013ApJ...778L..36L} further compared the NLFFF extrapolations
before and after the event (see Fig.~\ref{fig:liu2012}).
The results provide direct evidence of the tether-cutting reconnection model.
There are four flare footpoints.
About 10\% of the flux ($\sim 3\times 10^{19}\ {\rm Mx}$)
from the inner footpoints (e.g., FP2 and FP3 of loops FP2--FP1 and FP3--FP4)
undergoes a footpoint exchange to create shorter loops of FP2--FP3.
This result presents the rapid/irreversible changes
of the transverse field and corresponding 3-D  field changes in corona. 
A more comprehensive investigation including
the 3D magnetic field restructuring and flare energy release
as well as the helioseismic response
of two homologous flares,
the 2011 September 6 X2.1 and September 7 X1.8 flares in AR NOAA 11283,
was performed by \citet{2014ApJ...795..128L}.
Their observational and modeling results depicted a coherent picture
of coronal implosions, in which the central field collapses
while the peripheral field turns vertical,
consistent with what was found  by \citet{2005ApJ...622..722L}.

\begin{figure*}
  \centering
  \includegraphics[width=0.7\textwidth]{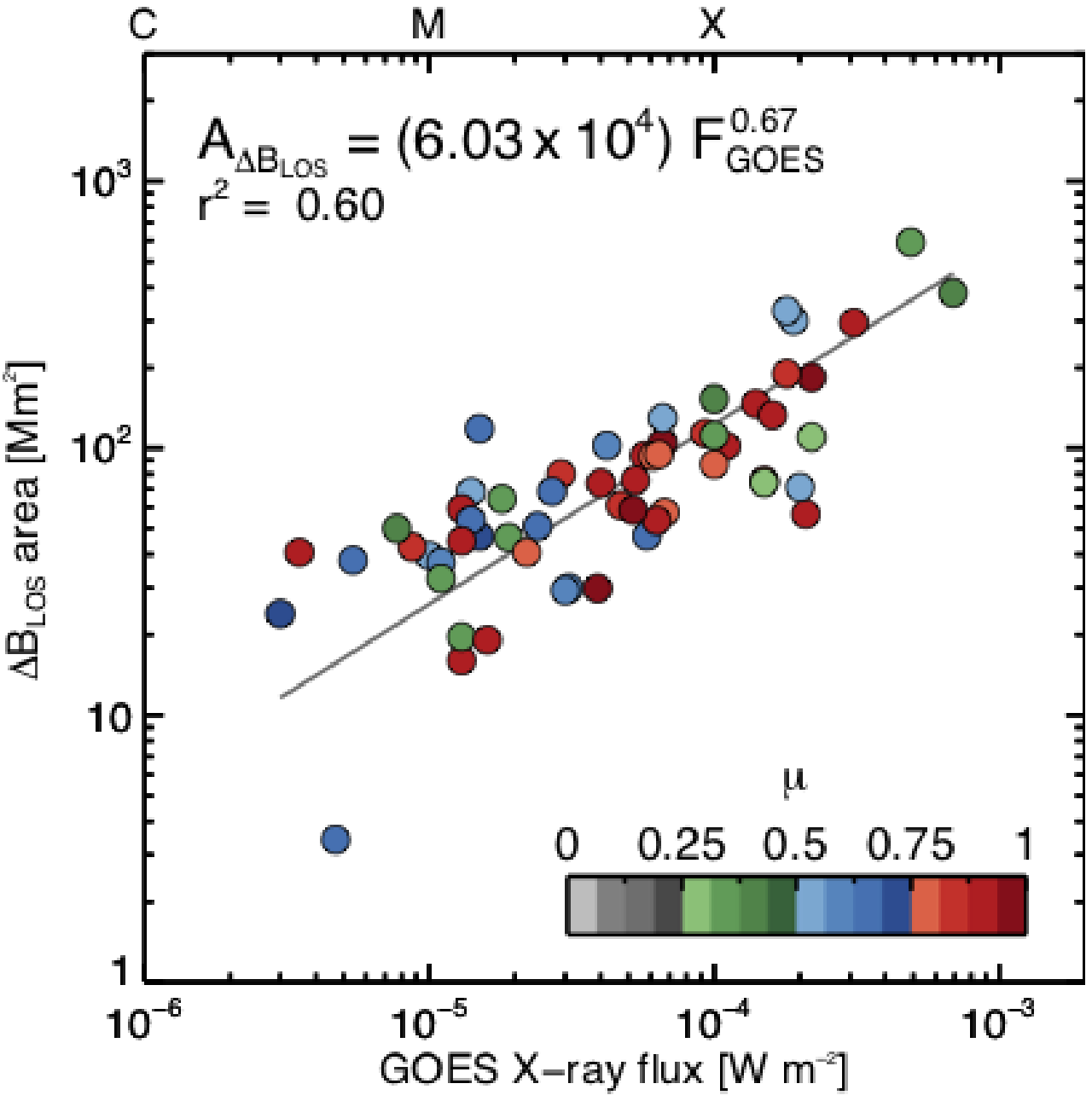}
  \caption{Area affected by rapid field changes corrected
    for foreshortening of LOS magnetic field
    as a function of the peak GOES soft X-ray flux of 75 events.
    Color-coded circles denote the center-to-limb distance
    $\mu$ (cosine of the heliocentric angle) of each event.
    The line is the best fit to a power law
    with a correlation coefficient of 0.6.
    {Image reproduced by permission from \citet{2018ApJ...852...25C},
    copyright by AAS.}
    }
\label{fig:castellanos2018}
\end{figure*}

There are two research directions that are particularly worth mentioning here.
\begin{itemize}
\item Joint analysis of photospheric magnetic fields and coronal topology.
\citet{2016SoPh..291..791P} studied two X-class flares observed by SDO and
the Solar Terrestrial Relations Observatory \citep[STEREO;][]{2008SSRv..136....5K}.
They found that the rapid changes of magnetic fields at the PIL
is associated with coronal loop contraction.
\citet{2017A&A...602A..60G} analyzed VTT (Vacuum Tower Telescope) data
covering an M-class flare and found an enhancement
of the transverse magnetic field of approximately 550 G.
This transverse field was found to bridge the PIL
and connect umbrae of opposite polarities in the $\delta$-spot.
At the same time, a newly formed system of loops appeared co-spatially
in the corona as seen in 171 {\AA} passband images of SDO/AIA.
Therefore, the rapid photospheric magnetic field evolution
is a part of 3D magnetic field re-structuring.

\item Statistical study of a large number of events.
\citet{2018ApJ...852...25C} carried out a statistical analysis of
permanent LOS magnetic field changes
during 18 X-, 37 M-, 19 C-, and 1 B-class flares using data from SDO/HMI.
They investigated the properties of permanent changes,
such as frequency, areas, and locations.
They detected changes of LOS field in 59 out of 75 flares
and found that the strong flares are more likely to show changes.
Figure~\ref{fig:castellanos2018} demonstrates the correlation
between the affected LOS field change area
and the peak GOES flux.
It is apparent that larger flare produces more prominent field changes.
\end{itemize}

\subsection{Sudden sunspot rotation and flow field changes}

The evolution of magnetic fields is closely associated
with photospheric flow motions.
Obviously, the studies of the flow fields
along with the magnetic field evolution is very important.
Several methods of flow tracking have been developed
as summarized and compared by \citet{2007ApJ...670.1434W}.
One particular method is the differential affine velocity estimator
\citep[DAVE;][]{2005ApJ...632L..53S,2006ApJ...646.1358S}
that uses the induction equation to derive flow fields.
A substantially improved version,
DAVE for vector magnetograms \citep[DAVE4VM;][]{2008ApJ...683.1134S},
derives not only the horizontal but also the vertical component of the flows,
which thus can analyze the flux emergence (i.e., vertical motions)
in addition to the horizontal motions.

\citet{2014ApJ...782L..31W} showed some initial results
of the flare-related acceleration of sunspot rotation
that is derived by DAVE using SDO/HMI observations of AR NOAA 11158.
The rotational speeds of the two sunspots increase significantly
during and right after the X2.2 flare.
Moreover, the direction of the enhanced sunspot rotation
agrees with that of the change of the horizontal Lorentz force.
Using the estimated torque and moment of inertia,
\citet{2014ApJ...782L..31W} estimated the angular acceleration of the sunspots.
Although there are some uncertainties in the measurements and assumptions,
the values agree with the observed angular acceleration
of suddenly rotating sunspot immediately after the flare.

\begin{figure*}
  \centering
  \includegraphics[width=1.0\textwidth]{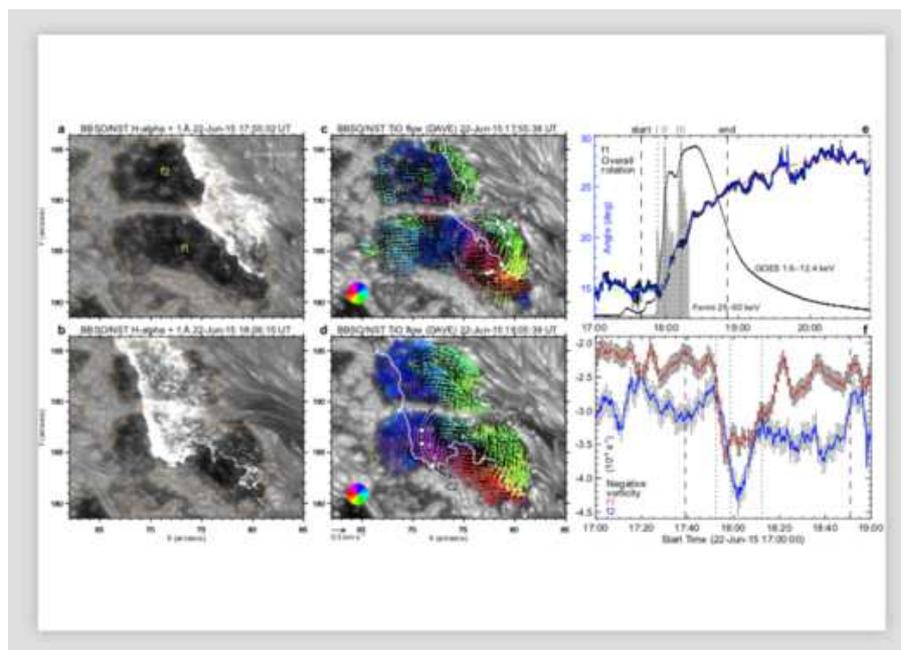}
  \caption{BBSO/GST chromospheric H$\alpha +1$ {\AA} images
    showing flare ribbons (a and b)
    and the corresponding photospheric TiO images (c and d).
    In panel (a), sunspots are labeled as f1 and f2,
    with the dotted lines contouring the vertical magnetic field at 1300 G.
    In panels (c) and (d), the superimposed arrows
    (color-coded by direction; see the color wheel)
    depict the differential sunspot rotation tracked with DAVE.
    The thick white curves are the co-temporal flare ribbon.
    (e) Temporal evolution of overall sunspot rotation,
    showing the orientation angle of f1 from an ellipse fit (blue)
    and its approximation using an acceleration plus a deceleration function.
    (f) Temporal evolution of vorticity derived based on DAVE velocity vectors
    indicating the accelerated sunspot rotation.
    {Image reproduced by permission from \citet{2016NatCo...713104L},
    copyright by ***.}
    }
\label{fig:liu2016}
\end{figure*}

\citet{2016NatCo...713104L} used GST data to analyze
the flow motions of the 2015 June 22 M6.6 flare.
It is particularly striking that the rotation is not uniform over the sunspot:
as the flare ribbon sweeps across, its different portions accelerate
(up to $50^{\circ}\ {\rm hr}^{-1}$) at different times
corresponding to peaks of the flare hard X-ray emission.
Associated with the rotation, the intensity and magnetic field
of the sunspot change significantly,
and the Poynting and helicity fluxes temporarily reverse their signs,
indicating that the energy propagation that causes the rotation is
from the higher atmosphere down to the photosphere.
Figure~\ref{fig:liu2016} demonstrates the key results of that study
(see also the corresponding movie).

\begin{figure*}
  \centering
  \includegraphics[width=0.8\textwidth]{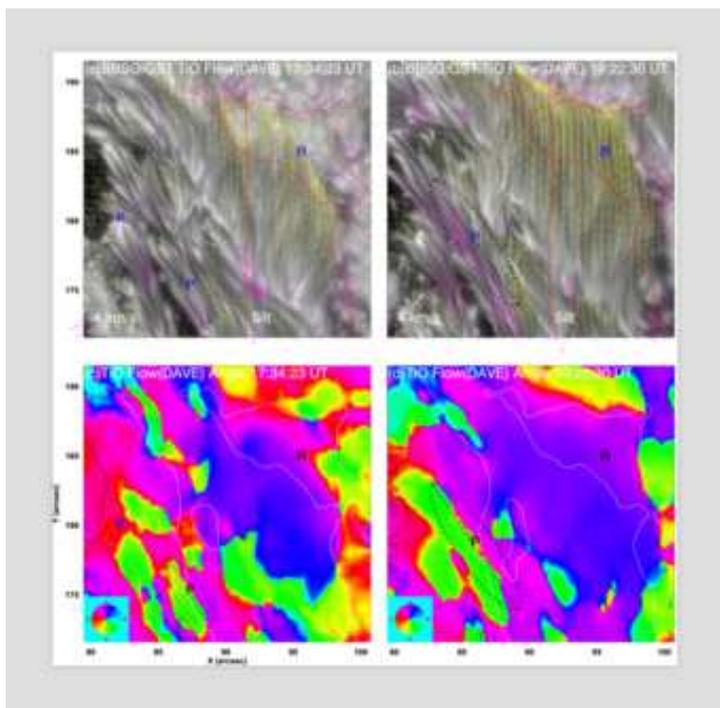}
  \caption{Flow field in the BBSO/GST TiO band.
    (a and b) Pre-flare (at 17:34:23 UT) and post-flare (at 19:22:30 UT) TiO images
    overplotted with arrows illustrating the flow vectors derived with DAVE.
    For clarity, arrows pointing northward (southward) are coded yellow (magenta).
    (c and d) Azimuth maps of corresponding flow vectors in panels (a) and (b),
    also overplotted with the PIL, precursor kernel,
    and region R contours.
    The shear flow region P showing the most obvious flare-related enhancement
    is outlined using the dashed ellipse,
    with its major axis quasi-parallel to the PIL.
    {Image reproduced by permission from \citet{2018ApJ...853..143W},
    copyright by AAS.}
    }
\label{fig:wangj2018}
\end{figure*}

\cite{2018ApJ...853..143W} analyzed the same AR with GST and HMI data.
For a penumbral segment in the negative field adjacent to the PIL,
an enhancement of penumbral flows
(up to an unusually high value of $2\ {\rm km\ s}^{-1}$)
and extension of penumbral fibrils
after the first peak of the flare hard X-ray emission.
They also found an area at the PIL,
which is co-spatial with a precursor brightening kernel,
that exhibits a gradual increase of shear flow velocity
(up to $0.9\ {\rm km\ s}^{-1}$) after the flare.
The enhancing penumbral
and shear flow regions are also accompanied by an increase
of horizontal field and decrease of magnetic inclination angle
measured from the horizontal.
These results further confirm the concept of back reaction of coronal restructuring
on the photosphere as a result of flare energy release.
Figure~\ref{fig:wangj2018} shows the evolution of the flow fields covering the flare.

\subsection{Theoretical interpretations}
\label{subsec:change_num}

The modeling efforts of ARs and related eruptions are summarized in Sect.~\ref{sec:num}.
Here we review certain points in explaining magnetic field restructuring following flares.
\citet{2014SoPh..289.2091L} reviewed solar eruption models
and classified them into three categories,
tether-cutting, break-out and loss-of-equilibrium, all of which can be catastrophic.
The tether-cutting model assumes a two-step reconnection that leads to eruption
in the form of flares and CMEs, in particular, for sigmoid ARs \citep[e.g.,][]{1980IAUS...91..207M, 2001ApJ...552..833M, 2006GMS...165...43M}.
The first-stage reconnection occurs near the solar surface at the onset of the eruption and produces a low-lying shorter loop across the PIL
and thus explains the observed enhancement of transverse fields after flare.
It also produces a much longer twisted flux rope
connecting the two far ends of a sigmoid
that triggers the second stage of eruption:
the twisted flux rope becomes unstable and erupts outward to form a full CME.

It is possible that in the earlier phase of the eruption,
contraction of the shorter flare loop occurs.
This has received increasing attention recently
\citep[e.g.,][]{2006ApJ...636L.173J}
and possibly corresponds to the first stage of the tether cutting.
The ribbon separation described in the standard flare models
such as the CSHKP model (Sect.~\ref{subsec:flares}) manifests the second stage.
This model may explain other observational findings such as
(1) transverse magnetic field at flaring PILs increases
rapidly/persistently immediately following the flares
\citep{2002ApJ...576..497W,2004ApJ...605..931W,2010ApJ...716L.195W};
(2) penumbral decay occurs in the peripheral penumbral areas of $\delta$-spots,
indicating that the magnetic field lines turn more vertical
after a flare in these areas
\citep{2004ApJ...601L.195W,2005ApJ...622..722L};
and (3) hard X-ray images of the Reuven Ramaty High Energy Solar Spectroscopic Imager
\citep[RHESSI;][]{2002SoPh..210....3L}
show four footpoints,
two inner ones and two outer ones,
and sometimes the hard X-ray emitting sources change
from confined footpoint structure to an elongated ribbon-like structure
after the flare reaches intensity maximum
\citep{2007ApJ...658L.127L,2007ApJ...669.1372L}.

In an attempt to quantitatively compare observations and modeling,
\citet{2011ApJ...727L..19L} compared idealized MHD simulation
of emerging flux in flare triggering with observation.
They selected a lower level in the simulation to examine
the near-surface magnetic structure evolution.
Changes of magnetic field orientation and strength in the photosphere 
after flares/CMEs are indeed found in the simulation.
The most obvious match is at the flaring PIL,
where field lines in the simulation are found to be more inclined
towards the horizontal,
and transverse field strength increased after the eruption.
At the outer side of the simulated sunspot penumbral area,
field lines turn to a more vertical direction
with a decreased transverse field strength.
These are consistent with the observed penumbral enhancement at the PIL
and decay of peripheral penumbrae \citep{2005ApJ...622..722L}.
The simulation also shows the downward net Lorentz force
pressing onto the photosphere, confirming the related observations.

\begin{figure*}
  \centering
  \includegraphics[width=0.8\textwidth]{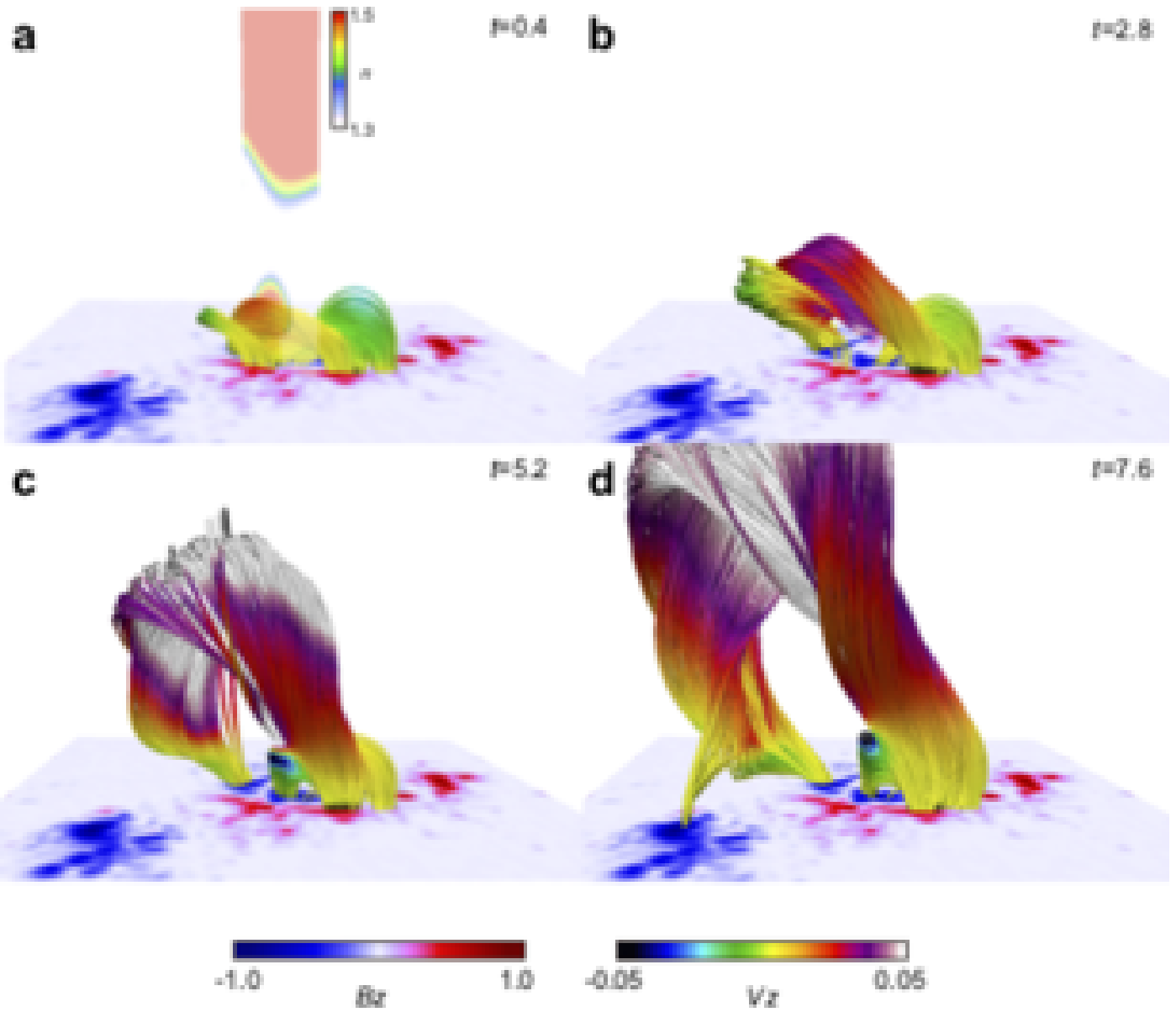}
  \includegraphics[width=0.9\textwidth]{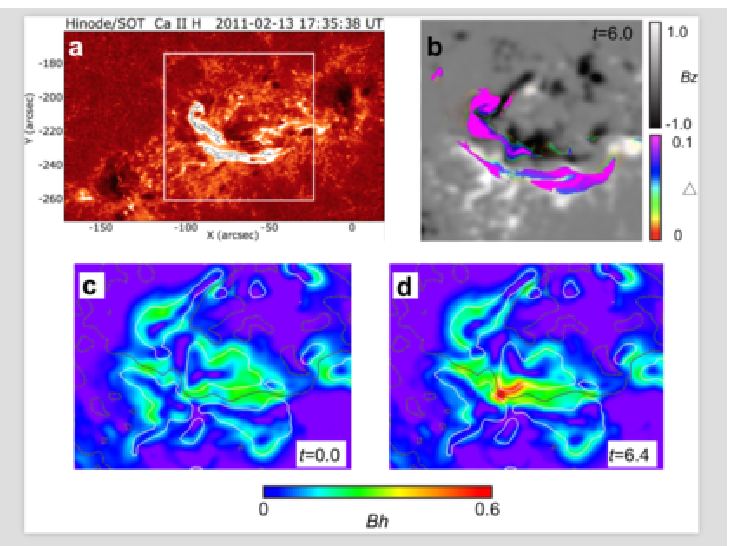}
  \caption{(Top) Temporal evolution of the modeled 3D dynamics
    of the eruptive flux rope on 2011 February 13 in AR NOAA 11158,
    together with the $B_{z}$ distribution at the bottom.
    (Bottom) Comparison of simulation results with observations.
    (a) Flare ribbons during the M6.6 flare, observed by Hinode at 17:35 UT.
    (b) Synthetic flare ribbons measured from total displacement of the field line
    superimposed on the $B_{z}$ distribution.
    The area corresponds to one surrounded by white square in panel (a).
    (c and d) $B_{h}$ distributions obtained from the simulation,
    just prior to and during the eruption, respectively.
    $B_{h}$ increased prominently across the main PIL
    (marked by the black lines).
    {Image reproduced by permission from \citet{2018NatCo...9..174I},
    copyright by ***.}
    }
\label{fig:inoue2018_2}
\end{figure*}

Recently, \citet{2018NatCo...9..174I} performed an MHD simulation that takes into account the observed photospheric magnetic field to reveal the dynamics of a solar eruption in a realistic magnetic environment. In this simulation, they confirmed that the tether-cutting reconnection occurring locally above the PIL creates a twisted flux tube, which is lifted into a toroidal unstable area where it loses equilibrium, destroys the force-free state, and drives the eruption. Figure~\ref{fig:inoue2018_2} shows that the simulation not only reproduces the flare ribbons well but also demonstrates the irreversible transverse field enhancement at the photospheric PIL. Although the authors did not emphasize this point, the peripheral penumbral decay is also apparent in the simulated data. The same event has been analyzed in detail observationally by \citet{2012ApJ...745L...4L,2013ApJ...778L..36L}. Note that \citet{2015ApJ...803...73I} demonstrated similar field changes for the X2.2 flare in the same AR. The rapid field change coincides with the onset of the flare.

As we mentioned earlier,
\citet{2008ASPC..383..221H} and \citet{2012SoPh..277...59F}
introduced the back reaction concept.
The authors made the prediction that after flares, at the flaring PIL,
the photospheric magnetic fields become more horizontal.
The analysis is based on the simple principle of energy
and momentum conservation:
the upward erupting momentum must be compensated by the downward momentum
as the back reaction.
In addition,
the field change should be stepwise (i.e. permanent)
because it results from the removal of magnetic energy and magnetic pressure
from the corona.
This is one of the few models
that specifically predict the rapid and permanent
changes of photospheric magnetic fields associated with flares
and support the observed Lorentz force change
\citep[e.g.,][]{2012ApJ...745L..17W,2012ApJ...757L...5W,2012ApJ...745L...4L,2012ApJ...748...77S,2013SoPh..287..415P,2014SoPh..289.3663P,2019ApJS..240...11P}.

As a more recent study,
\citet{2018ApJ...859...25W} analyzed four flare events using SDO/AIA and STEREO
and demonstrated the existence of real contractions of loops.
They identified two categories of implosion, which are
(1) a rapid contraction at the beginning of the flare impulsive phase,
as magnetic free energy is removed rapidly by a filament eruption;
and (2) a continuous loop shrinkage during the entire flare impulsive phase
that corresponds to ongoing conversion of magnetic free energy in a coronal volume.

\begin{figure*}
  \centering
  \includegraphics[width=1.0\textwidth]{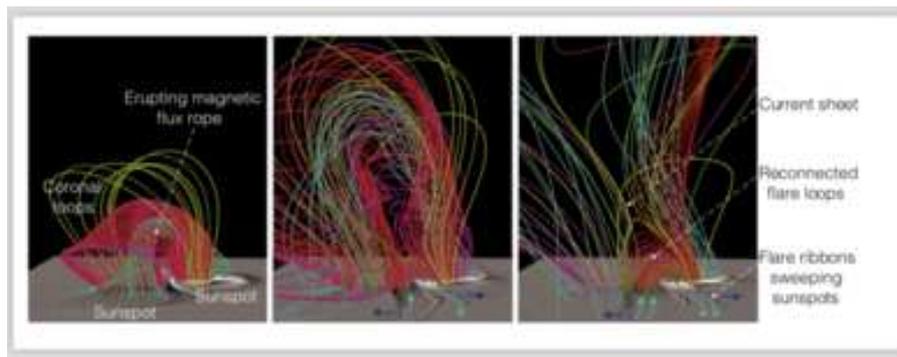}
  \caption{Series of snapshots, from left to right,
    of a realistic numerical simulation of an eruptive flare.
    The colored lines show representative coronal magnetic field lines
    plotted from fixed footpoints in the photosphere:
    the cyan field lines represent the erupting flux rope,
    and the red (green) field lines are
    those that eventually reconnect with pink (yellow) field lines.
    The gray scale plane shows the time-varying electric current densities
    in the photosphere.
    The blue arrows show the displacement of the ribbons
    and cyan curved arrows indicate how sunspot rotation
    is initiated as flare ribbons move across sunspots.
    {Image reproduced by permission from \citet{2016NatPh..12..998A},
    copyright by ***.}
    }
\label{fig:aulanier2016}
\end{figure*}

Finally, in \citet{2016NatPh..12..998A},
the sudden sunspot rotation is somehow demonstrated
in their simulation (see Fig.~\ref{fig:aulanier2016}).
Note that these simulations usually assume the line-tying condition,
i.e., the footpoint motions are not allowed
(see Sect.~\ref{subsec:num_fc} for details).
Nevertheless, the observed trend slightly above the photosphere
can demonstrate the direction for the rotational force,
although quantitative comparison is very difficult.

\section{Summary}
\label{sec:summary}

How close have we reached to the complete picture
of the formation and evolution of flare-producing ARs?
Thanks to the advancement of observation techniques and modeling efforts,
we have acquired a substantial amount of knowledge
that may set the grounds for a more complete understanding.
In this section,
we summarize our current understanding of
the genesis and evolution and key observational features of these ARs.

\subsection{The era with Hinode, SDO, and GST}

To a greater degree,
our understanding of the flaring ARs has been
pushed forward by the ceaseless improvement of observation instruments,
and the progress in the last decade has been made in particular
by Hinode, SDO, and GST.
In fact,
many parts of this review article
are based on the outcome of these missions.

Since launch in September 2006,
the Hinode spacecraft has sent us various important observables.
By virtue of seeing-free condition from space,
one of its trio of instruments, SOT,
has acquired high-resolution vector magnetograms,
revealed the detailed structure of flaring PILs,
and showed us its importance in triggering flares and CMEs
(Sect.~\ref{subsubsec:photo_pil}).
With the vector magnetograms,
though not quite satisfactorily,
now we can extrapolate the coronal field by the NLFFF techniques,
which is used as the initial condition of data-constrained simulations
(Sect.~\ref{subsec:num_data}).
Moreover, through simultaneous multi-wavelength observation
in concert with XRT and EIS,
Hinode realized even more comprehensive tracing of the dynamical evolution
over the different atmospheric layers.
The flux rope formation due to the photospheric shear motion
and the non-thermal broadening of EUV lines in response to the helicity injection
are good examples of Hinode's multi-wavelength probing of flare-producing ARs
(Sects.~\ref{subsubsec:sigmoids} and \ref{subsubsec:nonthermal}).

Everyday, tons of observational data are ceaselessly poured to the ground
from SDO (launched in February 2010).
They include photospheric intensitygram,
Dopplergram, (vector) magnetogram, and (E)UV images.
Its constant full-disk observation enables us
to statistically investigate the evolution of ARs
from appearance to eventual flare eruption with unprecedented details.
Together with EIS and XRT,
the multi-filter (multi-temperature) observation of AIA
provided the thermal diagnostics of ARs
such as DEM inversions (Sect.~\ref{subsubsec:sigmoids}).
The steady supply of vector magnetogram by HMI revealed
the rapid changes of not only the LOS field but also the transverse field
in time scales of down to $\sim 10$ minutes (Sect.~\ref{subsec:vector}).
Several new attempts to utilize vector data have started.
For instance, the series of vector magnetograms are used in data-driven simulations
to sequentially update the boundary condition of coronal field models
(Sect.~\ref{subsubsec:num_data_driven}).
Various photospheric parameters calculated from the vector data
are used for predicting the flares and CMEs
(see discussion in Sect.~\ref{subsubsec:dis_prediction}).

Thanks to the high spatial resolution with the 1.6-m aperture
and the longer duty cycle,
BBSO/GST (scientific observation initiated in January 2009) has played a key role
in obtaining insights into the rapid changes of photospheric
(high-$\beta$)
fields in response to dynamical evolution of coronal
(low-$\beta$)
fields during the course of flares and CMEs (Sect.~\ref{sec:change}).
The most important science outputs made by BBSO/GST
related to the flare-AR science include
(1) the detailed structure, development, and destabilization of a flux rope,
(2) the sudden flare-induced rotation of sunspots and evolution of photospheric flow fields,
and (3) the tiny and transient flare precursors in the lower atmosphere.
Through these discoveries,
now we know that the answer to the ``tail wags the dog'' problem,
i.e., whether the coronal eruption can cause changes in the photospheric field,
is yes.

The advancement of instruments
has also motivated the development of numerical modeling.
For instance,
the long-term monitoring of flare-productive ARs
by Hinode and SDO from birth to eruption
inspired the flux emergence models
and gave a clue to the formation mechanisms of $\delta$-spots
(e.g., NOAA 11158 in February 2011: Sect.~\ref{subsubsec:multibuoyant}).
Fine-scale flare-triggering fields
and rapid magnetic changes during the flares,
which are observable only with advanced instruments,
have been compared with the results of the flare simulations
(Sects.~\ref{subsec:num_fc} and \ref{subsec:change_num}).
Filtergram images of various wavelengths by XRT and AIA
provide the means to diagnose the coronal fields
(e.g., XRT image and NLFFF extrapolation of sigmoids:
Sect.~\ref{subsubsec:sigmoids}).
All of these results underscore the importance
of direct comparison of observation and modeling
in unraveling the formation and evolution of flare-producing ARs.

\subsection{From birth to eruption}
\label{subsec:sum_story}

In this subsection, we summarize some of the key aspects
related to the genesis of flare-producing ARs and eventual energy release,
which have been uncovered by the observational and theoretical studies
presented in this review article.

(1) Subsurface evolution:
The dynamo-generated toroidal flux loops start rising in the convection zone
(Sect.~\ref{subsec:fe}).
Subject to the background turbulent convection,
some of them may lose a simple $\Omega$-shape
and deform into a helical structure,
a top-dent configuration,
bifurcated multiple branches,
or collide with other flux systems
(Sect.~\ref{subsec:num_fe}).
Through these processes, the rising flux systems gain non-potentiality
that is represented by free magnetic energy and magnetic helicity.

(2) Formation of $\delta$-spots:
On their appearance in the photosphere,
some of these
rising flux loops form $\delta$-sunspots,
in which umbrae of positive and negative polarities
are so close to share a common penumbra
(Sect.~\ref{subsec:spots}).
Most of the $\delta$-spots are generated by multiple emerging loops
rather than a single $\Omega$-loop
and the diversity of polarity layout stems
from the difference in the subsurface history,
but strong flares also emanate from non-$\delta$ sunspots
such as the Inter-AR case
(Sect.~\ref{subsec:deltaspots}).

(3) Development of flaring PIL and photospheric features:
Due to shearing and converging motion,
the PIL between the opposite polarities obtains a strong transverse field
with high gradient and shear
(Sect.~\ref{subsec:photo}).
This is the outcome of the Lorentz force,
and this force also causes the rotational motion of sunspots
(Sect.~\ref{subsec:num_fe}).

(4) Formation of flux rope: The coronal fields lying above the PIL become sheared
in sync with the photospheric driving,
cancel against each other, and form a magnetic flux rope.
This helical structure is observed as a sigmoid in soft X-rays
and as a filament (prominence) in H$\alpha$
(Sects.~\ref{subsec:atm} and \ref{subsec:num_fc}).

(5) Flare occurrence and CME eruption:
When the energy is sufficiently accumulated,
the solar flare is eventually initiated
(Sect.~\ref{subsec:flares}).
The flux rope becomes destabilized and erupts,
often as a CME into the interplanetary space,
leaving behind a variety of remarkable observational features on the Sun.
The drastic evolution of coronal fields causes rapid and profound changes
in magnetic and flow fields even in the photosphere
(Sect.~\ref{sec:change}).
If the confinement of the overlying arcade in an AR is too strong, however,
the flux rope may not develop into a CME.

As is obvious from the fact that
helical structures are seen in many parts in the story above,
the whole process of AR formation, flare eruption, and CME propagation
appears to be, overall,
the large-scale transport of magnetic helicity and energy
from the solar interior all the way to outer space
\citep{1996SoPh..167..217L,2002ESASP.505...35L,2007AdSpR..39.1674D}.
In this sense, the formation of $\delta$-spots,
where abundant evidence of non-potentiality is observed,
is accepted as a natural consequence
of the helicity that is delivered from the interior.

\subsection{Key observational features and quantities}
\label{subsec:sum_param}

In the long history of observation of ARs producing strong flares and CMEs,
various features have been investigated.
Perhaps these features can be summarized into three important factors,
which are (a) the size, (b) complexity, and (c) evolution.
Given the large magnetic energy accumulated in the ARs,
it is reasonable that these ARs are larger in spot area,
or naturally in total magnetic flux.
However, as we saw in Sect.~\ref{subsec:flares},
the largest spot in history, RGO 14886, was not flare active,
probably because this AR had a simple bipolar (i.e., potential) magnetic field.
To increase free magnetic energy that is released through flare eruptions,
ARs need to contain morphological and magnetic complexity,
which is manifested as the dispersed polarities (i.e., $\gamma$-spots),
strong-field, strong-gradient, highly-sheared PILs in $\delta$-spots,
magnetic tongues, flux ropes, sigmoids, etc.
These complex structures manifest
during the course of AR evolution,
observed as flux emergence of various scales,
shearing motion on both sides of a PIL,
and rotational motion of the sunspots.
Of course, such evolutionary processes may serve
as a trigger of eventual flare eruption.

\begin{table}
\caption{Some selected parameters in the literature
    that address the productivity of X-class flares.}
\label{tab:xflare}
\footnotesize
\begin{tabular}{lll}
\hline\noalign{\smallskip}
Parameter & Production of X-class flares & Reference \\
\noalign{\smallskip}\hline\noalign{\smallskip}
Spot area & 40\% of $\geq$1000 MSH $\beta\gamma\delta$-spots & \citet{2000ApJ...540..583S} \\
PIL total unsigned flux ($R$-value) & 20\% of $\log{(R)}=5.0$ & \citet{2007ApJ...655L.117S}\\
& (within the next 24 hours) & \\
Fractal dimension & $\geq 1.25$ & \citet{2005ApJ...631..628M}\\
Power-law index & $>2.0$ & \citet{2005ApJ...629.1141A}\\
Peak helicity injection rate & $\geq 6\times 10^{36}\ {\rm Mx}^{2}\ {\rm s}^{-1}$ & \citet{2007ApJ...671..955L}\\
Total non-neutralized current & $\geq 4.6\times 10^{12}\ {\rm A}$ & \citet{2017SoPh..292..159K}\\
Maximum non-neutralized current & $\geq 8\times 10^{11}\ {\rm A}$ & \citet{2017SoPh..292..159K}\\
Normalized helicity gradient variance & 1.13 (1 day before the flare) & \citet{2010ApJ...710L.121R} \\
\noalign{\smallskip}\hline
\end{tabular}
\end{table}

As we have seen in many parts in this review,
there is a multitude of statistical investigations
that reveal the quantitative differences
between flaring and quiescent ARs.
In Table~\ref{tab:xflare},
we pick up several parameters from the literature
that are suggested to differentiate (and may subsequently predict) X-class flares.

One may notice from this table and other references in this article that
many of the variables that have been investigated so far
are snapshot parameters,
i.e., those derived from observation at a single moment.
However, since it is the AR evolution that drives the flaring activities,
we need to understand the importance of dynamic parameters,
i.e., those that describe the temporal change of magnetic fields.
One of the most striking examples is the very fast flux emergence
in the super-flaring AR NOAA 12673 (Fig.~\ref{fig:yang2017}).
\citet{2017RNAAS...1a..24S} showed that
the flux growth rate
(i.e., time derivative of unsigned total magnetic flux)
in this AR was greater than any values reported in the literature,
and its X9.3 flare occurred a couple of days
after this remarkable emergence was detected.
Therefore,
such time derivative quantities might be key
to predict flares and CMEs
\citep[Sect.~\ref{subsubsec:dis_prediction}; see also][]{2003ApJ...595.1296L,2007ApJ...656.1173L}.

\section{Discussion}
\label{sec:discussion}

Despite the remarkable progress made to date,
many outstanding questions remain.
However, some of them will be answered
if observational and numerical techniques are improved more
in the near future.
In this section,
we list some of the important questions
and discuss the possibilities
to utilize our knowledge of flare-productive ARs in related science fields.

\subsection{Outstanding questions and future perspective}
\label{subsec:dis_questions}

Observationally, we still do not have a ``visual'' image
of the subsurface emerging flux
and thus we cannot establish whether the complex 3D configuration of flaring ARs
deduced from the surface evolution is real or not.
In a statistical sense, on average,
these ARs show enhanced vorticity before they cause flare eruptions
(Sect.~\ref{subsubsec:seismology}).
However, we still do not have robust methods of imaging the rising flux
because the (local) helioseismic probing is hampered
by the fast emergence and the low signal-to-noise ratio.
The existence of strong flux may not be treated as a small perturbation,
which is assumed when solving the linear inverse problem in seismology.
Advancement in helioseismology techniques,
probably with the support of numerical modeling,
is desired to overcome this difficulty.

Turbulent convection plays a crucial role
in producing the morphological and magnetic complexity of these ARs.
The generation of $\Omega$-loops from the magnetic wreath
in the global anelastic simulations begins to establish the concept of the ``spot-dynamo''
\citep[Fig.~\ref{fig:toriumi2012}: see][]{2013ApJ...762...73N,2015SSRv..196..101B}.
However, due to the limitation of the anelastic approximation,
it is difficult to trace the story
after the flux loops pass through the uppermost convection zone
(about $-20\ {\rm Mm}$ and upward).
Compressible simulations that enable access to (very close to) the solar surface,
such as by \citet{2014ApJ...786...24H},
may reveal the dynamical interaction between the magnetic field and turbulent convection
in much greater detail.
The genesis of magnetic helicity,
namely, the twist and writhe of emerging flux
(observed in the form of magnetic shear, spot rotations, magnetic tongues, sigmoids, etc.:
Sect.~\ref{sec:longterm}),
is still a big mystery \citep{1999GMS...111...93L}.
Regarding the formation of flaring ARs,
it is also an interesting question
how and why super strong transverse field appears at the PIL in a $\delta$-spot
instead of at the core of sunspot umbra.
These issues may be solved by an advancement of numerical models.

There has been a dichotomy of theory
whether a magnetic flux rope is created well before the eruption
or at the very moment of it \citep[see, e.g.,][p. 266]{2006SSRv..123..251F}.
Thanks to the NLFFF, data-constrained, and data-driven models,
now the flux rope appears to be created from before eruption,
at least in the flare-productive ARs,
through the continued shearing along the PIL.
These numerical methods may be advanced even more
and provide a conclusive answer.
For example,
vector field measurements in higher atmospheric layers
may realize more accurate extrapolations.
In the current force-free methods,
it is assumed that the input photospheric vector field is in force-free
(Sect.~\ref{subsubsec:num_data_extrapolation}).
However, this is apparently not the case
because the photosphere is in the realm of high-$\beta$ plasma
(i.e., the photospheric plasma is largely affected
by the non-magnetic forces such as pressure gradient),
which requires a smoothing of the photospheric vector field
before the extrapolation is applied.
Chromospheric low-$\beta$ fields,
obtained by future instruments such as
the Daniel K. Inouye Solar Telescope (DKIST),
may give better boundary conditions for the force-free extrapolations,
data-constrained and data-driven models.
Moreover, magnetic information at multiple altitudes
allows us to calculate the partial derivatives in the vertical direction
(i.e., $\partial B_{x}/\partial z$ and $\partial B_{y}/\partial z$)
and may provide better estimates of
the total (vector) current density,
horizontal velocity, electric field,
and Lorentz force density.

Stereoscopic monitoring of the Sun from multiple vantage points,
for instance by spacecrafts around the Earth and at the Lagrangian L5 point
or by off-ecliptic explorers like Solar Orbiter,
is helpful in various aspects
\citep{2005AdSpR..35...65A,2015AdSpR..55.2745S,2018FrASS...5...32G}.
Apart from the early warning of space weather events
like Earth-directed CMEs and violent ARs beyond the east limb,
it may help probing the deeper interior with local helioseismology,
resolving the ambiguity of magnetic measurements,
and assessing the topology of entangled coronal fields
(see results from STEREO).
With advanced spectroscopic and imaging instruments,
atmospheric evolution such as build-up and eruption of flux ropes
and non-thermal broadening of EUV lines
(Sect.~\ref{subsec:atm})
may be revealed in further detail.
All these new capabilities will greatly improve our understanding
of the nature of flare-productive ARs.

The detection of flare-related activities from ground-based large-aperture telescopes
has been, in most cases, done by GST (Sect.~\ref{sec:change}).
To better understand the fine-scale dynamics in AR build-up and flare eruption,
it is necessary to increase the detection rate of these events
by enhancing the observing time.
One possible idea is to organize an international network
of high-resolution telescopes,
such as DKIST (4-m aperture in Maui),
New Vacuum Solar Telescope (NVST; 1-m aperture in Yunnan),
Swedish Solar Telescope (SST; 1-m aperture in La Palma),
GREGOR (1.5-m aperture in Tenerife),
and European Solar Telescope (EST; 4-m aperture under contemplation),
and conduct a long-running monitoring of a target AR.
Several key observations of dynamic activities in flaring ARs were already made
with NVST \citep{2016NatCo...711837X,2017ApJ...840L..23X}
and SST \citep{2016ApJ...819..157G,2018A&A...609A..14R}.
Therefore, the combination of these stations may open up unexplored discovery space
and provide insights into the evolution of small-scale magnetic features
in the very long run (days to weeks).

\subsection{Broader impacts on related science fields}

\subsubsection{Prediction and forecasting of solar flares and CMEs}
\label{subsubsec:dis_prediction}

Probably one of the most practical applications
of the knowledge of flaring ARs we have acquired
is the prediction of flares and CMEs.
Statistical investigations of various events
that introduce parameters such as those in Table~\ref{tab:xflare}
characterized the flare-productive ARs.
In the last decades,
the knowledge-based flare predictions using these quantities
have been significantly developed.

\begin{table}
\caption{13 flare-predictive parameters derived from the SDO/HMI vector data \citep{2015ApJ...798..135B}. $F$-score indicates the scoring of the parameter.}
\label{tab:sharp}
\footnotesize
\begin{tabular}{lll}
\hline\noalign{\smallskip}
Description & Formula & $F$-Score \\
\noalign{\smallskip}\hline\noalign{\smallskip}
Total unsigned current helicity &
$H_{c_{\rm total}}\propto \sum |B_{z}\cdot J_{z}|$ &
3560 \\
Total magnitude of Lorentz force &
$F\propto \sum B^{2}$ &
3051 \\
Total photospheric magnetic free energy density &
$\rho_{\rm tot}\propto \sum ({\vec B}^{\rm Obs}-{\vec B}^{\rm Pot})^{2}dA$ &
2996 \\
Total unsigned vertical current &
$J_{z_{\rm total}}=\sum |J_{z}|dA$ &
2733 \\
Absolute value of the net current helicity &
$H_{c_{\rm abs}}\propto \left| \sum B_{z}\cdot J_{z}\right|$ &
2618 \\
Sum of the modulus of the net current per polarity &
$J_{z_{\rm sum}}\propto \left|\sum^{B_{z}^{+}} J_{z}dA\right|+\left|\sum^{B_{z}^{-}} J_{z}dA\right|$ &
2448 \\
Total unsigned flux &
$\Phi=\sum |B_{z}|dA$ &
2437 \\
Area of strong field pixels in the active region &
${\rm Area}=\sum {\rm Pixels}$ &
2047 \\
Sum of $z$-component of Lorentz force &
$F_{z}\propto \sum(B_{x}^{2}+B_{y}^{2}-B_{z}^{2})dA$ &
1371 \\
Mean photospheric magnetic free energy &
$\overline{\rho}\propto \frac{1}{N}\sum ({\vec B}^{\rm Obs}-{\vec B}^{\rm Pot})^{2}$ &
1064 \\
Sum of flux near polarity inversion line &
$\Phi=\sum|B_{LoS}|dA$ within $R$ mask &
1057 \\
Sum of $z$-component of normalized Lorentz force &
$\delta F_{z}\propto \frac{\sum(B_{x}^{2}+B_{y}^{2}-B_{z}^{2})}{\sum B^{2}}$ &
864.1 \\
Fraction of Area with shear $> 45^{\circ}$ &
Area with shear $> 45^{\circ}$ / total area &
740.8 \\
\noalign{\smallskip}\hline
\end{tabular}
\end{table}

Nowadays,
these methods employ machine-learning algorithms.
For example, \citet{2015ApJ...798..135B} extracted various photospheric parameters
from the SDO/HMI vector magnetograms for individual ARs,
trained the machine,
and obtained a good predictive performance for $\geq$M1.0 flares.
The parameters investigated are listed in Table~\ref{tab:sharp},
which are basically the previously suggested variables
\citep{2003ApJ...595.1296L,2012SoPh..277...59F,2007ApJ...655L.117S},
It should be noted that most of them are ``extensive,''
where a given parameter increases with AR size
\citep{2007ApJ...665.1460T,2009ApJ...705..821W,2015ApJ...804L..28S,2017ApJ...850...39T}.

Many of the parameters listed in Table~\ref{tab:sharp} are, again, snapshot ones
(see Sect.~\ref{subsec:sum_param}),
and the inclusion of dynamic parameters may be helpful in flare predictions
\citep{2003ApJ...595.1296L,2007ApJ...656.1173L}.
For instance,
to the flare-predictive parameters in Table~\ref{tab:sharp},
\citet{2017ApJ...835..156N} added additional information
that indicates flare history and chromospheric pre-flare brightening
and also time derivatives of various observables.
By training the machine with three different algorithms,
the authors successfully obtained a prediction score higher than
that of \citet{2015ApJ...798..135B}.
This study clearly highlights the usage of dynamic parameters.

However, it is worth noting that increasing
the number of parameters does not necessarily improve the prediction performance.
In fact, \citet{2007ApJ...656.1173L} and \citet{2015ApJ...798..135B} found that
there was little value to add parameters more than a few.
This is because the model with many parameters
(i.e. large degrees of freedom) tends to overfit the training data
and, in that case,
the model may perform worse on the validation data.

Today,
while there remains a view that
the occurrence of flares is a ``stochastic'' process
\citep[e.g., the avalanche model by][]{1991ApJ...380L..89L}
and therefore the ``deterministic'' forecasting might be fundamentally impossible
\citep{2009AdSpR..43..739S},
the knowledge-based prediction is growing much more rapidly than ever before
\citep[e.g.,][]{2007SoPh..241..195Q,2009SpWea...7.6001C,2010ApJ...709..321Y,2013SoPh..283..157A,2015SpWea..13..778M,2016ApJ...821..127B,2017ApJ...843..104L,2018SoPh..293...48J,2018ApJ...856....7H,2018ApJ...858..113N}.
Together with the attempts to build up physics-based (i.e., modeling-based) algorithms
(Sects.~\ref{subsubsec:num_data_constrained} and \ref{subsubsec:num_data_driven}),
the recent development of this field may tell us that
the real-time space weather forecasting will come true
in the very near future.

\subsubsection{Investigating extreme space-weather events in history}

The strongest flare activity ever observed with an estimated GOES class of $\sim$X45
is the Carrington flare in September 1859 (see Sect.~\ref{subsec:flares}).
To understand the mechanisms and trends
of such extreme space weather events that may affect the Earth
\citep[like the occurrence frequency;][]{2012JGRA..117.8103S,2012SpWea..10.2012R,2016JSWSC...6A..23C},
it is crucial to increase the sample number
by surveying the greatest events in history.
However, often these events do not have
observations of sufficient data quality for scientific analysis.
In the modern age,
the data analyzed are often digitized intensity images of various wavelengths
and LOS or vector magnetograms.
For the historical events, however,
available records can be photographic plates or perhaps only sunspot drawings.
But still, there are several ways to elucidate
how and why the strong events occurred.

For instance,
there are several attempts to achieve magnetic information
from historical sunspot drawings. 
For the great storm of May 1921
\citep{2001JASTP..63..523S,2006AdSpR..38..188K},
\citet{2015AnGeo..33..109L} reconstructed ``magnetograms''
by applying their torus model
to the daily Mount Wilson drawings of sunspot magnetic fields
and studied the development of the target AR.
They found that
spot rotations and flux emergence occurred in the AR.
They pointed out the close association between
the drastic spot evolutions and the eventual magnetic storm.

Another approach is to reconstruct vector magnetogram from existing LOS magnetogram
by applying one of the machine-learning methods
called transfer learning \citep{pan2010}.
One of the purposes of this method is to convert some source data to target data
and, with this method,
one may use SDO/HMI vector magnetograms (for Cycle 24)
and SOHO/MDI LOS magnetograms (for Cycle 23)
as the source data and target data, respectively,
and reproduce ``vector magnetograms'' for ARs of Cycle 23.
Because there were many more stronger flares in Cycle 23,
such vector data may help investigate
the driving mechanisms of extreme events.

In many respects,
studying historical records
is beneficial in understanding the activity of the Sun.
It may tell us how strong events the Sun can produce,
how frequently these events occur,
and how they make an impact on our magnetic circumstances.
Although it is not easy to derive useful information from such records,
we can still take advantage of the current knowledge of flaring ARs.
Attempts to examine drastic spot evolution
and reconstruct magnetograms
may give us clues to understand the nature of severe space-weather events.

\subsubsection{Connection with stellar flares and CMEs}

The production of stellar flares and CMEs are now of great importance,
not only from the viewpoint of mass and angular momentum loss rates
especially of the active young stars \citep[e.g.,][]{2012ApJ...760....9A},
but also in the search for habitability of orbiting exoplanets.
The type II radio burst,
which is believed to be produced
by MHD shocks in front of the CME
propagating into the interplanetary space \citep{1995SSRv...72..243G},
is currently the best way
of detecting the stellar CMEs \citep{2017IAUS..328..243O}.

In this regard,
\citet{2018ApJ...856...39C,2018ApJ...862..113C} attempted to detect type II bursts
on nearby, magnetically-active, well-characterized M dwarf star EQ Peg.
During 20 hours of simultaneous radio and optical observation,
they detected four optical flare signatures
but no radio features identifiable as type II bursts.
Two radio bursts were found during the additional 44 hours of radio-only observation.
However, their characteristics were not consistent with
that of type II events.
From the statistics of the solar flares and CMEs \citep{2006ApJ...650L.143Y},
all the four detected flares are empirically predicted to have associated CMEs,
but none was detected at radio wavelengths in this data set.

As an independent analysis,
\citet{2014MNRAS.443..898L} searched for flares and CMEs
on 28 young late-type (K to M) stars in the open cluster Blanco-1.
From the five hour observation,
they found four H$\alpha$ flares from three M stars and one K star.
Interestingly, however,
they also did not detect any clear indications of CMEs
such as spectral asymmetries of the H$\alpha$ line
caused by large Doppler velocities.

Although we cannot rule out the possibility that
the signals were less than the detection sensitivity,
it is worth discussing the reason of the ``failed'' eruptions
by employing the knowledge of flare-productive ARs of the Sun.
As we saw, for instance, in Sects.~\ref{subsec:flares} and \ref{subsubsec:unified},
the flare eruption tends to fail
when the overlying coronal loops are strong and slowly decaying over height
\citep{2017ApJ...843L...9W,2018ApJ...860...58V,2018ApJ...864..138J}.
Observations and numerical modeling of flaring ARs show that,
for the failed events,
a magnetic flux rope is often trapped in the AR core
and does not have an access to open fields
\citep{2017ApJ...834...56T,2017ApJ...850...39T,2018ApJ...861..131D}.
As the Zeeman Doppler Imaging by \citet{2008MNRAS.390..567M} suggests,
active M dwarfs tend to be covered by strong magnetic patches
over the entire stellar surface.
Due to the strong confinement by coronal loops extending from these patches,
we may expect less successful CME eruptions
even if energetic stellar flares occur \citep{2016IAUS..320..196D}.
The confinement may also be due to the strong large-scale dipolar field,
as numerically modeled by \citet{2018ApJ...862...93A}.

Thanks to the advancement of observational capabilities,
many more ``superflares'' are now detected
on solar-like G-type stars
\citep{2012Natur.485..478M,2013ApJS..209....5S}.
Indications of huge starspots with large magnetic energy are seen
in these stars \citep[e.g.,][]{2013ApJ...771..127N}.
By conducting spectroscopic and polarimetric observations
on the properties of superflares and starspots,
and by comparing them with numerical models of solar-stellar flares and ARs,
the production mechanisms,
similarities and diversities,
and their stellar space-weather impacts
may be revealed in detail in the near future.

\begin{acknowledgements}
S.T. benefited from fruitful discussions held in the series of Flux Emergence Workshops, the Project for Solar-Terrestrial Environment Prediction (PSTEP), the solar-stellar team sponsored by the International Space Science Institute (ISSI), and Nagoya University ISEE/CICR International Workshop on Data-driven Models.
S.T. would like to thank Mark C.M. Cheung, Yuhong Fan, George H. Fisher, Manuel G\"{u}del, Hiroki Kurokawa, Mark G. Linton, Rachel A. Osten, and Takashi Sekii, for providing valuable comments, discussions, and continuous supports.
H.W. thanks Chang Liu for his contribution in writing the 2015 RAA review paper that prepared some knowledge for this review.
We thank the anonymous referees and the editor Carolus J. Schrijver
for very helpful comments.
Data are courtesy of the science teams of Hinode, SOHO, and SDO.
Hinode is a Japanese mission developed and launched by ISAS/JAXA, with NAOJ as domestic partner and NASA and STFC (UK) as international partners. It is operated by these agencies in cooperation with ESA and NSC (Norway).
SOHO is a project of international cooperation between ESA and NASA.
HMI and AIA are instruments on board SDO, a mission for NASA's Living With a Star program.
We thank Sian Prosser, the Royal Astronomical Society, for providing the sunspot drawing by Carrington.
The work was supported by JSPS KAKENHI Grant Numbers JP16K17671 (PI: S. Toriumi) and JP15H05814 (PI: K. Ichimoto)
and by the NINS program for cross-disciplinary study (Grant Numbers 01321802 and 01311904) on Turbulence, Transport, and Heating Dynamics in Laboratory and Astrophysical Plasmas: ``SoLaBo-X''.
H.W. acknowledges the support of US NSF under grant AGS-1821294
and US NASA under grants 80NSSC17K0016, 80NSSC18K1133, and 80NSSC18K1705.
\end{acknowledgements}

\appendix
\normalsize 
\section{Appendix: Original advocates of the kink instability}
\label{app:kink}

Little has been known about who first proposed the helical kink instability
as the formation mechanism for the $\delta$-spots.
Almost certainly, it is \citet{1996ApJ...469..954L}
who for the first time investigated this instability
in the context of $\delta$-spot formation.
However, the authors did not clearly claim in their paper
that they were the first to propose this idea.
Before that, from the observed proper motions of $\delta$-spots,
\citet{1991SoPh..136..133T} suggested in his posthumous publication
that the rising of a knotted twisted flux tube
creates the $\delta$-spots (twisted knot model).
Although his illustration,
adopted as Fig.~\ref{fig:tanaka1991}(a) in this review,
is highly evocative of the kink instability,
the term ``kink instability'' was not used at all in his paper.
It is now almost impossible to find out
whether Katsuo Tanaka and his longtime collaborator Harold Zirin
held an idea of the kink instability at that time
because both of them are deceased.
Here we show a brief history
between \citet{1991SoPh..136..133T} and \citet{1996ApJ...469..954L},
which George H. Fisher, Mark G. Linton, and Yuhong Fan gave to us.

While Fisher was working on the thin flux tube model
(Sect.~\ref{subsubsec:fe_interior_theory}) with Fan in the University of Hawaii,
he conceived an idea to add magnetic twist to the thin flux tube,
inspired by \citet{1978ApJ...220..675M},
who proposed magnetic twist as a driver of the flux rope eruption.
Although the concept of \citet{1978ApJ...220..675M} was
based more on the lateral kink instability,
in which the displacement of a twisted flux tube is
in a single perpendicular direction (i.e., an $\Omega$-loop in a 2D plane),
Fisher had the misconception that the driving mechanism
was the helical kink mode,
where the direction of the displacement rotates along the tube axis
\citep[see][Sect. 7.5.3 for the two modes]{2014masu.book.....P}.
When Fisher and Linton started working on this issue
after Fisher moved to the University of California, Berkeley in 1992,
Dana W. Longcope showed that the thin flux tube model cannot represent
the helical kink instability because this model, in principle, assumes
that any physical value is uniform over the tube's cross section
and thus does not include internal motion within the tube. 

In the meantime,
they studied the textbook of \citet{1988assu.book.....Z},
especially on the ``island $\delta$'' sunspot,\footnote{This part is based on \citet{1987SoPh..113..267Z}.}
as well as the seminal work by \citet{1991SoPh..136..133T}.
These two publications stimulated them to propose the helical kink instability
as the origin for the island $\delta$-spot.
Over 1994 and 1995,
Linton performed energy principle analysis on the instability with Longcope
and, eventually, the work resulted in \citet{1996ApJ...469..954L}.
Moreover, the presentation by Linton et al.,
probably in the 26th AAS/SPD meeting in 1995,
evolved into a collaboration with Russell B. Dahlburg on MHD simulations,
which was published later as \citet{1998ApJ...507..404L}.

Therefore, it is not easy to narrow down the originator of the idea to one.
However, the above story should be a good example of
a coincidental misconception serendipitously producing fruit.

\bibliographystyle{spbasic}      
\bibliography{lrsp_flarear}   

\begin{thebibliography}{639}
\bibitem[{{Carrington}(1858)}]{1858MNRAS..19....1C}
{Carrington} RC (1858) {On the Distribution of the Solar Spots in Latitudes
  since the Beginning of the Year 1854, with a Map}. \mnras 19:1--3,
  \doi{10.1093/mnras/19.1.1}
\end{thebibliography}

\end{document}